\documentclass[11pt]{article}

\usepackage[preprint]{acl}
 
\usepackage{times}
\usepackage{latexsym}

\usepackage[T1]{fontenc}

\usepackage[utf8]{inputenc}

\usepackage{microtype}

\usepackage{inconsolata}

\usepackage{graphicx}

\usepackage{booktabs}

\usepackage{amsmath}
\usepackage{amssymb}
\usepackage{mathtools}
\usepackage{amsthm}

\usepackage{multirow, caption, subcaption, wasysym, nicefrac, todonotes, float, capt-of, array, cleveref}

\title{\textsc{VeriTaS}: The First Dynamic Benchmark for\\Multimodal Automated Fact-Checking}

\author{Mark Rothermel \hspace{0.5cm} Marcus Kornmann \hspace{0.5cm} Marcus Rohrbach \hspace{0.5cm} Anna Rohrbach \\
  Multimodal AI Lab \\ Technical University of Darmstadt \\ hessian.AI \\
  \small{
    \textbf{Correspondence:} \href{mailto:mark.rothermel@tu-darmstadt.de}{mark.rothermel@tu-darmstadt.de}
  }}

\usepackage{xspace, makecell}

\newcommand{\method}{\textsc{VeriTaS}\xspace}

\newcommand{\averitec}{\textsc{AVeriTeC}\xspace}

\newcommand{\qwen}{\textsc{Qwen~3.5~(397B)}\xspace}
\newcommand{\opus}{\textsc{Claude~Opus~4.6}\xspace}
\newcommand{\gemma}{\textsc{Gemma~4~(31B)}\xspace}

\newcommand{\gpt}{\textsc{GPT-5.2}\xspace}
\newcommand{\gptmini}{\textsc{GPT-5 mini}\xspace}
\newcommand{\gptnano}{\textsc{GPT-5 nano}\xspace}
\newcommand{\gptfour}{\textsc{GPT-4o}\xspace}

\newcommand{\geminipro}{\textsc{Gemini 2.5 Pro}\xspace}
\newcommand{\geminiflash}{\textsc{Gemini 2.5 Flash}\xspace}

\newcommand{\claude}{\textsc{Claude Sonnet 4.5}\xspace}

\newcommand{\llama}{\textsc{Llama 4 Maverick}\xspace}

\definecolor{2b}{HTML}{0083CC}
\definecolor{8b}{HTML}{EC6500}

\definecolor{green_supported}{HTML}{09C479}
\definecolor{red_refuted}{HTML}{E03440}
\definecolor{yellow_conflicting}{HTML}{F5A300}
\definecolor{gray_nei}{HTML}{434343}

\newcommand{\highlight}[1]{\textbf{\textcolor{8b}{#1}}}

\newcommand{\myparagraph}[1]{\noindent\textbf{#1}}

\newcolumntype{C}[1]{>{\centering\arraybackslash\hspace{0pt}}p{#1}}

\usepackage{adjustbox, array}
\newcolumntype{R}[2]{%
    >{\adjustbox{angle=#1,lap=\width-(#2)}\bgroup}%
    c%
    <{\egroup}%
}
\newcommand*\rot{\multicolumn{1}{R{30}{1em}}}

\newcommand{\iconsupported}{\raisebox{-0.13em}{\includegraphics[height=0.9em]{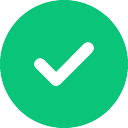}}}
\newcommand{\iconrefuted}{\raisebox{-0.13em}{\includegraphics[height=0.9em]{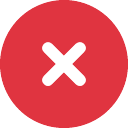}}}
\newcommand{\iconnei}{\raisebox{-0.13em}{\includegraphics[height=0.9em]{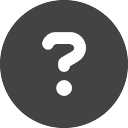}}}

\newcommand{\nei}{\iconnei\,\textcolor{gray_nei}{\small \textbf{\sffamily NEI}}\xspace}

\newcommand{\pristine}{\iconsupported\,\textcolor{green_supported}{\small \textbf{\sffamily Pristine}}\xspace}
\newcommand{\fabricated}{\iconrefuted\,\textcolor{red_refuted}{\small \textbf{\sffamily Fabricated}}\xspace}

\newcommand{\true}{\iconsupported\,\textcolor{green_supported}{\small \textbf{\sffamily True}}\xspace}
\newcommand{\false}{\iconrefuted\,\textcolor{red_refuted}{\small \textbf{\sffamily False}}\xspace}

\newcommand{\intact}{\iconsupported\,\textcolor{green_supported}{\small \textbf{\sffamily Intact}}\xspace}
\newcommand{\compromised}{\iconrefuted\,\textcolor{red_refuted}{\small \textbf{\sffamily Compromised}}\xspace}
\newcommand{\compr}{\iconrefuted\,\textcolor{red_refuted}{\small \textbf{\sffamily Compr.}}\xspace}

\newcommand{\positive}{\iconsupported\,\textcolor{green_supported}{\small \textbf{\sffamily Positive}}\xspace}
\newcommand{\negative}{\iconrefuted\,\textcolor{red_refuted}{\small \textbf{\sffamily Negative}}\xspace}

\usepackage{scalefnt}
\usepackage{algorithm}
\usepackage{algorithmic}
\DeclareTextFontCommand{\texttt}{\ttfamily \scalefont{0.93}}

\usepackage{calc}

\usepackage{tcolorbox}
\tcbset{
    graybox/.style={
        colback=gray!10,
        colframe=gray!50,
        arc=5mm,
        auto outer arc,
        left=2mm,
        right=2mm,
        top=2mm,
        bottom=2mm,
        boxrule=0.1mm,
        fontupper=\sffamily
    }
}

\begin{document}

\makeatletter
\g@addto@macro\@maketitle{
    \begin{figure}[H]
        \vspace{-2.5em}
        \setlength{\linewidth}{\textwidth}
        \setlength{\hsize}{\textwidth}
        \centering
        \makebox[\textwidth][c]{
            \includegraphics[width=1.05\textwidth]{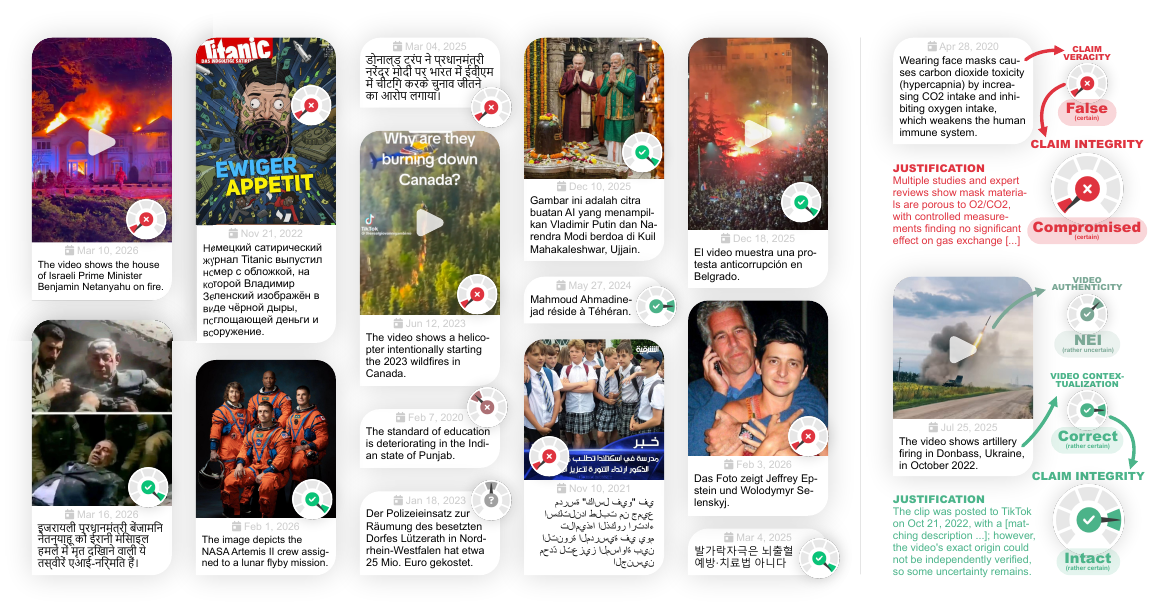}
        }
        \vspace{-2.5em}
        \caption{The \method benchmark: Example claims including media, claim date, and claim integrity score. The two claims on the right showcase lower-level annotations of media/claim properties used to infer the overall integrity. Each annotation contains a justification. New claims are added quarterly via an automated pipeline. 
        }
        \label{fig:teaser_figure}
    \end{figure}
}
\makeatother

\maketitle

\begin{abstract}
    The growing scale of online misinformation urgently demands Automated Fact-Checking (AFC). Existing benchmarks for evaluating AFC systems, however, are largely limited in terms of task scope, modalities, domain, language diversity, realism, or coverage of misinformation types. Critically, they are \emph{static}, thus subject to data leakage as their claims enter the pretraining corpora of LLMs. As a result, benchmark performance no longer reliably reflects the actual ability to verify claims.
We introduce \emph{Verified Theses and Statements (\method)}, the first \emph{dynamic} benchmark for multimodal AFC, designed to remain robust under ongoing large-scale pretraining of foundation models. \method currently comprises 25,000 real-world claims from 104 professional fact-checking organizations across 54 languages, covering textual and audiovisual content. Claims are added \emph{quarterly} via a fully automated seven-stage pipeline that normalizes claim formulation, retrieves original media, and maps heterogeneous expert verdicts to a novel, standardized, and disentangled scoring scheme with textual justifications.
Through human evaluation, we demonstrate that the automated annotations closely match human judgments.
We commit to updating \method in the future, establishing a leakage-resistant benchmark, supporting meaningful AFC evaluation in the era of rapidly evolving foundation models.
The code and data are publicly available under \href{https://veritas.mai.informatik.tu-darmstadt.de/}{veritas.mai.informatik.tu-darmstadt.de}.
\end{abstract}
\section{Introduction}
\label{sec:intro}

Mis- and disinformation lead the current short-term risks for global society \cite{elsner2025GlobalRisksReport}.
Humans perceive multimodal information as more credible than text alone \cite{newman2012NonprobativePhotographsWords}, making images and videos powerful tools for persuasion \cite{hameleers2020PicturePaintsThousand,greifeneder2020PsychologyFakeNews}. As a result, multimodal content achieves higher engagement and spreads faster, particularly in the context of misinformation \cite{li2020PictureWorthThousand,zannettou2018OriginsMemesMeans,wang2021UnderstandingUseFauxtography}. 
\citet{dufour2024AMMeBaLargeScaleSurvey} found that visual misinformation accounts for up to $80\%$ of fact-checked claims. At the same time, generative methods like \textsc{Nano Banana} \cite{google2025IntroducingNanoBanana} or \textsc{Sora 2} \cite{openai2025Sora2Here} evolve rapidly, producing realistic deepfakes, outpacing countermeasures. Manual fact-checking, however, is a very laborious task \cite{hassan2015QuestAutomateFactChecking}, which gives rise to robust \emph{multimodal Automated Fact-Checking (AFC)}.

To evaluate AFC methods, numerous benchmarks have been proposed. However, many lack full modality coverage, missing images or videos \cite{schlichtkrull2023AVeriTeCDatasetRealworld,cao2025AVerImaTeCDatasetAutomatic,geng2025M4FCMultimodalMultilingual}, restrict the task scope to isolated sub-problems \cite{papadopoulos2024VERITERobustBenchmark,tonglet2024ImageTellMe}, or rely heavily on synthetically generated claims \cite{xu2024M3AMultimodalMisinformation,xu2025MDAM3MisinformationDetection}, raising concerns about realism. Additionally, practically all AFC benchmarks model the task as a single-label classification problem, conflating orthogonal claim properties into coarse labels, often ignoring uncertainty and finer-grained properties, and treating severe confusions (\textit{True} vs.\ \textit{False}) equally to less severe ones (e.g., \textit{True} vs.\ \textit{Not Enough Information}).

More fundamentally, all existing AFC benchmarks are \emph{static} and therefore vulnerable to \emph{data leakage}. State-of-the-art AFC systems rely on large language models (LLMs) that are continually pretrained on public web data. Since benchmark claims originate from the same public sources, models may implicitly encode both claims and verdicts in their parametric knowledge, undermining meaningful evaluation. We empirically substantiate this effect in Fig.~\ref{fig:results_longitudinal}, demonstrating a performance drop for claims published after the models' knowledge cutoff date.

To address these limitations, we introduce \emph{Verified Theses and Statements (\method)}, the first \emph{dynamic} AFC benchmark. \method is extended quarterly with newly emerging real-world claims, effectively mitigating data leakage in future splits through fully automated data acquisition. It is also the first benchmark to combine real-world multimodal claims featuring both images and videos. Unlike the vast majority of previous work, \method also provides multilingual, open-domain, and balanced claim data with verdict annotations accompanied by textual justifications for the rulings, derived from expert ground truth. Fig.~\ref{fig:teaser_figure} shows example claims from \method.

Moreover, we contribute a novel rating scheme that (1) disentangles previously coarse-grained rulings into separate fine-grained claim and media properties, (2) incorporates uncertainty by modeling each of these properties on a scale from $-1$ to $1$ allowing to (3) employ Mean Squared Error (MSE) as an evaluation metric to reward near-correct predictions and strongly penalize flipped decisions.

Finally, we benchmark state-of-the-art multimodal AFC systems and strong LLM baselines, revealing substantial room for improvement. We made the benchmark data and pipeline code available 
at \href{https://veritas.mai.informatik.tu-darmstadt.de/}{veritas.mai.informatik.tu-darmstadt.de}
and commit to extending \method in each subsequent quarter until at least Q4 2028.


\section{Related Work}
\label{sec:related_work}

\begin{table*}[ht]
    \resizebox{\linewidth}{!}{
    \begin{tabular}{l|C{1.2cm}C{0.8cm}C{0.8cm}C{0.8cm}C{0.8cm}C{0.8cm}C{0.8cm}C{0.8cm}C{1.0cm}C{0.8cm}C{0.8cm}C{1.2cm}c}
        \toprule
        \textbf{Benchmark} & \rot{\textbf{Primary task}} & \rot{\textbf{Images}} & \rot{\textbf{Videos}} & \rot{\textbf{Dynamic}} & \rot{\textbf{Automated}} & \rot{\textbf{Real claims}} & \rot{\textbf{Misinfo coverage}} & \rot{\textbf{\# Languages}} & \rot{\textbf{Open-domain}} & \rot{\textbf{Justifications}} & \rot{\textbf{Balanced}} & \rot{\textbf{\# Instances}} & \textbf{Annotation} \\
        \midrule
        \textsc{M³A} \hfill \cite{xu2024M3AMultimodalMisinformation} 
        & CV & \checkmark & \checkmark & - & \checkmark & - & - & 1 & News & - & - & 7.3\,M & 4 Classes \\
        \textsc{MDAM³-DB} \hfill \cite{xu2025MDAM3MisinformationDetection} 
        & CV & \checkmark & \checkmark & - & \checkmark & - & - & 1 & News & - & - & 90\,K  & 6 Classes \\
        \textsc{XFacta} \hfill \cite{xiao2025XFactaContemporaryRealWorld} 
        & CV & \checkmark & - & \RIGHTcircle & - & \checkmark & \RIGHTcircle & 1 & SM & - & \checkmark & 2.4\,K & 4 Classes \\
        \textsc{MuMiN} \hfill \cite{nielsen2022MuMiNLargeScaleMultilingual} 
        & CV & \checkmark & - & - & \checkmark & \checkmark & \checkmark & 41 & \checkmark & - & - & 13\,K & 3 Classes \\
        \textsc{Khatiwada et al.} \hfill \cite{khatiwada2025MultimodalMultiLabelElectionContext} 
        & CV & \checkmark & - & - & \checkmark & \checkmark & \RIGHTcircle & 1 & SM & - & - & 77\,K & 5 Labels \\
        \textsc{ClaimReview2024+} \hfill \cite{braun2025DEFAMEDynamicEvidencebased}
        & CV & \checkmark & - & - & - & \checkmark & \checkmark & 1 & \checkmark & - & - & 300 & 4 Classes \\
        \textsc{M4FC} \hfill \cite{geng2025M4FCMultimodalMultilingual} 
        & CV & \checkmark & - & - & - & \checkmark & \checkmark & 10 & \checkmark & - & \checkmark & 7.0\,K & 4 Classes \\
        \textsc{RealFactBench} \hfill \cite{yang2025RealFactBenchBenchmarkEvaluating} 
        & CV & \checkmark & - & - & - & \checkmark & \checkmark & 1 & \checkmark & - & - & 6.0\,K & 2 Classes \\
        \textsc{MR$^2$} \hfill \cite{hu2023MR2BenchmarkMultimodal} 
        & CV & \checkmark & - & - & - & \checkmark & \RIGHTcircle & 2 & \checkmark & - & - & 15\,K & 3 Classes \\
        \textsc{VLDBench} \hfill \cite{raza2025VLDBenchVisionLanguage} 
        & CV & \checkmark & - & - & - & \checkmark & \RIGHTcircle & 1 & News & - & \checkmark & 63\,K & 2 Classes \\
        \textsc{AVerImaTeC} \hfill \cite{cao2025AVerImaTeCDatasetAutomatic} 
        & CV & \checkmark & - & - & - & \checkmark & - & 1 & \checkmark & \checkmark & - & 1.3\,K & 4 Classes \\
        \textsc{FACTIFY} \hfill \cite{mishra2022FACTIFYMultiModalFact} 
        & CV & \checkmark & - & - & - & \RIGHTcircle & \RIGHTcircle & 1 & \checkmark & - & \checkmark & 50\,K & 5 Classes \\
        \textsc{FACTIFY 3M} \hfill \cite{chakraborty2023FACTIFY3MBenchmarkMultimodala} 
        & CV & \checkmark & - & - & - & \RIGHTcircle & \RIGHTcircle & 1 & \checkmark & - & - & 3\,M & 5 Classes \\
        \textsc{MMFakeBench} \hfill \cite{liu2025MMFakeBenchMixedSourceMultimodal} 
        & CV & \checkmark & - & - & - & - & \RIGHTcircle & 1 & \checkmark & - & - & 11\,K & 3 Classes \\
        \textsc{OmniFake} \hfill \cite{li2025UnifiedMultimodalMisinformation} 
        & CV & \checkmark & - & - & - & - & - & 1 & SM & - & - & 127\,K & 3 Classes \\
        \textsc{TRUE} \hfill \cite{niu2025PioneeringExplainableVideo} 
        & CV & - & \checkmark & - & \checkmark & \checkmark & \checkmark & 1 & \checkmark & \checkmark & - & 2.9\,K & 2 Classes \\
        \textsc{GroundLie360} \hfill \cite{yang2025NewDatasetBenchmark}
        & CV & - & \checkmark & - & - & \checkmark & \checkmark & 1 & \checkmark & - & \checkmark & 2.0\,K & 6 Classes \\
        \midrule
        \textsc{MMOOC} \hfill \cite{xu2024MMOOCMultimodalMisinformation} 
        & OOCD & \checkmark & \checkmark & - & \checkmark & - & - & 1 & News & - & - & 455\,K & 2 Classes \\
        \textsc{VERITE} \hfill \cite{papadopoulos2024VERITERobustBenchmark} 
        & OOCD & \checkmark & - & - & \checkmark & \RIGHTcircle & - & 1 & \checkmark & - & \checkmark & 1.0\,K & 3 Classes \\
        \textsc{NewsCLIPpings} \hfill \cite{luo2021NewsCLIPpingsAutomaticGeneration} 
        & OOCD & \checkmark & - & - & \checkmark & \RIGHTcircle & - & 1 & News & - & \checkmark & 988\,K & 2 Classes \\
        \textsc{5Pils-OOC} \hfill \cite{tonglet2025COVECOntextVEracity} 
        & OOCD & \checkmark & - & - & - & \checkmark & - & 1 & \checkmark & - & \checkmark & 1.2\,K & 2 Classes \\
        \textsc{COSMOS} (test split) \hfill \cite{aneja2021COSMOSCatchingOutofContext} 
        & OOCD & \checkmark & - & - & - & \RIGHTcircle & - & 1 & News & - & \checkmark & 1.7\,K & 2 Classes \\
        \midrule
        \textsc{DGM$^4$} \hfill \cite{shao2023DetectingGroundingMultiModal} 
        & MD & \checkmark & - & - & \checkmark & - & \checkmark & 1 & News & - & - & 230\,K & 2 Classes \\
        \midrule
        \textbf{\method (Ours)} 
        & CV & \checkmark & \checkmark & \checkmark & \checkmark & \checkmark & \checkmark & 54+ & \checkmark & \checkmark & \checkmark & 25\,K+ & 5 Scores \\
        \bottomrule
    \end{tabular}
    }
    \caption{Overview of related multimodal AFC benchmarks. \textbf{Primary task}: CV = Claim Verification, OOCD = Out-of-Context Detection, MD = Manipulation Detection. \textbf{Dynamic}: Whether the authors highlight the benchmark as extensible with new claims (\RIGHTcircle) and if the authors committed to update it regularly (\checkmark). \textbf{Automated}: Whether the construction pipeline is fully autonomous. \textbf{Real claims}: Whether the benchmark is entirely made from real-world claims (and perhaps close derivations) (\checkmark) or contains some (\RIGHTcircle) or mostly (-) synthetic, i.e., invented claims. \textbf{Misinfo coverage}: Whether the claims cover the full (\checkmark) or broad (\RIGHTcircle) spectrum of contemporary misinformation types for the given modalities, or just selected types (-). \textbf{Open-domain}: SM = Social Media. \textbf{Justifications}: Whether the benchmark contains explanations in addition to the annotations.}
    \label{tab:prior_work}
\end{table*}

\paragraph{Benchmarks by Task Scope and Domain}
Automated Fact-Checking (AFC) remains a challenging and unsolved problem, particularly in realistic, open-domain settings \cite{akhtar2023MultimodalAutomatedFactChecking,dmonte2025ClaimVerificationAge,schlichtkrull2024AutomatedVerificationTextual}. To tackle the complexity, most benchmarks restrict the task scope, either by decomposing AFC into sub-tasks, such as
image contextualization \cite{tonglet2024ImageTellMe},
out-of-context detection \cite{xu2024MMOOCMultimodalMisinformation, papadopoulos2024VERITERobustBenchmark, luo2021NewsCLIPpingsAutomaticGeneration, tonglet2025COVECOntextVEracity, aneja2021COSMOSCatchingOutofContext},
manipulation detection \cite{shao2023DetectingGroundingMultiModal},
check-worthiness estimation  \cite{meer2025HintsOfTruthMultimodalCheckworthinessa},
fact-check retrieval \cite{papadopoulos2025MultimodalMultilingualFactChecked}, 
content interpretation \cite{jin2024MMSOCBenchmarkingMultimodala}, 
deepfake detection \cite{skoularikis2025HumorArtMisinformation}, 
claim disambiguation \cite{staliunaite2025Dis2DisExplainingAmbiguity, glockner2024AmbiFCFactCheckingAmbiguous},
claim detection \cite{cheema2022MMClaimsDatasetMultimodal}, 
claim normalization \cite{sundriyal2023ChaosClarityClaim}, or 
claim matching \cite{pisarevskaya2025ZeroshotFewshotLearning}. Some benchmarks limit the domain to specific platforms like Reddit \cite{nakamura2020FakedditNewMultimodal}, or topics like elections \cite{khatiwada2025MultimodalMultiLabelElectionContext}, finance \cite{rangapur2025FinFactBenchmarkDataset}, climate charts \cite{su2025ClimateVizBenchmarkStatistical}, or the Ukraine-Russia war \cite{bondielli2024DatasetMultimodalFake}.
In contrast, \method targets the holistic and open-domain, end-to-end task of \textbf{claim verification}, where the goal is to predict expert ratings.

\paragraph{Benchmarks by Modality and Annotation.}
Early AFC benchmarks are predominantly text-only, including \textsc{FEVER} \cite{thorne2018FEVERLargescaleDataset} and \textsc{LIAR} \cite{wang2017LiarLiarPants}, with later publications such as \averitec \cite{schlichtkrull2023AVeriTeCDatasetRealworld} addressing previous issues like temporal leakage \cite{glockner2022MissingCounterEvidenceRenders}. However, only multimodal benchmarks can capture the important role of visual content---see Tab.~\ref{tab:prior_work} for a detailed comparison. We consider multimodal AFC benchmarks to incorporate claims with associated images and/or videos, unlike datasets that use multimodal evidence but retain text-only claims as proposed by \citet{tang2024M3DMultiModalMultiDocument,yao2023EndtoEndMultimodalFactChecking, wang2025PiecingItAll}.
Multimodal AFC benchmarks are often limited in size \cite{zlatkova2019FactCheckingMeetsFauxtography}, or miss either images or videos entirely. Only a few datasets support both images and videos \cite{xu2024M3AMultimodalMisinformation,xu2025MDAM3MisinformationDetection}. In contrast, \method covers \textit{all} common modalities. Additionally, among all multimodal AFC benchmarks, only a few provide justifications for explanation \cite{cao2025AVerImaTeCDatasetAutomatic, niu2025PioneeringExplainableVideo}. Almost all benchmarks use a single-label $n$-class annotation scheme, often entangling different, orthogonal properties (such as media authenticity and claim veracity) in the same coarse category with unclear separation. The only exception is \citet{khatiwada2025MultimodalMultiLabelElectionContext} who employ a multi-labeling scheme. \method explicitly models claim and media properties with disentangled, uncertainty-aware scores and textual justifications.

\paragraph{Dynamic Claim Datasets.}
Dynamic benchmarking has mainly been explored in adversarial settings where data collection and model training co-evolve \cite{shirali2023TheoryDynamicBenchmarks}, most prominently in \textsc{DynaBench} \cite{kiela2021DynabenchRethinkingBenchmarking}. By contrast, \method is updated independently of evaluated models. Sustaining a dynamic dataset requires \textit{automation}: Many benchmarks rely on human annotation \cite{xiao2025XFactaContemporaryRealWorld} or synthetic claim generation \cite{xu2024M3AMultimodalMisinformation, xu2025MDAM3MisinformationDetection}, raising concerns about scalability or realism. \textsc{MuMiN} \cite{nielsen2022MuMiNLargeScaleMultilingual} and TRUE \cite{niu2025PioneeringExplainableVideo} are the only ones to accomplish both realism and automation. However, \textsc{MuMiN} lacks videos, and TRUE lacks images. Two dynamic claim repositories exist: \textsc{ClaimsKG} \cite{tchechmedjiev2019ClaimsKGKnowledgeGraph} and \textsc{CimpleKG} \cite{burel2025CimpleKGContinuouslyUpdated}. Both aggregate fact-checked claims daily, but are text-only and not designed as evaluation benchmarks. \citet{xiao2025XFactaContemporaryRealWorld} express intent to ``continuously update'' their benchmark, \textsc{XFacta}, yet the most recent change is more than $8$ months ago. \citet{fatahibayat2025FactBenchDynamicBenchmark} propose a dynamic, text-only benchmark that consists not of claims but of prompts of varying complexity to probe the factuality of LLMs.

In a nutshell, and as can be seen in Tab.~\ref{tab:prior_work}, prior work always lacks multiple important properties. In contrast, \method combines real-world multimodal claims, fine-grained annotations derived from expert-provided rulings with justifications, and a fully automated construction pipeline. Its dynamic nature makes \method the first benchmark specifically designed for long-term, leakage-resistant evaluation of multimodal AFC systems.

\section{The \method Construction Pipeline}
\label{sec:method}

\method (\textit{Verified Theses and Statements}) is a dynamic, multimodal, and multilingual benchmark for evaluating Multimodal Automated Fact-Checking (MAFC) systems. It gets extended with recent, real-world claims each quarter, combining text, images, and videos. The data is gathered via a seven-stage pipeline enabling large-scale, standardized claim acquisition and annotation. 

\begin{figure}[t]
    \centering
    \includegraphics[width=\linewidth]{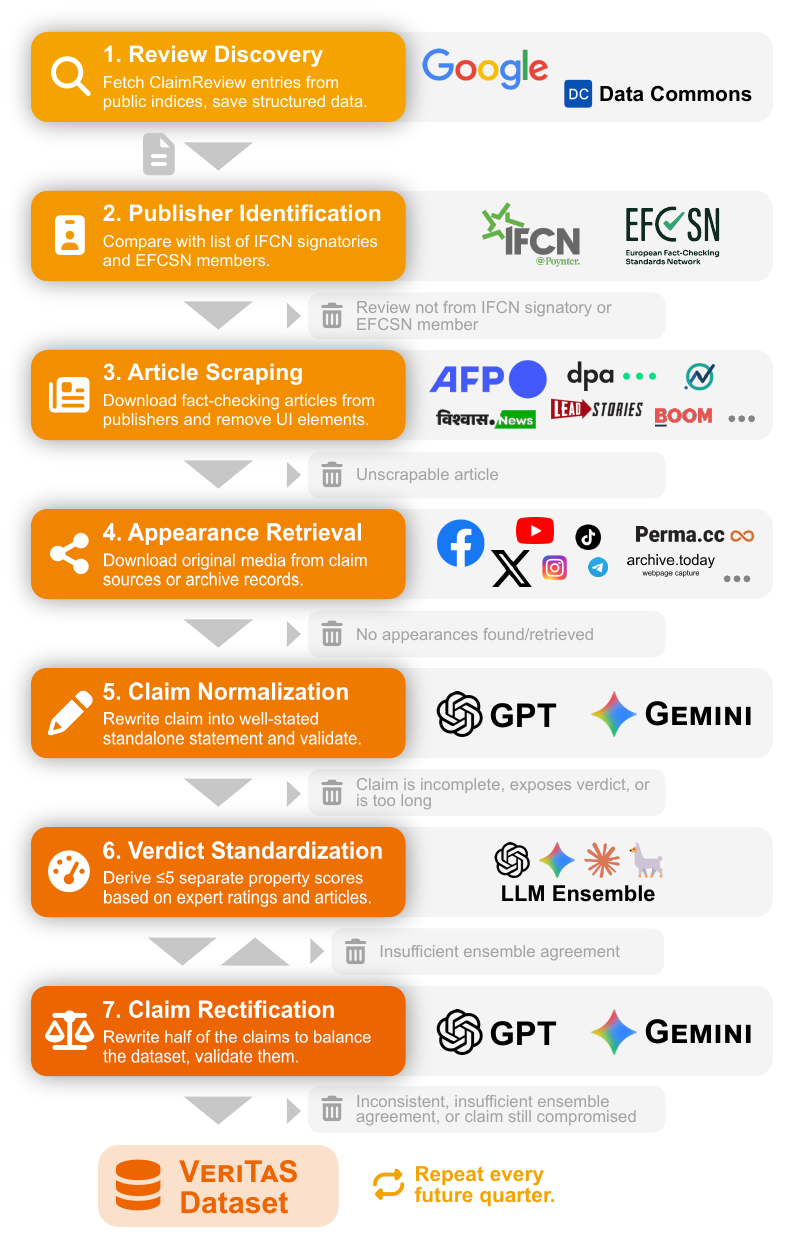}
    \caption{The seven stages of \method repeated on a quarterly basis.}
    \label{fig:concept_figure}
\end{figure}

LLMs have proven to be reliable enough for automated annotation~\cite{khatiwada2025MultimodalMultiLabelElectionContext}.
Across stages, we employ LLMs from the GPT~5~\cite{openai2025GPT5SystemCard} and \textsc{Gemini}~\cite{google2025Gemini3Pro} families as these represent the state of the art in multimodal language modeling at the time of data collection, achieving leading performance in image and video understanding tasks, respectively. We select model sizes based on task complexity and model availability (see App.~\ref{app:llms} for the used LLM versions). For multi-step reasoning, we apply chain-of-thought prompting; for generation tasks, we use few-shot in-context examples to constrain output style. Please refer to the code release for the prompts. Fig.~\ref{fig:concept_figure} summarizes the seven autonomous stages we are going to describe next---App.~\ref{app:method_details} contains additional details.


\subsection{Stage 1: Review Discovery}
\textbf{Goal.} Retrieve expert-annotated ground truth claims.  
We automatically collect fact-check reviews via the ClaimReview\footnote{\href{https://www.claimreviewproject.com/}{claimreviewproject.com}} schema and, where missing, we determine the language of each claim.
\textbf{Output.} Parsed ClaimReview data of about $398\,K$ reviews published between Jan 1, 2016 and Mar 31, 2026, including claim text, rating, review URL, date, language, and appearance URLs.

\subsection{Stage 2: Publisher Identification}
\textbf{Goal.} Ensure review credibility.
We identified $848$ distinct publishers via review URLs, and retain reviews only if they originate from credible fact-checkers (cf.\ App.~\ref{app:method_details}), dismissing about $64\,K$ ($16\%$) of the reviews not meeting this criterion. 
\textbf{Output.} $335\,K$ reviews from professional fact-checking organizations meeting international standards.

\subsection{Stage 3: Article Scraping}
\textbf{Goal.} Obtain full fact-check context.
Since ClaimReview provides no justifications for the ratings, we scrape the original fact-checking articles incl.\ media. With \gptnano we extract the main textual body via span detection, preserving the original text but removing cookie notices, ads, and other UI noise. We discard $8.8\,K$ ($4.2\%$) inaccessible articles as well as $7.3\,K$ ($3.5\%$) unrealistically short articles.
\textbf{Output.} $208\,K$ reviews\footnotemark{} with cleaned article content.

\subsection{Stage 4: Appearance Retrieval}
\textbf{Goal.} Recover original claim sources and media.
An \emph{appearance} denotes the original location where a claim surfaced, e.g., a social media post. ClaimReview provides appearance URLs in only $13.3\%$ of cases. For reviews with missing appearances, we prompt \gptmini to extract appearance URLs from the article text, including archived versions (e.g., Perma.cc; see App.~\ref{app:app_stats}), which are resolved to original sources when possible.
If original content cannot be downloaded, archived versions are used. We retain up to two successfully scraped appearances, omitting videos longer than $5$ minutes or larger than $128\,MB$ to maintain a reasonable claim scope. $105\,K$ ($52.8\%)$ Reviews without any valid appearance are discarded.
\textbf{Output.} $94\,K$ reviews\footnotemark[\value{footnote}] with one or two downloaded appearances.

\footnotetext{At the time of writing, there remain unprocessed reviews at this stage since the gathered data was already sufficient.}

\subsection{Stage 5: Claim Normalization}
The raw claim text provided by ClaimReview is often malformed, e.g., exposing the verdict, constituting incomplete sentences, or including unnecessary information like ``Social media posts claiming...'' and omitting media completely.
\textbf{Goal.} Produce precise, self-contained claims with relevant media.

\paragraph{Media Filtering.}
We retain only media inherent to the claim, removing duplicates via cosine similarity $>0.85$ in the embedding space of the CLIP variant \texttt{Qdrant/clip-ViT-B-32-vision}. Videos are inspected with \geminiflash, images with \gptmini, to exclude irrelevant content. Media precedence follows: original appearance $>$ archived appearance $>$ article media. At most $4$ media are kept per claim, prioritizing videos.

\paragraph{Claim Reformulation.}
Using \geminipro and \gpt for video claims and non-video claims, respectively, raw claim texts are rewritten into about $72\,K$ concise, self-contained statements, conditioned on article content, appearances, and metadata, e.g., claimant and date. Reformulated claims explicitly reference associated media when present and preserve the original language.

\paragraph{Claim Validation.}
$480$ ($0.1\%$) claims are discarded by LLM validation with \gptmini and \geminiflash as they (i) expose the verdict, (ii) lack required media, or (iii) exceed $70$ words to limit claim scope.
\textbf{Output.} $72\,K$ validated, well-formed claims\footnotemark[\value{footnote}] with associated media, if relevant.

\subsection{Stage 6: Verdict Standardization}
\label{sec:stage_6}

\begin{figure}
    \centering
    \includegraphics[width=0.75\linewidth]{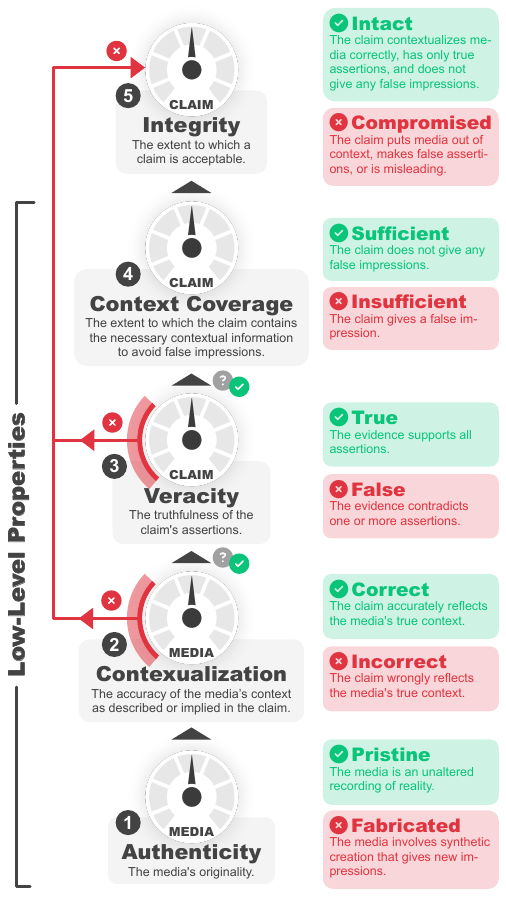}
    \caption{Verdict derivation (Stage 6), assessing properties (1) to (4). Negative decisions in (2) or (3) result in early termination. Definitions are shown on the right.}
    \label{fig:stage_6}
\end{figure}

\textbf{Goal.} Standardize heterogeneous fact-check ratings.
Fact-checking organizations employ heterogeneous rating schemes that often conflate multiple dimensions, using labels such as “half true,” “legend,” or “three Pinocchios” (see Fig.~\ref{fig:raw_ratings}). To obtain a consistent and interpretable representation, we decompose claim assessment into four independent properties that are evaluated sequentially (Fig.~\ref{fig:stage_6}): (1) \textit{Media Authenticity}, (2) \textit{Media Contextualization}, (3) \textit{Veracity}, and (4) \textit{Context Coverage}. To retain a single scalar notion of overall acceptability, we further introduce a higher-order property termed \emph{Integrity} (5). Integrity captures whether a claim is acceptable as presented: Low integrity indicates that at least one lower-level property is violated, whereas high integrity corresponds to positive assessments of Properties (2–4). Integrity constitutes the primary evaluation target, while the lower-level properties serve as explanatory factors that localize the source of compromise and simplify the annotation task. Media-specific properties (1–2) are assessed per media item; if no media is present, evaluation begins at veracity, targeting the textual part. The \textit{Context Coverage} property explicitly captures claims that are technically true yet misleading due to omitted crucial context, a phenomenon known as ``cherry-picking'' \cite{schlichtkrull2023AVeriTeCDatasetRealworld}.

Inspired by \citet{glockner2024AmbiFCFactCheckingAmbiguous}, who propose to avoid a rigid class-like categorization of claims to incorporate uncertainty, we model each property on a scale from $-1$ to $+1$, where $0$ denotes full uncertainty (\nei, Not Enough Information). A score of $<-\nicefrac{1}{3}$ is considered a \negative decision, analogously for \positive. The integrity score is determined directly by the worst-scoring property among (2--4), which we refer to as the \emph{compromising} property. Consequently, a claim is \intact if all media (if any) are correctly contextualized, the claim veracity is true, and no false impressions arise from missing context; otherwise, it is \compromised. If a prediction at step (2) or (3) results in a \negative decision, the follow-up properties will not significantly impact the integrity. Thus, we terminate evaluation for the claim early for \negative decisions at steps (2) and (3).

To increase robustness, we use an ensemble of four LLMs (\gpt, \geminipro, \claude \cite{anthropic2025ClaudeSonnet45}, and \llama \cite{metaai2025Llama4Herd}, aggregating predictions by their mean. We discard $318$ ($0.9\%$) claims that receive an ensemble prediction with an internal score difference exceeding $1$ to keep only instances with high inter-model agreement. Additionally, each ensemble member provides a one-paragraph justification for its decision, reciting the core arguments from the fact-checking article. We instruct \gptmini to summarize the four justifications into a single one. The justification for the rating of the compromising property serves as justification for the integrity.
\textbf{Output.} $36\,K$ claims\footnotemark[\value{footnote}] with high-agreement verdicts and justifications.

\subsection{Stage 7: Claim Rectification}
\textbf{Goal.} Balance \intact and \compromised claims.  
Fact-checkers primarily verify compromised claims since these are the most harmful, cf.\ Fig.~\ref{fig:quarter_stats/natural/claims/integrity}. Only $\sim0.3\%$ of ClaimReviews yield \intact claims. Thus, we generate about $32\,K$ ``corrected'' (\intact) versions of the \compromised claims using \gpt and \geminipro, guided by the justification of the compromising property, retaining media.

Consistency between corrected text and media is validated using \gptmini or \geminiflash. To avoid stylistic shortcuts, we use the same LLMs to validate if rectified claims are \textit{shareable}, i.e., relevant, not overly specific, and worth sharing, like original claims. This way, we identify $7.3\,K$ ($23.1\%$) rectified claims for exclusion. Finally, rectified claims are reevaluated using Stage~6. Claims with integrity $> \nicefrac{1}{3}$ are kept and replace their originals; all $178$ ($0.6\%$) others are discarded along with $3.1\,K$ ($9.9\%$) claims with insufficient ensemble agreement.
\textbf{Output.} $17\,K$ consistent, shareable, intact rectified claims.

\section{Results}
\label{sec:results}

\subsection{\method Statistics}

\begin{figure*}
    \centering
    \includegraphics[width=\linewidth]{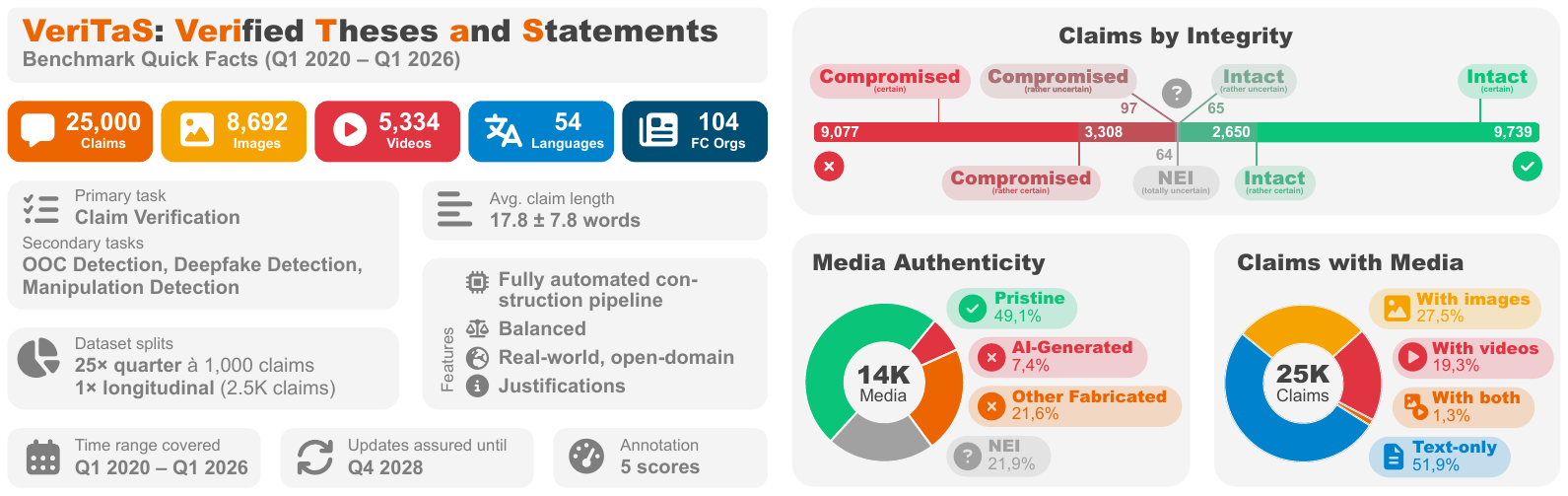}
    \caption{\method dataset statistics, see also Tab.~\ref{tab:verdict_stats}.}
    \label{fig:quick_facts}
\end{figure*}

\method currently contains $25\,K$ claims spanning $25$ quarterly splits from Q1 2020 to Q1 2026. Each quarter comprises $1\,K$ claims, balanced so that the number of \intact and \compromised instances is equal. Earlier periods are excluded due to insufficient review coverage. The most recent quarter is generally the only one suitable for reliable benchmarking. In addition to the quarterly splits, we release a budget-friendly \emph{longitudinal split} containing 2,500 balanced claims ($100$ per quarter), enabling temporal analysis across the full time span. Summary statistics are reported in Fig.~\ref{fig:quick_facts} and App.~\ref{app:verdict_stats}.

\method contains $8,692$ images and $5,334$ videos while it covers $54$ languages, with $37$ occurring at least $50$ times. English is most frequent ($39.0\%$), followed by Spanish ($10.9\%$) and Hindi ($5.8\%$), see Fig.~\ref{fig:quarter_stats/release/claims/claim_languages} for more languages. Claim appearances are predominantly sourced from Facebook ($38.4\%$) and X/Twitter ($20.5\%$), while all other platforms contribute less than $6\%$ each. AFP Fact Checking accounts for the largest share of reviewed claims ($24.8\%$), with all other organizations contributing below $6\%$ each.

\subsection{Validation through Human Evaluation}
We validate the outputs from stages~6 and 7 (Sec.~\ref{sec:stage_6}) via human evaluation. Each claim is independently annotated by $\geq 2$ native or C1-level speakers of the claim’s language. Annotators sequentially assess all properties, following the exact same annotation procedure as in stage 6, including scoring and providing written justifications.

\begin{table}[]
    \centering
    \resizebox{\linewidth}{!}{
    \begin{tabular}{l|c|cccc}
        \toprule
        \textbf{Integrity} & & \multicolumn{2}{c}{\textbf{Error Rates (↓)}} & \multicolumn{2}{c}{\textbf{Accuracy (↑)}} \\
         & $N$ & \highlight{MSE} & MAE & $7$-bin & $3$-bin \\
        \midrule
        \method                            & 204 & \textbf{0.034} & 0.102 & 69.1 & \textbf{97.5} \\
        - w/ ensemble, w/o filtering       & 207 & \textbf{0.035} & 0.105 & 68.6 & \textbf{97.6} \\
        - \gpt alone                       & 207 & 0.076 & 0.184 & 51.2 & 95.7 \\
        - \textsc{Gemini 3.1 Pro} alone    & 207 & 0.071 & 0.099 & \textbf{73.4} & 95.7 \\
        - \textsc{Claude Sonnet 4} alone   & 207 & 0.048 & \textbf{0.091} & 72.5 & 97.1 \\
        - \textsc{Llama 4 Maverick} alone  & 207 & 0.042 & 0.103 & 66.7 & 97.1 \\
        \bottomrule
    \end{tabular}
    }
    \caption{Human evaluation results over $N$ claims. The scores reflect agreement between human annotation and automated \method annotations for integrity for six different settings. \textbf{MSE} = Mean Squared Error (the main metric), \textbf{MAE} = Mean Absolute Error, \textbf{Accuracy} = share of matches of $n$-bin discretized scores (in \%).}
    \label{tab:human_eval}
\end{table}

We report Mean Squared Error (MSE), Mean Absolute Error (MAE), and accuracy after discretizing scores into three or seven bins, respectively. MSE is our primary metric, as it penalizes severe confusions more strongly (App.~\ref{app:metrics}). $\text{MSE} \leq 0.04$ indicates very high agreement (App.~\ref{app:metrics}). Results are summarized in Tab.~\ref{tab:human_eval}, where we also include five ablated \method variants: \textit{w/ ensemble, w/o filtering} (keeping claims with high ensemble disagreement) and just one of the four ensemble members as the sole predictor. The evaluation shows a very strong agreement with an MSE below $0.04$, corresponding to roughly one flipped prediction among $100$ otherwise correct judgments. 
The observations are confirmed by confusion matrices (Fig.~\ref{fig:human_eval_confusion}) and the results for low-level properties (Tab.~\ref{tab:human_eval_extra}).

Overall, these results confirm that the automated verdict mapping closely aligns with human judgments, validating the \method construction pipeline. Additionally, the increased error rates for the ablated variants in Tab.~\ref{tab:human_eval} indicate the benefits of the ensemble approach and the agreement filtering. Moreover, evaluators were asked to flag problematic claims. Only a small minority ($\sim5.1\%$) exhibited quality concerns (primarily missing media, otherwise clarity issues), indicating that the normalized/rectified claims are generally well-formed and realistic. See App.~\ref{app:human_eval} for more details on the evaluation.

\subsection{Benchmarking AFC Methods on \method latest quarter}
We analyze baselines and current AFC methods on the Q1 2026 split to assess performance on the most recent data---particularly after the knowledge cutoff of all tested methods. We consider various foundation models: \textsc{Claude~Opus~4.6}, \gpt and \textsc{Gemini~3.1~Pro} as state-of-the-art multimodal models; \gptfour as an earlier-generation model to study knowledge-cutoff effects; \textsc{Gemini~3~Flash} as a mid-sized LLM, as well as \llama, \textsc{Gemma~4~(31B)}, and \textsc{Qwen~3.5~(397B)} as representative open-source models. Each model is evaluated once using parametric knowledge only and once with web search augmentation, see App.~\ref{app:web-search-tool} for the tool implementation details. Note that \textit{all} evaluations on \method require excluding evidence sources published after the claim's release date to avoid temporal leakage \cite{glockner2022MissingCounterEvidenceRenders}.

\textsc{Gemini~3.1~Pro} and \textsc{Gemini~3~Flash} process videos natively. For all other models, we represent videos via five evenly spaced frames and a speech transcript. Additionally, we compare against two recent fact-checking systems, \textsc{DEFAME} \cite{braun2025DEFAMEDynamicEvidencebased} and \textsc{Loki} \cite{li2025LokiOpenSourceTool}, both with three different backbones in Tab.~\ref{tab:results_quarter}. 

\begin{table*}[t]
    \centering
    \resizebox{\linewidth}{!}{
    \begin{tabular}{lc|cc|cc|ccc|cc|cc}
    \toprule
        \textbf{Method} & \textbf{Search} & \multicolumn{4}{c|}{\textbf{Overall Results}} & \multicolumn{7}{c}{\textbf{MSE on Subsets (↓)}} \\
        & & \multicolumn{2}{c|}{\textbf{Errors (↓)}} & \multicolumn{2}{c|}{\textbf{Accuracy (↑)}} & \multicolumn{3}{c|}{\textbf{By Media}} & \multicolumn{2}{c|}{\textbf{By Language}} & \multicolumn{2}{c}{\textbf{By Integrity}} \\
         & & \highlight{MSE} & MAE & $7$-bin & $3$-bin & w/ images & w/ videos & text-only & English & non-English & \intact & \compr \\
        \midrule
        \textsc{Gemini~3~Flash}  & -         & \underline{0.632} & \textbf{0.409} & \textbf{60.3} & \textbf{81.4} & 0.553 & \underline{0.743} & 0.625 & \underline{0.604} & 0.648 & 1.097 & \underline{0.167} \\
        \textsc{Gemini~3.1~Pro}  & -         & 0.636 & \underline{0.412} & \underline{58.9} & \underline{81.1} & \underline{0.508} & 0.751 & 0.659 & 0.617 & \underline{0.647} & 1.210 & \textbf{0.049} \\
        \textsc{Claude~Opus~4.6} & -         & \textbf{0.453} & 0.471 & 30.1 & 62.2 & \textbf{0.451} & \textbf{0.509} & \textbf{0.361} & \textbf{0.450} & \textbf{0.455} & \textbf{0.572} & 0.337 \\
        \gptfour                 & -         & 0.826 & 0.690 & 25.4 & 45.3 & 1.050 & 0.759 & \underline{0.560} & 0.967 & 0.760 & 1.418 & 0.247 \\
        \gpt                     & -         & 0.701 & 0.691 & 12.8 & 42.0 & 0.644 & 0.820 & 0.571 & 0.678 & 0.710 & 1.179 & 0.240 \\
        \llama                   & -         & 0.943 & 0.767 & 18.1 & 44.2 & 1.127 & 0.949 & 0.583 & 0.966 & 0.931 & 1.270 & 0.635 \\
        \textsc{Gemma~4~(31B)}   & -         & 0.675 & 0.465 & 51.5 & 77.2 & 0.590 & 0.791 & 0.703 & 0.632 & 0.697 & \underline{1.059} & 0.290 \\
        \textsc{Qwen~3.5~(397B)} & -         & 0.887 & 0.563 & 51.3 & 69.8 & 0.828 & 1.070 & 0.587 & 0.911 & 0.881 & 1.404 & 0.373 \\
        \midrule
        \textsc{Gemini~3~Flash} & \checkmark & \underline{0.275} & \underline{0.237} & \textbf{67.2} & \textbf{90.5} & \underline{0.250} & \underline{0.332} & \textbf{0.158} & 0.268 & \underline{0.279} & 0.371 & 0.173 \\
        \textsc{Gemini~3.1~Pro} & \checkmark & 0.388 & 0.316 & \underline{61.1} & 82.9 & 0.335 & 0.443 & 0.277 & 0.397 & 0.386 & 0.643 & \textbf{0.132} \\
        \textsc{Claude~Opus~4.6} & \checkmark & \textbf{0.183} & \textbf{0.221} & 60.8 & \underline{89.6} & \textbf{0.134} & \textbf{0.210} & \underline{0.196} & \textbf{0.172} & \textbf{0.188} & \textbf{0.161} & 0.202 \\
        \gptfour                & \checkmark & 0.841 & 0.615 & 38.4 & 59.6 & 1.141 & 0.702 & 0.547 & 0.993 & 0.770 & 1.362 & 0.333 \\
        \gpt                    & \checkmark & 0.353 & 0.422 & 22.8 & 77.1 & 0.271 & 0.437 & 0.313 & \underline{0.261} & 0.393 & 0.544 & \underline{0.166} \\
        \llama                  & \checkmark & 0.863 & 0.712 & 22.0 & 50.2 & 0.934 & 0.904 & 0.535 & 0.811 & 0.878 & 1.032 & 0.707 \\
        \textsc{Gemma~4~(31B)}  & \checkmark & 0.360 & 0.295 & 60.7 & 87.2 & 0.328 & 0.429 & 0.236 & 0.331 & 0.374 & \underline{0.277} & 0.438 \\
        \textsc{Qwen~3.5~(397B)} & \checkmark & 0.318 & 0.296 & 57.8 & 86.7 & 0.263 & 0.354 & 0.254 & 0.237 & 0.354 & 0.393 & 0.240 \\
        \midrule
        DEFAME w/ \textsc{Gemini~3~Flash} & \checkmark & \underline{0.450} & \underline{0.320} & \textbf{62.2} & \underline{87.2} & \underline{0.318} & \underline{0.589} & 0.319 & 0.387 & \underline{0.472} & 0.507 & 0.391 \\
        DEFAME w/ \textsc{Claude~Opus~4.6}  & \checkmark & \textbf{0.282} & \textbf{0.277} & 58.1 & \textbf{88.3} & \textbf{0.157} & \textbf{0.411} & \textbf{0.283} & \textbf{0.213} & \textbf{0.314} & \textbf{0.440} & \underline{0.123} \\
        DEFAME w/ \textsc{Gemma~4~(31B)}   & \checkmark & 0.465 & 0.349 & \underline{59.3} & 83.6 & 0.344 & 0.599 & \underline{0.316} & \underline{0.329} & 0.527 & \underline{0.491} & 0.441 \\
        \textsc{Loki} w/ \textsc{Gemini~3~Flash}  & \checkmark & 1.487 & 0.852 & 39.4 & 54.8  & 1.422 & 1.541 & 1.659 & 1.295 & 1.561 & 2.821 & 0.151\\
        \textsc{Loki} w/ \textsc{Claude~Opus~4.6}  & \checkmark & 1.638 & 0.920 & 38.0 & 52.0 & 1.648 & 1.797 & 1.359 & 1.657 & 1.623 & 3.195 & \textbf{0.080} \\
        \textsc{Loki} w/ \textsc{Gemma~4~(31B)}  & \checkmark & 1.109 & 0.800 & 23.0 & 42.5 & 1.025 & 1.214 & 0.894 & 0.871 & 1.215 & 0.892 & 1.336 \\ 
        \bottomrule
    \end{tabular}
    }
    \caption{Model evaluation on the \textbf{Q1 2026 split} with(out) search tool, single runs, with \textbf{best} and \underline{second best}.}
    \label{tab:results_quarter}
\end{table*}

Across all baselines, \textsc{Claude~Opus~4.6} achieves the strongest MSE in both evaluation settings, even surpassing the AFC-specialized models DEFAME and \textsc{Loki}. The second-best model, \textsc{Gemini~3~Flash}, even surpassing \textsc{Gemini~3.1~Pro}, trails by $+0.18$ and $+0.09$ MSE, depending on retrieval access. Open-source \llama performs the worst among LLMs, also under search augmentation. In contrast, open-source \textsc{Gemma~4~(31B)} and \textsc{Qwen~3.5~(397B)}both perform on par or better than top-tier \textsc{Gemini~3.1~Pro}. Surprisingly, despite its specialization, \textsc{Loki} underperforms all other models. Subset results reveal several observations: (1) Video claims are the most difficult to fact-check for almost all models, (2) only \gpt and \qwen seem to noticeably struggle with non-English claims, and (3) many models exhibit a strong bias towards rating claims as \compromised. 

\subsection{The Role of Knowledge Cutoff}
\begin{figure}[t]
    \centering
    \includegraphics[width=\linewidth]{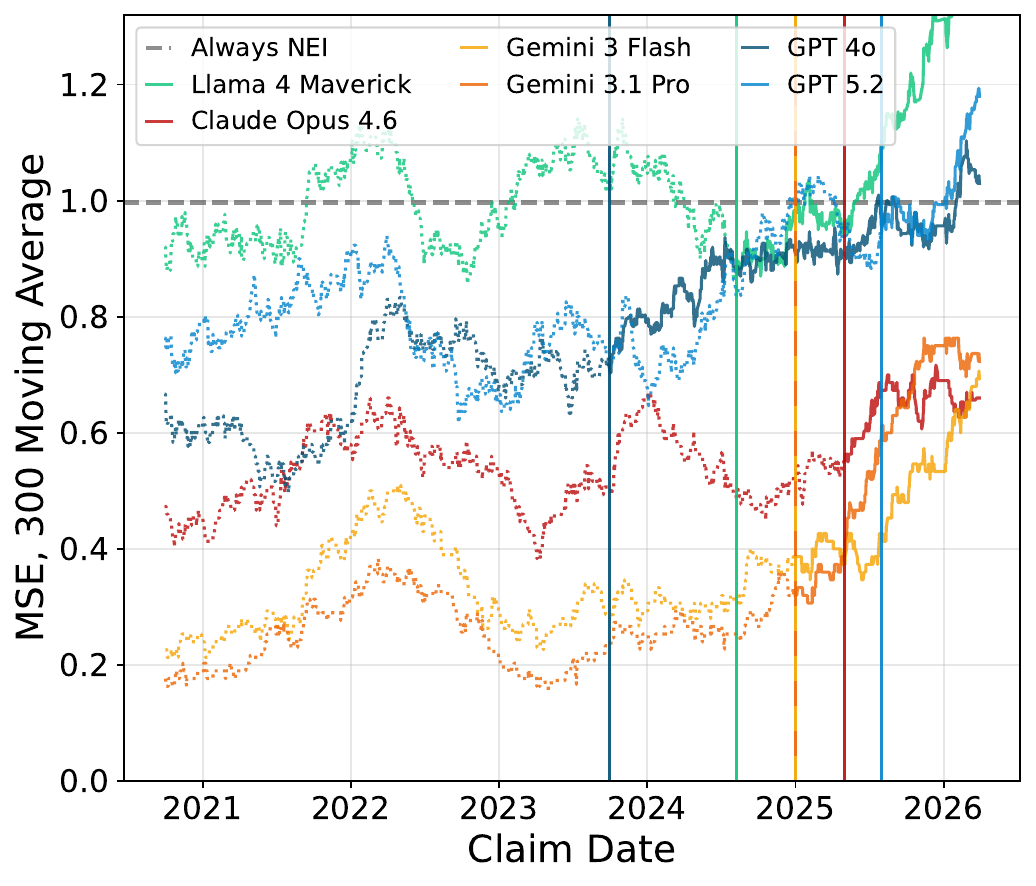}
    \caption{Baseline MSE performance without web search on the \textbf{longitudinal split}, single runs, smoothed with moving average by claim date. Lower is better. Vertical lines indicate knowledge cutoff dates, solid curves post-cutoff performance.}
    \label{fig:results_longitudinal}
\end{figure}

We evaluate the same models on the longitudinal \method split. Fig.~\ref{fig:results_longitudinal} shows the results. All evaluated models exhibit a marked increase in MSE after their respective Knowledge Cutoff Dates (KCDs), as one can see for the consistently increasing solid curves. On average, MSE rises from roughly $0.6$ to above $0.8$, substantially reducing the margin to trivial \textit{always \nei} behavior. The effect is particularly pronounced for \textsc{Gemini~3.1~Pro}, whose MSE increases from around $0.3$ to over $0.7$. We can rule out the increasing share of videos as a confounding factor, since the same phenomenon is observed for text-only claims, see Fig.~\ref{fig:results_longitudinal_per_modality} for the breakdown.

\subsection{Real-World Insights}
The data obtained from stages 1–6 roughly reflect the natural distribution of real-world fact-checked claims and reveal significant temporal trends, despite the dismissal of numerous reviews and claims. The share of AI-generated media steadily increases over time, exceeding $25\%$ of claims by Q4 2025, while the proportion of pristine media correspondingly declines (Fig.~\ref{fig:natural_claim_stats_2}). The fraction of claims containing media—particularly videos—also increases over time (Fig.~\ref{fig:natural_claim_stats_1}). Platform coverage shifts are observable, with a declining share of Facebook and a growing share of Instagram (Fig.~\ref{fig:natural_claim_stats_2}). Finally, the volume of all published ClaimReviews rises until Q1 2025 and decreases thereafter (Fig.~\ref{fig:quarter_stats/reviews}).

\section{Discussion}
\label{sec:discussion}

\myparagraph{Data leakage severely undermines static AFC benchmarks.}
The increase in error rate of models after their Knowledge Cutoff Date (KCD) provides strong empirical evidence that LLMs partially encode real-world claims and verdicts from before their KCD. Vice versa, benchmarks dominated by pre-KCD data systematically overestimate model performance. This ``\textit{pre-cutoff inflation}'' phenomenon is concerning since realistic fact-checking usually targets newly emerging claims. Notably, the most recent KCD among evaluated models (\gpt: Aug~31,~2025) postdates a large fraction of existing AFC benchmarks, rendering their evaluation increasingly unreliable.


\myparagraph{Nature and Validity of Rectified Claims.}
High-quality AFC benchmarks should be grounded in real claims, as purely synthetic data fails to capture the diversity and realism of misinformation and raises ethical concerns. In contrast, \method generates \intact claims conditioned on \textit{expert ground truth} through rectification. Although roughly $50\%$ of claims are generated this way, they are fundamentally different from synthetic claims in prior work. \method' rectified claims are directly grounded in real-world fact-checking articles and their associated evidence, rather than being artificially constructed or designed to mislead. Moreover, they undergo a multi-stage validation process, including filtering, where a substantial fraction is discarded, as well as an additional \textit{Integrity} check. Finally, human evaluation indicates that only a small minority of all claims ($\sim5.1\%$) exhibit quality concerns. Taken together, these results suggest that rectified claims are not synthetic artifacts in the conventional sense, but rather evidence-grounded corrections of real-world claims that preserve realism while enabling dataset balancing.

\myparagraph{Substantial room for improvement.}
No evaluated model approaches an MSE of $0.1$, which we consider a reasonable threshold for acceptable AFC. Even the best-performing model (\opus, MSE $=0.18$) corresponds to approximately one flipped prediction in $22$ otherwise perfect ratings. Specialized AFC systems appear to degrade the performance of the tested LLMs when used as backbones. These results underscore that multimodal AFC remains far from solved. Finally, the fact that \gemma performs on par with substantially larger proprietary models such as \gpt and \textsc{Gemini~3.1~Pro} suggests that model scale alone is not required to achieve decent performance.

\myparagraph{Reliance on Proprietary LLMs.}
Dependence on proprietary models raises concerns regarding reproducibility. To assess sensitivity to the underlying backbone, we partially regenerated a subset of Q4 2025 using a substantially different LLM family (see App.~\ref{app:llm_dependence}). The resulting claims preserve the same core assertions with only minor wording differences, and ensemble-based verdicts remain highly consistent. While this does not yet fully demonstrate reproducibility with open-weight models, these findings suggest that \method is largely invariant to the specific proprietary backbone. 

\myparagraph{Outlook.}
The dynamic design of \method will capture emerging misinformation trends through quarterly updates. Future work must address potential paradigm shifts, such as increased prevalence of audio-only content or platform restrictions that limit content access. Scaling claim retrieval beyond the current $25\,\mathrm{K}$ claims toward the full set of roughly $400\,\mathrm{K}$ reviews would further improve coverage and analytical power. Since claims can be ambiguous, future work should find strategies to address multiple interpretations of claims. Another avenue of research is required to incorporate the claimant's intent: Satirical content, while most often being \compromised, could be considered \intact, given its humorous intention. Finally, since \method only evaluates verdict scores, future work would need to develop evaluation approaches for the presented evidence. 

\section{Conclusion}
\label{sec:conclusion}

We introduced \method, the first dynamic benchmark for multimodal automated fact-checking, designed to remain robust under continual large-scale pretraining of foundation models. By collecting real-world, multilingual, multimodal claims via a fully automated pipeline and updating the benchmark quarterly, \method mitigates data leakage that increasingly invalidates static AFC benchmarks. Human evaluation validates both the annotation scheme and the construction process. Baseline experiments reveal that current state-of-the-art multimodal LLMs fall far short of reliable fact-checking performance, while performance on pre–knowledge-cutoff data substantially overestimates real-world capability. \method establishes a principled foundation for meaningful, future-proof evaluation of multimodal AFC systems.

\newpage

\section{Limitations}
\label{sec:limitations}

Despite its scope and level of automation, \method has several limitations. First, we do not perform rigorous cross-lingual claim deduplication. While the share of duplicates in the release data appears to be very low (App.~\ref{app:duplicate_analysis}), semantically equivalent claims appear across languages, particularly when fact-checking organizations publish parallel versions of the same claim (e.g., AFP). While this reflects real-world dissemination, it introduces a minor redundancy within splits. Future \method iterations should introduce textual and visual claim deduplication for sampling.

Second, although rectified claims are generated using targeted prompting and automated validation, their linguistic style is not always optimal. Some rectified claims may remain verbose or insufficiently natural, which could expose exploitable shortcuts for inference. While analysis shows no noticeable advantage on rectified claims (App.~\ref{app:stylistic_shortcuts}), explicit prompting or finetuning strategies might still exploit shortcuts.

Third, \method is designed around current modalities, platforms, and publication practices. Potential paradigmatic shifts—such as increased prevalence of audio-only content, emergence of new communication platforms, restrictions on media download, or the discontinuation of fact-checking infrastructures, e.g., ClaimReview by Google or Data Commons or fact-checking organizations themselves—could impair future data collection. While the pipeline has operated reliably for data from the past six years and successfully demonstrated its extension in Q1 2026, its long-term robustness cannot be guaranteed.

Finally, although fully automated, the construction of \method incurs non-trivial API and computational costs. Creating the full benchmark required approximately \$$16.5$\,K so far, corresponding to about \$$660$ per quarter split. Additionally, hosting \llama in-house required roughly $2.9$\,K GPU hours on $8$ NVIDIA H100 GPUs, i.e., about $116$\,h per quarter. Note that these costs do not affect users of \method, which is freely available at  \href{https://veritas.mai.informatik.tu-darmstadt.de/}{veritas.mai.informatik.tu-darmstadt.de}.

\section{Ethical Considerations}
\label{sec:ethical_considerations}

With the introduction of \method, we aim to support society in addressing mis- and disinformation by enabling more reliable and meaningful evaluation of multimodal AFC systems. By solving issues of previous benchmarks, \method is intended to accelerate responsible AFC research sustainably.

\method relies exclusively on real-world claims that have already been fact-checked by professional organizations. This design choice mitigates ethical risks associated with \textit{generating} synthetic misinformation, which could otherwise contribute to the creation or dissemination of harmful content. Using expert-derived judgments promotes transparency, accountability, and reproducibility, while preserving the realism necessary for valid evaluation.

At the same time, potential misuse risks exist. Models evaluated on \method, when used to monitor social networks, could be repurposed for surveillance, censorship, or selective moderation, particularly in politically sensitive contexts. Moreover, although \method reduces evaluation bias caused by data leakage, it may still reflect geographic, linguistic, or institutional biases inherent in the global fact-checking ecosystem.

All data are sourced from publicly available materials; therefore, we did not anonymize individual claims or media. However, since some content may have been removed from the web after our collection, for copyright compliance, and to avoid automated scraping, access to \method is restricted to eligible researchers via a gated request process. Finally, the dataset may contain offensive or disturbing content, which we intentionally retain, as such material is an intrinsic part of real-world fact-checking and necessary for faithful evaluation.

\section*{Acknowledgments}
\label{sec:acknowledgments}

We thank \textbf{Tobias Wieczorek}, \textbf{Florian Schröter}, \textbf{Aritra Marik}, and the other evaluators for their great work, each annotating dozens of claims and, thus, enabling the validation of \method.

We gratefully acknowledge support from the hessian.AI Service Center (funded by the German \textbf{Federal Ministry of Research, Technology and Space (BMFTR)}, grant no.\ \textbf{16IS22091}) and the \textbf{hessian.AI Innovation Lab} (funded by the Hessian Ministry for Digital Strategy and Innovation, grant no.\ \textbf{S-DIW04/0013/003}).

The research was partially funded by a LOEWE Start-Professur (\textbf{LOEWE/4b//519/05.01.002-(0006)/94}), LOEWE-Spitzen-Professur (\textbf{LOEWE/4a//519/05.00.002-(0010)/93}) and an Alexander von Humboldt Professorship in Multimodal Reliable AI, sponsored by the BMFTR and has benefited from the Excellence Cluster “Reaonsable AI” by the German Research Foundation (Deutsche Forschungsgemeinschaft - DFG) under Germany's Excellence Strategy – \textbf{EXC-3057}.

\bibliography{bibliography}

@inproceedings{akhtar2023MultimodalAutomatedFactChecking,
  title = {Multimodal {{Automated Fact-Checking}}: {{A Survey}}},
  shorttitle = {Multimodal {{Automated Fact-Checking}}},
  booktitle = {The 2023 {{Conference}} on {{Empirical Methods}} in {{Natural Language Processing}} ({{EMNLP}}): {{Findings}}},
  author = {Akhtar, Mubashara and Schlichtkrull, Michael and Guo, Zhijiang and Cocarascu, Oana and Simperl, Elena and Vlachos, Andreas},
  year = 2023,
  month = oct,
  eprint = {2305.13507},
  primaryclass = {cs},
  publisher = {arXiv},
  urldate = {2024-01-22},
  archiveprefix = {arXiv},
  langid = {english}
}

@misc{aneja2021COSMOSCatchingOutofContext,
  title = {{{COSMOS}}: {{Catching Out-of-Context Misinformation}} with {{Self-Supervised Learning}}},
  shorttitle = {{{COSMOS}}},
  author = {Aneja, Shivangi and Bregler, Chris and Nie{\ss}ner, Matthias},
  year = 2021,
  month = apr,
  number = {arXiv:2101.06278},
  eprint = {2101.06278},
  primaryclass = {cs},
  publisher = {arXiv},
  doi = {10.48550/arXiv.2101.06278},
  urldate = {2024-05-10},
  archiveprefix = {arXiv}
}

@techreport{anthropic2025ClaudeSonnet45,
  type = {System {{Card}}},
  title = {Claude {{Sonnet}} 4.5 {{System Card}}},
  author = {Anthropic},
  year = 2025,
  month = sep,
  url = {https://www.anthropic.com/claude-sonnet-4-5-system-card},
  urldate = {2026-01-05},
  note = {Accessed on Jan 5, 2026.}
}

@article{bondielli2024DatasetMultimodalFake,
  title = {Dataset for Multimodal Fake News Detection and Verification Tasks},
  author = {Bondielli, Alessandro and Dell'Oglio, Pietro and Lenci, Alessandro and Marcelloni, Francesco and Passaro, Lucia},
  year = 2024,
  month = jun,
  journal = {Data in Brief},
  volume = {54},
  pages = {110440},
  issn = {2352-3409},
  doi = {10.1016/j.dib.2024.110440},
  urldate = {2024-09-06}
}

@inproceedings{braun2025DEFAMEDynamicEvidencebased,
  title = {{{DEFAME}}: {{Dynamic Evidence-based FAct-checking}} with {{Multimodal Experts}}},
  shorttitle = {{{DEFAME}}},
  booktitle = {Proceedings of the 42nd {{International Conference}} on {{Machine Learning}}},
  author = {Braun, Tobias and Rothermel, Mark and Rohrbach, Marcus and Rohrbach, Anna},
  year = 2025,
  month = oct,
  pages = {5383--5417},
  publisher = {PMLR},
  issn = {2640-3498},
  doi = {10.48550/arXiv.2412.10510},
  urldate = {2026-01-03},
  langid = {english}
}

@inproceedings{burel2025CimpleKGContinuouslyUpdated,
  title = {{{CimpleKG}}: {{A Continuously Updated Knowledge Graph}} on~{{Misinformation}}, {{Factors}} and~{{Fact-Checks}}},
  shorttitle = {{{CimpleKG}}},
  booktitle = {The {{Semantic Web}} -- {{ISWC}} 2024},
  author = {Burel, Gr{\'e}goire and Mensio, Martino and Peskine, Youri and Troncy, Raphael and Papotti, Paolo and Alani, Harith},
  editor = {Demartini, Gianluca and Hose, Katja and Acosta, Maribel and Palmonari, Matteo and Cheng, Gong and {Skaf-Molli}, Hala and Ferranti, Nicolas and Hern{\'a}ndez, Daniel and Hogan, Aidan},
  year = 2025,
  pages = {97--114},
  publisher = {Springer Nature Switzerland},
  address = {Cham},
  doi = {10.1007/978-3-031-77847-6_6},
  isbn = {978-3-031-77847-6},
  langid = {english}
}

@inproceedings{cao2025AVerImaTeCDatasetAutomatic,
  title = {{{AVerImaTeC}}: {{A Dataset}} for {{Automatic Verification}} of {{Image-Text Claims}} with {{Evidence}} from the {{Web}}},
  shorttitle = {{{AVerImaTeC}}},
  booktitle = {{{NeurIPS}} 2025},
  author = {Cao, Rui and Ding, Zifeng and Guo, Zhijiang and Schlichtkrull, Michael and Vlachos, Andreas},
  year = 2025,
  month = may,
  doi = {10.48550/arXiv.2505.17978},
  urldate = {2025-06-02}
}

@inproceedings{chakraborty2023FACTIFY3MBenchmarkMultimodala,
  title = {{{FACTIFY3M}}: {{A}} Benchmark for Multimodal Fact Verification with Explainability through {{5W Question-Answering}}},
  shorttitle = {{{FACTIFY3M}}},
  booktitle = {Proceedings of the 2023 {{Conference}} on {{Empirical Methods}} in {{Natural Language Processing}}},
  author = {Chakraborty, Megha and Pahwa, Khushbu and Rani, Anku and Chatterjee, Shreyas and Dalal, Dwip and Dave, Harshit and G, Ritvik and Gurumurthy, Preethi and Mahor, Adarsh and Mukherjee, Samahriti and Pakala, Aditya and Paul, Ishan and Reddy, Janvita and Sarkar, Arghya and Sensharma, Kinjal and Chadha, Aman and Sheth, Amit and Das, Amitava},
  editor = {Bouamor, Houda and Pino, Juan and Bali, Kalika},
  year = 2023,
  month = dec,
  pages = {15282--15322},
  publisher = {Association for Computational Linguistics},
  address = {Singapore},
  doi = {10.18653/v1/2023.emnlp-main.945},
  urldate = {2026-01-03}
}

@inproceedings{cheema2022MMClaimsDatasetMultimodal,
  title = {{{MM-Claims}}: {{A Dataset}} for {{Multimodal Claim Detection}} in {{Social Media}}},
  shorttitle = {{{MM-Claims}}},
  booktitle = {Findings of the {{Association}} for {{Computational Linguistics}}: {{NAACL}} 2022},
  author = {Cheema, Gullal Singh and Hakimov, Sherzod and Sittar, Abdul and {M{\"u}ller-Budack}, Eric and Otto, Christian and Ewerth, Ralph},
  editor = {Carpuat, Marine and {de Marneffe}, Marie-Catherine and Meza Ruiz, Ivan Vladimir},
  year = 2022,
  month = jul,
  pages = {962--979},
  publisher = {Association for Computational Linguistics},
  address = {Seattle, United States},
  doi = {10.18653/v1/2022.findings-naacl.72},
  urldate = {2026-01-04}
}

@misc{dmonte2025ClaimVerificationAge,
  title = {Claim {{Verification}} in the {{Age}} of {{Large Language Models}}: {{A Survey}}},
  shorttitle = {Claim {{Verification}} in the {{Age}} of {{Large Language Models}}},
  author = {Dmonte, Alphaeus and Oruche, Roland and Zampieri, Marcos and Calyam, Prasad and Augenstein, Isabelle},
  year = 2025,
  month = feb,
  number = {arXiv:2408.14317},
  eprint = {2408.14317},
  primaryclass = {cs},
  publisher = {arXiv},
  doi = {10.48550/arXiv.2408.14317},
  urldate = {2026-01-04},
  archiveprefix = {arXiv}
}

@misc{dufour2024AMMeBaLargeScaleSurvey,
  title = {{{AMMeBa}}: {{A Large-Scale Survey}} and {{Dataset}} of {{Media-Based Misinformation In-The-Wild}}},
  shorttitle = {{{AMMeBa}}},
  author = {Dufour, Nicholas and Pathak, Arkanath and Samangouei, Pouya and Hariri, Nikki and Deshetti, Shashi and Dudfield, Andrew and Guess, Christopher and Escayola, Pablo Hern{\'a}ndez and Tran, Bobby and Babakar, Mevan and Bregler, Christoph},
  year = 2024,
  month = may,
  number = {arXiv:2405.11697},
  eprint = {2405.11697},
  primaryclass = {cs},
  publisher = {arXiv},
  doi = {10.48550/arXiv.2405.11697},
  urldate = {2024-05-23},
  archiveprefix = {arXiv}
}

@techreport{elsner2025GlobalRisksReport,
  title = {Global {{Risks Report}} 2025},
  author = {Elsner, Mark and Atkinson, Grace and Zahidi, Saadia},
  year = 2025,
  month = jan,
  institution = {World Economic Forum},
  urldate = {2026-01-04},
  langid = {english},
  url = {https://www.weforum.org/publications/global-risks-report-2025/}
}

@inproceedings{fatahibayat2025FactBenchDynamicBenchmark,
  title = {{{FactBench}}: {{A Dynamic Benchmark}} for {{In-the-Wild Language Model Factuality Evaluation}}},
  shorttitle = {{{FactBench}}},
  booktitle = {Proceedings of the 63rd {{Annual Meeting}} of the {{Association}} for {{Computational Linguistics}} ({{Volume}} 1: {{Long Papers}})},
  author = {Fatahi Bayat, Farima and Zhang, Lechen and Munir, Sheza and Wang, Lu},
  editor = {Che, Wanxiang and Nabende, Joyce and Shutova, Ekaterina and Pilehvar, Mohammad Taher},
  year = 2025,
  month = jul,
  pages = {33090--33110},
  publisher = {Association for Computational Linguistics},
  address = {Vienna, Austria},
  doi = {10.18653/v1/2025.acl-long.1587},
  urldate = {2026-01-03},
  isbn = {979-8-89176-251-0}
}

@misc{geng2025M4FCMultimodalMultilingual,
  title = {{{M4FC}}: A {{Multimodal}}, {{Multilingual}}, {{Multicultural}}, {{Multitask Real-World Fact-Checking Dataset}}},
  shorttitle = {{{M4FC}}},
  author = {Geng, Jiahui and Tonglet, Jonathan and Gurevych, Iryna},
  year = 2025,
  month = oct,
  number = {arXiv:2510.23508},
  eprint = {2510.23508},
  primaryclass = {cs},
  publisher = {arXiv},
  doi = {10.48550/arXiv.2510.23508},
  urldate = {2025-12-27},
  archiveprefix = {arXiv}
}

@inproceedings{glockner2022MissingCounterEvidenceRenders,
  title = {Missing {{Counter-Evidence Renders NLP Fact-Checking Unrealistic}} for {{Misinformation}}},
  booktitle = {Proceedings of the 2022 {{Conference}} on {{Empirical Methods}} in {{Natural Language Processing}}},
  author = {Glockner, Max and Hou, Yufang and Gurevych, Iryna},
  editor = {Goldberg, Yoav and Kozareva, Zornitsa and Zhang, Yue},
  year = 2022,
  month = dec,
  pages = {5916--5936},
  publisher = {Association for Computational Linguistics},
  address = {Abu Dhabi, United Arab Emirates},
  doi = {10.18653/v1/2022.emnlp-main.397},
  urldate = {2026-01-04}
}

@article{glockner2024AmbiFCFactCheckingAmbiguous,
  title = {{{AmbiFC}}: {{Fact-Checking Ambiguous Claims}} with {{Evidence}}},
  shorttitle = {{{AmbiFC}}},
  author = {Glockner, Max and Stali{\=u}nait{\.e}, Ieva and Thorne, James and Vallejo, Gisela and Vlachos, Andreas and Gurevych, Iryna},
  year = 2024,
  journal = {Transactions of the Association for Computational Linguistics},
  pages = {1--18},
  publisher = {MIT Press},
  issn = {2307-387X},
  doi = {10.1162/tacl_a_00629},
  urldate = {2024-07-04},
  langid = {english}
}

@techreport{google2025Gemini3Pro,
  type = {Model {{Card}}},
  title = {Gemini 3 {{Pro Model Card}}},
  author = {Google},
  year = 2025,
  month = nov,
  urldate = {2026-01-05},
  url = {https://deepmind.google/models/model-cards/gemini-3-pro/},
  note = {Accessed on Jan 5, 2026.}
}

@misc{google2025IntroducingNanoBanana,
  title = {Introducing {{Nano Banana Pro}}},
  author = {Google},
  year = 2025,
  month = nov,
  journal = {Google},
  urldate = {2026-01-04},
  howpublished = {https://blog.google/technology/ai/nano-banana-pro/},
  langid = {american},
  url = {https://blog.google/technology/ai/nano-banana-pro/},
  note = {Accessed on Jan 5, 2026.}
}

@book{greifeneder2020PsychologyFakeNews,
  title = {The {{Psychology}} of {{Fake News}}: {{Accepting}}, {{Sharing}}, and {{Correcting Misinformation}}},
  shorttitle = {The {{Psychology}} of {{Fake News}}},
  author = {Greifeneder, Rainer and Jaffe, Mariela and Newman, Eryn and Schwarz, Norbert},
  year = 2020,
  month = aug,
  edition = {1},
  publisher = {Routledge},
  address = {London},
  doi = {10.4324/9780429295379},
  isbn = {978-0-429-29537-9}
}

@article{hameleers2020PicturePaintsThousand,
  title = {A Picture Paints a Thousand Lies? {{The}} Effects and Mechanisms of Multimodal Disinformation and Rebuttals Disseminated via Social Media},
  shorttitle = {A Picture Paints a Thousand Lies?},
  author = {Hameleers, M. and Powell, T. E. and Van Der Meer, T. G. L. A. and Bos, L.},
  year = 2020,
  journal = {Political Communication},
  volume = {37},
  doi = {10.1080/10584609.2019.1674979},
  urldate = {2024-07-04},
  langid = {english}
}

@inproceedings{hassan2015QuestAutomateFactChecking,
  title = {The {{Quest}} to {{Automate Fact-Checking}}},
  author = {Hassan, Naeemul and Adair, Bill and Hamilton, J. and Li, Chengkai and Tremayne, Mark and Yang, Jun and Yu, Cong},
  year = 2015,
  urldate = {2026-01-04},
  url = {https://www.semanticscholar.org/paper/The-Quest-to-Automate-Fact-Checking-Hassan-Adair/714a12ddf7d2c6ddb8a99bfb6663d2144880aa49}
}

@inproceedings{hu2023MR2BenchmarkMultimodal,
  title = {{{MR2}}: {{A Benchmark}} for {{Multimodal Retrieval-Augmented Rumor Detection}} in {{Social Media}}},
  shorttitle = {{{MR2}}},
  booktitle = {Proceedings of the 46th {{International ACM SIGIR Conference}} on {{Research}} and {{Development}} in {{Information Retrieval}}},
  author = {Hu, Xuming and Guo, Zhijiang and Chen, Junzhe and Wen, Lijie and Yu, Philip S.},
  year = 2023,
  month = jul,
  series = {{{SIGIR}} '23},
  pages = {2901--2912},
  publisher = {Association for Computing Machinery},
  address = {New York, NY, USA},
  doi = {10.1145/3539618.3591896},
  urldate = {2024-01-29},
  isbn = {978-1-4503-9408-6}
}

@inproceedings{jin2024MMSOCBenchmarkingMultimodala,
  title = {{{MM-SOC}}: {{Benchmarking Multimodal Large Language Models}} in {{Social Media Platforms}}},
  shorttitle = {{{MM-SOC}}},
  booktitle = {Findings of the {{Association}} for {{Computational Linguistics}}: {{ACL}} 2024},
  author = {Jin, Yiqiao and Choi, Minje and Verma, Gaurav and Wang, Jindong and Kumar, Srijan},
  editor = {Ku, Lun-Wei and Martins, Andre and Srikumar, Vivek},
  year = 2024,
  month = aug,
  pages = {6192--6210},
  publisher = {Association for Computational Linguistics},
  address = {Bangkok, Thailand},
  doi = {10.18653/v1/2024.findings-acl.370},
  urldate = {2026-01-03}
}

@article{khatiwada2025MultimodalMultiLabelElectionContext,
  title = {Towards a {{Multi-modal Multi-Label Election-Context Repository}} for {{Classifying Misinformation}}},
  author = {Khatiwada, Prerana and Wang, Qile and Barner, Kenneth E. and Mauriello, Matthew Louis},
  year = 2025,
  month = jun,
  journal = {Workshop Proceedings of the 19th International AAAI Conference on Web and Social Media},
  volume = {2025},
  pages = {26},
  doi = {10.36190/2025.26},
  urldate = {2026-01-04},
  langid = {american}
}

@inproceedings{kiela2021DynabenchRethinkingBenchmarking,
  title = {Dynabench: {{Rethinking Benchmarking}} in {{NLP}}},
  shorttitle = {Dynabench},
  booktitle = {Proceedings of the 2021 {{Conference}} of the {{North American Chapter}} of the {{Association}} for {{Computational Linguistics}}: {{Human Language Technologies}}},
  author = {Kiela, Douwe and Bartolo, Max and Nie, Yixin and Kaushik, Divyansh and Geiger, Atticus and Wu, Zhengxuan and Vidgen, Bertie and Prasad, Grusha and Singh, Amanpreet and Ringshia, Pratik and Ma, Zhiyi and Thrush, Tristan and Riedel, Sebastian and Waseem, Zeerak and Stenetorp, Pontus and Jia, Robin and Bansal, Mohit and Potts, Christopher and Williams, Adina},
  editor = {Toutanova, Kristina and Rumshisky, Anna and Zettlemoyer, Luke and {Hakkani-Tur}, Dilek and Beltagy, Iz and Bethard, Steven and Cotterell, Ryan and Chakraborty, Tanmoy and Zhou, Yichao},
  year = 2021,
  month = jun,
  pages = {4110--4124},
  publisher = {Association for Computational Linguistics},
  address = {Online},
  doi = {10.18653/v1/2021.naacl-main.324},
  urldate = {2026-01-03}
}

@article{li2020PictureWorthThousand,
  title = {Is a {{Picture Worth}} a {{Thousand Words}}? {{An Empirical Study}} of {{Image Content}} and {{Social Media Engagement}}},
  shorttitle = {Is a {{Picture Worth}} a {{Thousand Words}}?},
  author = {Li, Yiyi and Xie, Ying},
  year = 2020,
  month = feb,
  journal = {Journal of Marketing Research},
  volume = {57},
  number = {1},
  pages = {1--19},
  issn = {0022-2437, 1547-7193},
  doi = {10.1177/0022243719881113},
  urldate = {2024-03-10},
  langid = {english}
}

@inproceedings{li2025LokiOpenSourceTool,
  title = {Loki: {{An Open-Source Tool}} for {{Fact Verification}}},
  shorttitle = {Loki},
  booktitle = {Proceedings of the 31st {{International Conference}} on {{Computational Linguistics}}: {{System Demonstrations}}},
  author = {Li, Haonan and Han, Xudong and Wang, Hao and Wang, Yuxia and Wang, Minghan and Xing, Rui and Geng, Yilin and Zhai, Zenan and Nakov, Preslav and Baldwin, Timothy},
  editor = {Rambow, Owen and Wanner, Leo and Apidianaki, Marianna and {Al-Khalifa}, Hend and Eugenio, Barbara Di and Schockaert, Steven and Mather, Brodie and Dras, Mark},
  year = 2025,
  month = jan,
  pages = {28--36},
  publisher = {Association for Computational Linguistics},
  address = {Abu Dhabi, UAE},
  url = {https://aclanthology.org/2025.coling-demos.4/},
  urldate = {2026-01-06}
}

@misc{li2025UnifiedMultimodalMisinformation,
  title = {Towards {{Unified Multimodal Misinformation Detection}} in {{Social Media}}: {{A Benchmark Dataset}} and {{Baseline}}},
  shorttitle = {Towards {{Unified Multimodal Misinformation Detection}} in {{Social Media}}},
  author = {Li, Haiyang and Wang, Yaxiong and Tang, Shengeng and Wu, Lianwei and Cheng, Lechao and Zhong, Zhun},
  year = 2025,
  month = oct,
  number = {arXiv:2509.25991},
  eprint = {2509.25991},
  primaryclass = {cs},
  publisher = {arXiv},
  doi = {10.48550/arXiv.2509.25991},
  urldate = {2025-12-27},
  archiveprefix = {arXiv}
}

@misc{liu2025MMFakeBenchMixedSourceMultimodal,
  title = {{{MMFakeBench}}: {{A Mixed-Source Multimodal Misinformation Detection Benchmark}} for {{LVLMs}}},
  shorttitle = {{{MMFakeBench}}},
  author = {Liu, Xuannan and Li, Zekun and Li, Peipei and Huang, Huaibo and Xia, Shuhan and Cui, Xing and Huang, Linzhi and Deng, Weihong and He, Zhaofeng},
  year = 2025,
  month = feb,
  number = {arXiv:2406.08772},
  eprint = {2406.08772},
  primaryclass = {cs},
  publisher = {arXiv},
  doi = {10.48550/arXiv.2406.08772},
  urldate = {2025-03-25},
  archiveprefix = {arXiv}
}

@inproceedings{luo2021NewsCLIPpingsAutomaticGeneration,
  title = {{{NewsCLIPpings}}: {{Automatic Generation}} of {{Out-of-Context Multimodal Media}}},
  shorttitle = {{{NewsCLIPpings}}},
  booktitle = {Proceedings of the 2021 {{Conference}} on {{Empirical Methods}} in {{Natural Language Processing}}},
  author = {Luo, Grace and Darrell, Trevor and Rohrbach, Anna},
  editor = {Moens, Marie-Francine and Huang, Xuanjing and Specia, Lucia and Yih, Scott Wen-tau},
  year = 2021,
  month = nov,
  pages = {6801--6817},
  publisher = {Association for Computational Linguistics},
  address = {Online and Punta Cana, Dominican Republic},
  doi = {10.18653/v1/2021.emnlp-main.545},
  urldate = {2026-01-04}
}

@misc{meer2025HintsOfTruthMultimodalCheckworthinessa,
  title = {{{HintsOfTruth}}: {{A Multimodal Checkworthiness Detection Dataset}} with {{Real}} and {{Synthetic Claims}}},
  shorttitle = {{{HintsOfTruth}}},
  author = {van der Meer, Michiel and Korshunov, Pavel and Marcel, S{\'e}bastien and van der Plas, Lonneke},
  year = 2025,
  month = jun,
  number = {arXiv:2502.11753},
  eprint = {2502.11753},
  primaryclass = {cs},
  publisher = {arXiv},
  doi = {10.48550/arXiv.2502.11753},
  urldate = {2025-06-17},
  archiveprefix = {arXiv}
}

@misc{metaai2025Llama4Herd,
  title = {The {{Llama}} 4 Herd: {{The}} Beginning of a New Era of Natively Multimodal {{AI}} Innovation},
  shorttitle = {The {{Llama}} 4 Herd},
  author = {{{Meta AI}}},
  year = 2025,
  month = apr,
  journal = {Meta AI},
  urldate = {2026-01-05},
  howpublished = {https://ai.meta.com/blog/llama-4-multimodal-intelligence/},
  langid = {english},
  note = {Accessed on Jan 5, 2026.}
}

@inproceedings{mishra2022FACTIFYMultiModalFact,
  title = {{{FACTIFY}}: {{A Multi-Modal Fact Verification Dataset}}},
  shorttitle = {{{FACTIFY}}},
  booktitle = {{{DE-FACTIFY}}@{{AAAI}}},
  author = {Mishra, Shreyash and Suryavardan, S. and Bhaskar, Amrit and Chopra, P. and Reganti, Aishwarya N. and Patwa, Parth and Das, Amitava and Chakraborty, Tanmoy and Sheth, A. and Ekbal, Asif},
  year = 2022,
  url = {https://www.semanticscholar.org/paper/FACTIFY%3A-A-Multi-Modal-Fact-Verification-Dataset-Mishra-Suryavardan/c0532d8d69af3bc0836b88c5aae2ce6166ac5136},
  urldate = {2026-01-03}
}

@inproceedings{nakamura2020FakedditNewMultimodal,
  title = {Fakeddit: {{A New Multimodal Benchmark Dataset}} for {{Fine-grained Fake News Detection}}},
  shorttitle = {Fakeddit},
  booktitle = {Proceedings of the {{Twelfth Language Resources}} and {{Evaluation Conference}}},
  author = {Nakamura, Kai and Levy, Sharon and Wang, William Yang},
  editor = {Calzolari, Nicoletta and B{\'e}chet, Fr{\'e}d{\'e}ric and Blache, Philippe and Choukri, Khalid and Cieri, Christopher and Declerck, Thierry and Goggi, Sara and Isahara, Hitoshi and Maegaard, Bente and Mariani, Joseph and Mazo, H{\'e}l{\`e}ne and Moreno, Asuncion and Odijk, Jan and Piperidis, Stelios},
  year = 2020,
  month = may,
  pages = {6149--6157},
  publisher = {European Language Resources Association},
  address = {Marseille, France},
  url = {https://aclanthology.org/2020.lrec-1.755/},
  urldate = {2026-01-04},
  isbn = {979-10-95546-34-4},
  langid = {english}
}

@article{newman2012NonprobativePhotographsWords,
  title = {Nonprobative Photographs (or Words) Inflate Truthiness},
  author = {Newman, Eryn J. and Garry, Maryanne and Bernstein, Daniel M. and Kantner, Justin and Lindsay, D. Stephen},
  year = 2012,
  month = oct,
  journal = {Psychonomic Bulletin \& Review},
  volume = {19},
  number = {5},
  pages = {969--974},
  issn = {1069-9384, 1531-5320},
  doi = {10.3758/s13423-012-0292-0},
  urldate = {2024-03-10},
  langid = {english}
}

@inproceedings{nielsen2022MuMiNLargeScaleMultilingual,
  title = {{{MuMiN}}: {{A Large-Scale Multilingual Multimodal Fact-Checked Misinformation Social Network Dataset}}},
  shorttitle = {{{MuMiN}}},
  booktitle = {Proceedings of the 45th {{International ACM SIGIR Conference}} on {{Research}} and {{Development}} in {{Information Retrieval}}},
  author = {Nielsen, Dan S. and McConville, Ryan},
  year = 2022,
  month = jul,
  series = {{{SIGIR}} '22},
  pages = {3141--3153},
  publisher = {Association for Computing Machinery},
  address = {New York, NY, USA},
  doi = {10.1145/3477495.3531744},
  urldate = {2026-01-04},
  isbn = {978-1-4503-8732-3}
}

@article{niu2025PioneeringExplainableVideo,
  title = {Pioneering {{Explainable Video Fact-Checking}} with a {{New Dataset}} and {{Multi-role Multimodal Model Approach}}},
  author = {Niu, Kaipeng and Xu, Danni and Yang, Bingjian and Liu, Wenxuan and Wang, Zheng},
  year = 2025,
  month = apr,
  journal = {Proceedings of the AAAI Conference on Artificial Intelligence},
  volume = {39},
  number = {27},
  pages = {28276--28283},
  issn = {2374-3468},
  doi = {10.1609/aaai.v39i27.35048},
  urldate = {2026-01-03},
  copyright = {Copyright (c) 2025 Association for the Advancement of Artificial Intelligence},
  langid = {english}
}

@techreport{openai2025GPT5SystemCard,
  type = {System {{Card}}},
  title = {{{GPT-5 System Card}}},
  author = {OpenAI},
  year = 2025,
  month = aug,
  institution = {OpenAI},
  url = {https://cdn.openai.com/gpt-5-system-card.pdf},
  urldate = {2026-01-05},
  note = {Accessed on Jan 5, 2026.}
}

@misc{openai2025Sora2Here,
  title = {Sora 2 Is Here},
  author = {OpenAI},
  year = 2025,
  month = sep,
  urldate = {2026-01-04},
  howpublished = {https://openai.com/index/sora-2/},
  langid = {american},
  note = {Accessed on Jan 5, 2026.}
}

@article{papadopoulos2024VERITERobustBenchmark,
  title = {{{VERITE}}: A {{Robust}} Benchmark for Multimodal Misinformation Detection Accounting for Unimodal Bias},
  shorttitle = {{{VERITE}}},
  author = {Papadopoulos, Stefanos-Iordanis and Koutlis, Christos and Papadopoulos, Symeon and Petrantonakis, Panagiotis C.},
  year = 2024,
  month = jan,
  journal = {International Journal of Multimedia Information Retrieval},
  volume = {13},
  number = {1},
  pages = {4},
  issn = {2192-662X},
  doi = {10.1007/s13735-023-00312-6},
  urldate = {2024-01-26},
  langid = {english}
}

@inproceedings{papadopoulos2025MultimodalMultilingualFactChecked,
  title = {Multimodal and {{Multilingual Fact-Checked Article Retrieval}}},
  booktitle = {Proceedings of the 2025 {{International Conference}} on {{Multimedia Retrieval}}},
  author = {Papadopoulos, Stefanos-Iordanis and Be{\v n}ov{\'a}, Ivana and Kula, Sebastian and Gregor, Michal and Karantaidis, George and Jav{\r u}rek, Tom{\'a}{\v s} and {\v S}imko, Mari{\'a}n and Papadopoulos, Symeon},
  year = 2025,
  month = jun,
  series = {{{ICMR}} '25},
  pages = {1063--1071},
  publisher = {Association for Computing Machinery},
  address = {New York, NY, USA},
  doi = {10.1145/3731715.3733402},
  urldate = {2025-07-03},
  isbn = {979-8-4007-1877-9}
}

@inproceedings{pisarevskaya2025ZeroshotFewshotLearning,
  title = {Zero-Shot and {{Few-shot Learning}} with {{Instruction-following LLMs}} for {{Claim Matching}} in {{Automated Fact-checking}}},
  booktitle = {Proceedings of the 31st {{International Conference}} on {{Computational Linguistics}}},
  author = {Pisarevskaya, Dina and Zubiaga, Arkaitz},
  editor = {Rambow, Owen and Wanner, Leo and Apidianaki, Marianna and {Al-Khalifa}, Hend and Eugenio, Barbara Di and Schockaert, Steven},
  year = 2025,
  month = jan,
  pages = {9721--9736},
  publisher = {Association for Computational Linguistics},
  address = {Abu Dhabi, UAE},
  urldate = {2026-01-04},
  url = {https://aclanthology.org/2025.coling-main.650/}
}

@inproceedings{rangapur2025FinFactBenchmarkDataset,
  title = {Fin-{{Fact}}: {{A Benchmark Dataset}} for {{Multimodal Financial Fact-Checking}} and {{Explanation Generation}}},
  shorttitle = {Fin-{{Fact}}},
  booktitle = {Companion {{Proceedings}} of the {{ACM}} on {{Web Conference}} 2025},
  author = {Rangapur, Aman and Wang, Haoran and Jian, Ling and Shu, Kai},
  year = 2025,
  month = may,
  series = {{{WWW}} '25},
  pages = {785--788},
  publisher = {Association for Computing Machinery},
  address = {New York, NY, USA},
  doi = {10.1145/3701716.3715292},
  urldate = {2026-01-03},
  isbn = {979-8-4007-1331-6}
}

@article{raza2025VLDBenchVisionLanguage,
  title = {{{VLDBench}}: {{Vision Language Models Disinformation Detection Benchmark}}},
  shorttitle = {{{VLDBench}}},
  author = {Raza, Shaina and Vayani, Ashmal and Jain, Aditya and Narayanan, Aravind and Khazaie, Vahid Reza and Bashir, S. and Dolatabadi, Elham and Uddin, Gias and Emmanouilidis, Christos and Qureshi, Rizwan and Shah, Mubarak},
  year = 2025,
  month = feb,
  journal = {Information Fusion Journal},
  url = {https://doi.org/10.48550/arXiv.2502.11361},
  urldate = {2025-03-04}
}

@article{schlichtkrull2023AVeriTeCDatasetRealworld,
  title = {{{AVeriTeC}}: {{A Dataset}} for {{Real-world Claim Verification}} with {{Evidence}} from the {{Web}}},
  shorttitle = {{{AVeriTeC}}},
  author = {Schlichtkrull, Michael and Guo, Zhijiang and Vlachos, Andreas},
  year = 2023,
  month = dec,
  journal = {Advances in Neural Information Processing Systems},
  volume = {36},
  pages = {65128--65167},
  urldate = {2026-01-04},
  langid = {english},
  url = {https://proceedings.neurips.cc/paper_files/paper/2023/hash/cd86a30526cd1aff61d6f89f107634e4-Abstract-Datasets_and_Benchmarks.html}
}

@article{schlichtkrull2024AutomatedVerificationTextual,
  title = {The {{Automated Verification}} of {{Textual Claims}} ({{AVeriTeC}}) {{Shared Task}}},
  author = {Schlichtkrull, Michael and Chen, Yulong and Whitehouse, Chenxi and Deng, Zhenyun and Akhtar, Mubashara and Aly, Rami and Guo, Zhijiang and Christodoulopoulos, Christos and Cocarascu, Oana and Mittal, Arpit and Thorne, James and Vlachos, Andreas},
  year = 2024,
  journal = {FEVER Workshop at EMNLP 2024},
  doi = {10.18653/v1/2024.fever-1.1},
  langid = {english}
}

@article{shao2023DetectingGroundingMultiModal,
  title = {Detecting and {{Grounding Multi-Modal Media Manipulation}}},
  author = {Shao, Rui and Wu, Tianxing and Liu, Ziwei},
  year = 2023,
  month = jun,
  journal = {2023 IEEE/CVF Conference on Computer Vision and Pattern Recognition (CVPR)},
  pages = {6904--6913},
  publisher = {IEEE},
  address = {Vancouver, BC, Canada},
  doi = {10.1109/CVPR52729.2023.00667},
  urldate = {2024-03-26},
  copyright = {https://doi.org/10.15223/policy-029},
  isbn = {9798350301298}
}

@inproceedings{shirali2023TheoryDynamicBenchmarks,
  title = {A {{Theory}} of {{Dynamic Benchmarks}}},
  booktitle = {{{ICLR}} 2023},
  author = {Shirali, Ali and Abebe, Rediet and Hardt, Moritz},
  year = 2023,
  month = mar,
  eprint = {2210.03165},
  primaryclass = {cs},
  doi = {10.48550/arXiv.2210.03165},
  urldate = {2025-03-21},
  archiveprefix = {arXiv}
}

@inproceedings{skoularikis2025HumorArtMisinformation,
  title = {'{{Humor}}, {{Art}}, or {{Misinformation}}?': {{A Multimodal Dataset}} for {{Intent-Aware Synthetic Image Detection}}},
  shorttitle = {'{{Humor}}, {{Art}}, or {{Misinformation}}?},
  booktitle = {Proceedings of the 2nd {{International Workshop}} on {{Diffusion}} of {{Harmful Content}} on {{Online Web}}},
  author = {Skoularikis, Anastasios and Papadopoulos, Stefanos-Iordanis and Papadopoulos, Symeon and Petrantonakis, Panagiotis C.},
  year = 2025,
  month = oct,
  series = {{{DHOW}} '25},
  pages = {95--104},
  publisher = {Association for Computing Machinery},
  address = {New York, NY, USA},
  doi = {10.1145/3746275.3762215},
  urldate = {2026-01-04},
  isbn = {979-8-4007-2057-4}
}

@inproceedings{staliunaite2025Dis2DisExplainingAmbiguity,
  title = {{{Dis2Dis}}: {{Explaining Ambiguity}} in {{Fact-Checking}}},
  shorttitle = {{{Dis2Dis}}},
  booktitle = {Findings of the {{Association}} for {{Computational Linguistics}}: {{NAACL}} 2025},
  author = {Staliunaite, Ieva and Vlachos, Andreas},
  editor = {Chiruzzo, Luis and Ritter, Alan and Wang, Lu},
  year = 2025,
  month = apr,
  pages = {246--267},
  publisher = {Association for Computational Linguistics},
  address = {Albuquerque, New Mexico},
  doi = {10.18653/v1/2025.findings-naacl.14},
  urldate = {2025-06-17},
  isbn = {979-8-89176-195-7}
}

@misc{su2025ClimateVizBenchmarkStatistical,
  title = {{{ClimateViz}}: {{A Benchmark}} for {{Statistical Reasoning}} and {{Fact Verification}} on {{Scientific Charts}}},
  shorttitle = {{{ClimateViz}}},
  author = {Su, Ruiran and Si, Jiasheng and Guo, Zhijiang and Pierrehumbert, Janet B.},
  year = 2025,
  month = jun,
  number = {arXiv:2506.08700},
  eprint = {2506.08700},
  primaryclass = {cs},
  publisher = {arXiv},
  doi = {10.48550/arXiv.2506.08700},
  urldate = {2025-06-17},
  archiveprefix = {arXiv}
}

@inproceedings{sundriyal2023ChaosClarityClaim,
  title = {From {{Chaos}} to {{Clarity}}: {{Claim Normalization}} to {{Empower Fact-Checking}}},
  shorttitle = {From {{Chaos}} to {{Clarity}}},
  booktitle = {Findings of the {{Association}} for {{Computational Linguistics}}: {{EMNLP}} 2023},
  author = {Sundriyal, Megha and Chakraborty, Tanmoy and Nakov, Preslav},
  editor = {Bouamor, Houda and Pino, Juan and Bali, Kalika},
  year = 2023,
  month = dec,
  pages = {6594--6609},
  publisher = {Association for Computational Linguistics},
  address = {Singapore},
  doi = {10.18653/v1/2023.findings-emnlp.439},
  urldate = {2025-01-24}
}

@inproceedings{tang2024M3DMultiModalMultiDocument,
  title = {{{M3D}}: {{MultiModal MultiDocument Fine-Grained Inconsistency Detection}}},
  shorttitle = {{{M3D}}},
  booktitle = {Proceedings of the 2024 {{Conference}} on {{Empirical Methods}} in {{Natural Language Processing}}},
  author = {Tang, Chia-Wei and Chen, Ting-Chih and Nguyen, Kiet A. and Mehrab, Kazi Sajeed and Ishmam, Alvi Md and Thomas, Chris},
  editor = {{Al-Onaizan}, Yaser and Bansal, Mohit and Chen, Yun-Nung},
  year = 2024,
  month = nov,
  pages = {22270--22293},
  publisher = {Association for Computational Linguistics},
  address = {Miami, Florida, USA},
  doi = {10.18653/v1/2024.emnlp-main.1243},
  urldate = {2025-01-23}
}

@inproceedings{tchechmedjiev2019ClaimsKGKnowledgeGraph,
  title = {{{ClaimsKG}}: {{A Knowledge Graph}} of {{Fact-Checked Claims}}},
  shorttitle = {{{ClaimsKG}}},
  booktitle = {The {{Semantic Web}} -- {{ISWC}} 2019: 18th {{International Semantic Web Conference}}, {{Auckland}}, {{New Zealand}}, {{October}} 26--30, 2019, {{Proceedings}}, {{Part II}}},
  author = {Tchechmedjiev, Andon and Fafalios, Pavlos and Boland, Katarina and Gasquet, Malo and Zloch, Matth{\"a}us and Zapilko, Benjamin and Dietze, Stefan and Todorov, Konstantin},
  year = 2019,
  month = oct,
  pages = {309--324},
  publisher = {Springer-Verlag},
  address = {Berlin, Heidelberg},
  doi = {10.1007/978-3-030-30796-7_20},
  urldate = {2025-03-21},
  isbn = {978-3-030-30795-0}
}

@inproceedings{thorne2018FEVERLargescaleDataset,
  title = {{{FEVER}}: A {{Large-scale Dataset}} for {{Fact Extraction}} and {{VERification}}},
  shorttitle = {{{FEVER}}},
  booktitle = {Proceedings of the 2018 {{Conference}} of the {{North American Chapter}} of the {{Association}} for {{Computational Linguistics}}: {{Human Language Technologies}}, {{Volume}} 1 ({{Long Papers}})},
  author = {Thorne, James and Vlachos, Andreas and Christodoulopoulos, Christos and Mittal, Arpit},
  editor = {Walker, Marilyn and Ji, Heng and Stent, Amanda},
  year = 2018,
  month = jun,
  pages = {809--819},
  publisher = {Association for Computational Linguistics},
  address = {New Orleans, Louisiana},
  doi = {10.18653/v1/N18-1074},
  urldate = {2026-01-04}
}

@inproceedings{tonglet2024ImageTellMe,
  title = {``{{Image}}, {{Tell}} Me Your Story!'' {{Predicting}} the Original Meta-Context of Visual Misinformation},
  booktitle = {Proceedings of the 2024 {{Conference}} on {{Empirical Methods}} in {{Natural Language Processing}}},
  author = {Tonglet, Jonathan and Moens, Marie-Francine and Gurevych, Iryna},
  editor = {{Al-Onaizan}, Yaser and Bansal, Mohit and Chen, Yun-Nung},
  year = 2024,
  month = nov,
  pages = {7845--7864},
  publisher = {Association for Computational Linguistics},
  address = {Miami, Florida, USA},
  doi = {10.18653/v1/2024.emnlp-main.448},
  urldate = {2026-01-04}
}

@misc{tonglet2025COVECOntextVEracity,
  title = {{{COVE}}: {{COntext}} and {{VEracity}} Prediction for out-of-Context Images},
  shorttitle = {{{COVE}}},
  author = {Tonglet, Jonathan and Thiem, Gabriel and Gurevych, Iryna},
  year = 2025,
  month = feb,
  number = {arXiv:2502.01194},
  eprint = {2502.01194},
  primaryclass = {cs},
  publisher = {arXiv},
  doi = {10.48550/arXiv.2502.01194},
  urldate = {2025-02-12},
  archiveprefix = {arXiv}
}

@inproceedings{wang2017LiarLiarPants,
  title = {``{{Liar}}, {{Liar Pants}} on {{Fire}}'': {{A New Benchmark Dataset}} for {{Fake News Detection}}},
  shorttitle = {``{{Liar}}, {{Liar Pants}} on {{Fire}}''},
  booktitle = {Proceedings of the 55th {{Annual Meeting}} of the {{Association}} for {{Computational Linguistics}} ({{Volume}} 2: {{Short Papers}})},
  author = {Wang, William Yang},
  editor = {Barzilay, Regina and Kan, Min-Yen},
  year = 2017,
  month = jul,
  pages = {422--426},
  publisher = {Association for Computational Linguistics},
  address = {Vancouver, Canada},
  doi = {10.18653/v1/P17-2067},
  urldate = {2024-04-18}
}

@article{wang2021UnderstandingUseFauxtography,
  title = {Understanding the {{Use}} of {{Fauxtography}} on {{Social Media}}},
  author = {Wang, Yuping and Tahmasbi, Fatemeh and Blackburn, Jeremy and Bradlyn, Barry and Cristofaro, Emiliano De and Magerman, David and Zannettou, Savvas and Stringhini, Gianluca},
  year = 2021,
  month = may,
  journal = {Proceedings of the International AAAI Conference on Web and Social Media},
  volume = {15},
  pages = {776--786},
  issn = {2334-0770},
  doi = {10.1609/icwsm.v15i1.18102},
  urldate = {2026-01-04},
  copyright = {Copyright (c) 2021 Association for the Advancement of Artificial Intelligence},
  langid = {english}
}

@inproceedings{wang2025PiecingItAll,
  title = {Piecing {{It All Together}}: {{Verifying Multi-Hop Multimodal Claims}}},
  shorttitle = {Piecing {{It All Together}}},
  booktitle = {Proceedings of the 31st {{International Conference}} on {{Computational Linguistics}}},
  author = {Wang, Haoran and Rangapur, Aman and Xu, Xiongxiao and Liang, Yueqing and Gharwi, Haroon and Yang, Carl and Shu, Kai},
  editor = {Rambow, Owen and Wanner, Leo and Apidianaki, Marianna and {Al-Khalifa}, Hend and Eugenio, Barbara Di and Schockaert, Steven},
  year = 2025,
  month = jan,
  pages = {7453--7469},
  publisher = {Association for Computational Linguistics},
  address = {Abu Dhabi, UAE},
  url = {https://aclanthology.org/2025.coling-main.498/},
  urldate = {2026-01-04}
}

@misc{xiao2025XFactaContemporaryRealWorld,
  title = {{{XFacta}}: {{Contemporary}}, {{Real-World Dataset}} and {{Evaluation}} for {{Multimodal Misinformation Detection}} with {{Multimodal LLMs}}},
  shorttitle = {{{XFacta}}},
  author = {Xiao, Yuzhuo and Han, Zeyu and Wang, Yuhan and Jiang, Huaizu},
  year = 2025,
  month = aug,
  number = {arXiv:2508.09999},
  eprint = {2508.09999},
  primaryclass = {cs},
  publisher = {arXiv},
  doi = {10.48550/arXiv.2508.09999},
  urldate = {2026-01-04},
  archiveprefix = {arXiv}
}

@article{xu2024M3AMultimodalMisinformation,
  title = {{{M3A}}: {{A}} Multimodal Misinformation Dataset for Media Authenticity Analysis},
  shorttitle = {{{M3A}}},
  author = {Xu, Qingzheng and Chen, Huiqiang and Du, Heming and Zhang, Hu and {\L}ukasik, Szymon and Zhu, Tianqing and Yu, Xin},
  year = 2024,
  month = dec,
  journal = {Computer Vision and Image Understanding},
  volume = {249},
  pages = {104205},
  publisher = {Elsevier BV},
  issn = {1077-3142},
  doi = {10.1016/j.cviu.2024.104205},
  urldate = {2024-11-01},
  langid = {english}
}

@inproceedings{xu2024MMOOCMultimodalMisinformation,
  title = {{{MMOOC}}: {{A Multimodal Misinformation Dataset}} for~{{Out-of-Context News Analysis}}},
  shorttitle = {{{MMOOC}}},
  booktitle = {Information {{Security}} and {{Privacy}}},
  author = {Xu, Qingzheng and Du, Heming and Chen, Huiqiang and Liu, Bo and Yu, Xin},
  editor = {Zhu, Tianqing and Li, Yannan},
  year = 2024,
  pages = {444--459},
  publisher = {Springer Nature},
  address = {Singapore},
  doi = {10.1007/978-981-97-5101-3_24},
  isbn = {978-981-97-5101-3},
  langid = {english}
}

@inproceedings{xu2025MDAM3MisinformationDetection,
  title = {{{MDAM3}}: {{A Misinformation Detection}} and {{Analysis Framework}} for {{Multitype Multimodal Media}}},
  shorttitle = {{{MDAM3}}},
  booktitle = {Proceedings of the {{ACM}} on {{Web Conference}} 2025},
  author = {Xu, Qingzheng and Du, Heming and {\L}ukasik, Szymon and Zhu, Tianqing and Wang, Sen and Yu, Xin},
  year = 2025,
  month = apr,
  series = {{{WWW}} '25},
  pages = {5285--5296},
  publisher = {Association for Computing Machinery},
  address = {New York, NY, USA},
  doi = {10.1145/3696410.3714498},
  urldate = {2025-06-17},
  isbn = {979-8-4007-1274-6}
}

@inproceedings{yang2025NewDatasetBenchmark,
  title = {A {{New Dataset}} and {{Benchmark}} for {{Grounding Multimodal Misinformation}}},
  booktitle = {Proceedings of the 33rd {{ACM International Conference}} on {{Multimedia}}},
  author = {Yang, Bingjian and Xu, Danni and Niu, Kaipeng and Liu, Wenxuan and Wang, Zheng and Kankanhalli, Mohan},
  year = 2025,
  month = oct,
  series = {{{MM}} '25},
  pages = {12571--12577},
  publisher = {Association for Computing Machinery},
  address = {New York, NY, USA},
  doi = {10.1145/3746027.3758191},
  urldate = {2026-01-03},
  isbn = {979-8-4007-2035-2}
}

@inproceedings{yang2025RealFactBenchBenchmarkEvaluating,
  title = {{{RealFactBench}}: {{A Benchmark}} for {{Evaluating Large Language Models}} in {{Real-World Fact-Checking}}},
  shorttitle = {{{RealFactBench}}},
  booktitle = {Proceedings of the 33rd {{ACM International Conference}} on {{Multimedia}}},
  author = {Yang, Shuo and Dai, Yuqin and Wang, Guoqing and Zheng, Xinran and Xu, Jinfeng and Li, Jinze and Ying, Zhenzhe and Wang, Weiqiang and Ngai, Edith C. H.},
  year = 2025,
  month = oct,
  series = {{{MM}} '25},
  pages = {13435--13441},
  publisher = {Association for Computing Machinery},
  address = {New York, NY, USA},
  doi = {10.1145/3746027.3758307},
  urldate = {2026-01-04},
  isbn = {979-8-4007-2035-2}
}

@inproceedings{yao2023EndtoEndMultimodalFactChecking,
  title = {End-to-{{End Multimodal Fact-Checking}} and {{Explanation Generation}}: {{A Challenging Dataset}} and {{Models}}},
  shorttitle = {End-to-{{End Multimodal Fact-Checking}} and {{Explanation Generation}}},
  booktitle = {Proceedings of the 46th {{International ACM SIGIR Conference}} on {{Research}} and {{Development}} in {{Information Retrieval}}},
  author = {Yao, Barry Menglong and Shah, Aditya and Sun, Lichao and Cho, Jin-Hee and Huang, Lifu},
  year = 2023,
  month = jul,
  eprint = {2205.12487},
  primaryclass = {cs},
  pages = {2733--2743},
  doi = {10.1145/3539618.3591879},
  urldate = {2024-01-22},
  archiveprefix = {arXiv},
  langid = {english}
}

@inproceedings{zannettou2018OriginsMemesMeans,
  title = {On the {{Origins}} of {{Memes}} by {{Means}} of {{Fringe Web Communities}}},
  booktitle = {Proceedings of the {{Internet Measurement Conference}} 2018},
  author = {Zannettou, Savvas and Caulfield, Tristan and Blackburn, Jeremy and De Cristofaro, Emiliano and Sirivianos, Michael and Stringhini, Gianluca and {Suarez-Tangil}, Guillermo},
  year = 2018,
  month = oct,
  series = {{{IMC}} '18},
  pages = {188--202},
  publisher = {Association for Computing Machinery},
  address = {New York, NY, USA},
  doi = {10.1145/3278532.3278550},
  urldate = {2026-01-04},
  isbn = {978-1-4503-5619-0}
}

@inproceedings{zlatkova2019FactCheckingMeetsFauxtography,
  title = {Fact-{{Checking Meets Fauxtography}}: {{Verifying Claims About Images}}},
  shorttitle = {Fact-{{Checking Meets Fauxtography}}},
  booktitle = {Proceedings of the 2019 {{Conference}} on {{Empirical Methods}} in {{Natural Language Processing}} and the 9th {{International Joint Conference}} on {{Natural Language Processing}} ({{EMNLP-IJCNLP}})},
  author = {Zlatkova, Dimitrina and Nakov, Preslav and Koychev, Ivan},
  editor = {Inui, Kentaro and Jiang, Jing and Ng, Vincent and Wan, Xiaojun},
  year = 2019,
  month = nov,
  pages = {2099--2108},
  publisher = {Association for Computational Linguistics},
  address = {Hong Kong, China},
  doi = {10.18653/v1/D19-1216},
  urldate = {2024-01-25}
}

\clearpage
\appendix
\section{LLM Glossary}
\label{app:llms}
Tab.~\ref{tab:llms} summarizes the LLMs used in the \method construction pipeline. At the time of processing Q1 2020 -- Q4 2025, rate limits were too restrictive for the newly introduced \textsc{Gemini~3} family, therefore we defaulted to \geminipro and \geminiflash. We upgraded to \textsc{Gemini~3.1~Pro} and \textsc{Gemini~3~Flash} for Q1 2026.

\begin{table}[h]
    \centering
    \resizebox{\linewidth}{!}{
    \begin{tabular}{l|ll}
        \toprule
        \textbf{LLM name} & \textbf{Stages} & \textbf{Version specifier} \\
        \midrule
        \gpt & 5, 6 & \texttt{gpt-5.2-2025-12-11} \\
        \gptmini & 4 -- 7 & \texttt{gpt-5-mini-2025-08-07} \\
        \gptnano & 3 & \texttt{gpt-5-nano-2025-08-07} \\
        \geminipro & 5, 6, 7 & \texttt{gemini-2.5-pro} \\
        \textsc{Gemini~3.1~Pro} & 5, 6, 7 & \texttt{gemini-3.1-pro-preview} \\
        \geminiflash & 5, 7 & \texttt{gemini-2.5-flash} \\
        \textsc{Gemini~3~Flash} & 5, 7 & \texttt{gemini-3-flash-preview} \\
        \claude & 6 & \texttt{claude-sonnet-4-5-20250929} \\
        \llama & 6 & \texttt{Llama-4-Maverick-17B-128E-Instruct-FP8} \\
        \bottomrule
    \end{tabular}
    }
    \caption{The Large Language Models (LLMs) used in this work, along with the version specifier and the corresponding \method pipeline stage where they were used. Stages with \textsc{Gemini~2.5} models were upgraded to the \textsc{Gemini~3} family after it became available.}
    \label{tab:llms}
\end{table}

\section{\method Stage Implementation Details}
\label{app:method_details}

\paragraph{Stage 1: Review Discovery}
ClaimReviews are obtained via the Google Fact Check Tools API\footnote{\href{https://developers.google.com/fact-check/tools/api}{developers.google.com/fact-check/tools/api}} and DataCommons\footnote{\href{https://datacommons.org/factcheck/}{datacommons.org/factcheck}}. Extracted review instances are refreshed by re-downloading ClaimReview data directly from the publisher websites. Language is determined using \texttt{langdetect}\footnote{\href{https://pypi.org/project/langdetect/}{pypi.org/project/langdetect}}.

We observed several reviews that indicate a claim date after the review's publication date. Since it is not possible to fact-check ``future'' claims, we assume this is an error in the metadata provided by the publishing fact-checking organization. Therefore, we dismiss all of these $1.2$\,K ($0.3\%$) reviews.

\paragraph{Stage 2: Publisher Identification}
We consider a fact-checking organization as credible if it is a signatory of the International Fact-Checking Network (IFCN)\footnote{\href{https://ifcncodeofprinciples.poynter.org/signatories}{ifcncodeofprinciples.poynter.org/signatories}} or member of the European Fact-Checking Standards Network (EFCSN)\footnote{\href{https://members.efcsn.com/signatories}{members.efcsn.com/signatories}}.

\paragraph{Stage 3: Article Scraping}
We scrape article content for each review using a dynamic engine based on \textsc{Firecrawl}\footnote{\href{https://github.com/firecrawl/firecrawl}{github.com/firecrawl/firecrawl}} and \textsc{Decodo}\footnote{\href{https://decodo.com}{decodo.com}}. Missing review publication and modification dates are obtained via HTML metadata.

\paragraph{Stage 4: Appearance Retrieval}
Appearances are scraped via \texttt{scrapeMM}\footnote{\href{https://github.com/multimodal-ai-lab/scrapeMM}{github.com/multimodal-ai-lab/scrapeMM}} which downloads media and text through social media APIs and via HTTP requests using \texttt{yt-dlp}\footnote{\href{https://github.com/yt-dlp/yt-dlp}{github.com/yt-dlp/yt-dlp}}, \textsc{Firecrawl}, and \textsc{Decodo}.

\paragraph{Stage 6: Verdict Standardization}
To enable the ensemble LLMs to meaningfully return scoring values on the $[-1, 1]$ interval, we discretize it into $7$ bins as detailed in Fig.~\ref{fig:binning}.

For media authenticity, if the decision maps to \fabricated, LLMs (and human annotators) may provide additional tags for finer-grained fabrication annotation, including:
\begin{itemize}
    \item \emph{AI-generated}: The media was (partially or entirely) synthesized by AI.
    \item \emph{Manipulated}: The media is derived from a real recording, but was altered to change its meaning.
    \item \emph{Forged}: The media is mostly or entirely a manually created invention.
\end{itemize}
When aggregating tags from multiple predictors, majority voting is applied for each tag individually (present vs.\ not present).

\section{Score Discretization}
\label{app:score_discretization}
Instead of a finite set of labels, \method uses continuous scores to rate claims and media w.r.t.\ their specific properties. Scores range from $-1$ (the \negative extreme) to $1$ (the \positive extreme), where values in between indicate different degrees of uncertainty. This allows the use of distance-based metrics and straightforward ensemble rating aggregation.

\begin{figure}[t]
    \centering
    \includegraphics[width=\linewidth]{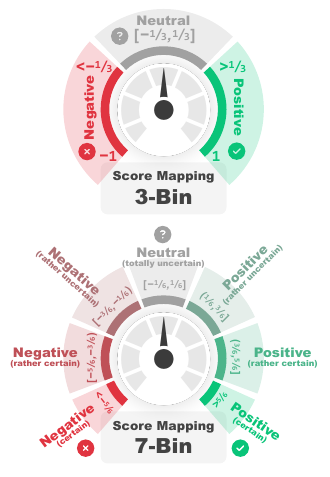}
    \caption{Discretization strategies for mapping from the continuous rating interval $[-1, 1]$ to $3$ or $7$ discrete labels, respectively.}
    \label{fig:binning}
\end{figure}

However, when computing accuracy or using LLMs to predict ratings, the scale must be discretized. We use two different discretization strategies: $3$-bin discretization (representing the conventional classification approach) and a more fine-grained $7$-bin approach that accounts for uncertainty. We use the $3$-bin solely for reporting accuracy, while the $7$-bin is applied to LLM predictions. Please refer to Fig.~\ref{fig:binning} for the exact scale-to-label binning mapping.

\section{On the Choice of the Metrics}
\label{app:metrics}

We report four metrics: Mean Squared Error (MSE), Mean Absolute Error (MAE), and Accuracy ($3$-bin and $7$-bin). Since \method operates on a continuous label space, error-based metrics are more suitable to capture graded deviations from the ground truth.

We adopt MSE as the primary metric for performance evaluation for two reasons. First, unlike Accuracy, MSE accounts for the magnitude of errors. Predictions close to the ground truth incur only a small penalty, whereas Accuracy assigns the same penalty to all misclassifications regardless of distance (cf.\ Fig.~\ref{fig:metric_comparison}). Second, compared to MAE, MSE penalizes large deviations more strongly: Confusing opposite extremes (e.g., \true vs.\ \false) incurs a substantially higher penalty than small errors. This behavior better reflects the practical cost of severe misjudgments while remaining more forgiving for near-correct predictions. Table~\ref{tab:metric_interpretation} shows how to interpret the MSE values in the context of \method.

\begin{table*}[]
    \centering
    \begin{tabular}{l|llp{7cm}}
        \toprule
        \textbf{MSE} & \textbf{Interpretation} & \textbf{Max.\ flipped} & \textbf{Equivalent to...} \\
        \midrule
        0.00 & Perfect & 0 & Exact match. \\
        0.04 & Very Good & 1 in 100 & 36 predictions being off by $1/3$ in 100 otherwise perfect predictions. \\
        0.10 & Good & 1 in 40 & Slightly better than being off by $1/3$ for all predictions. \\
        1.00 & Abstention & 1 in 4 & Being off by 1 always (roughly same as constantly predicting 0, i.e., \nei). \\
        \bottomrule
    \end{tabular}
    \caption{How to interpret the Mean Squared Error (MSE) when evaluating with \method. \textbf{Max.\ flipped} indicates, in a set of otherwise perfect predictions, the maximum number of flipped predictions, which are predictions that have a score difference of $2$ to the target (i.e., complete opposite).}
    \label{tab:metric_interpretation}
\end{table*}

\begin{figure}
    \centering
    \includegraphics[width=\linewidth]{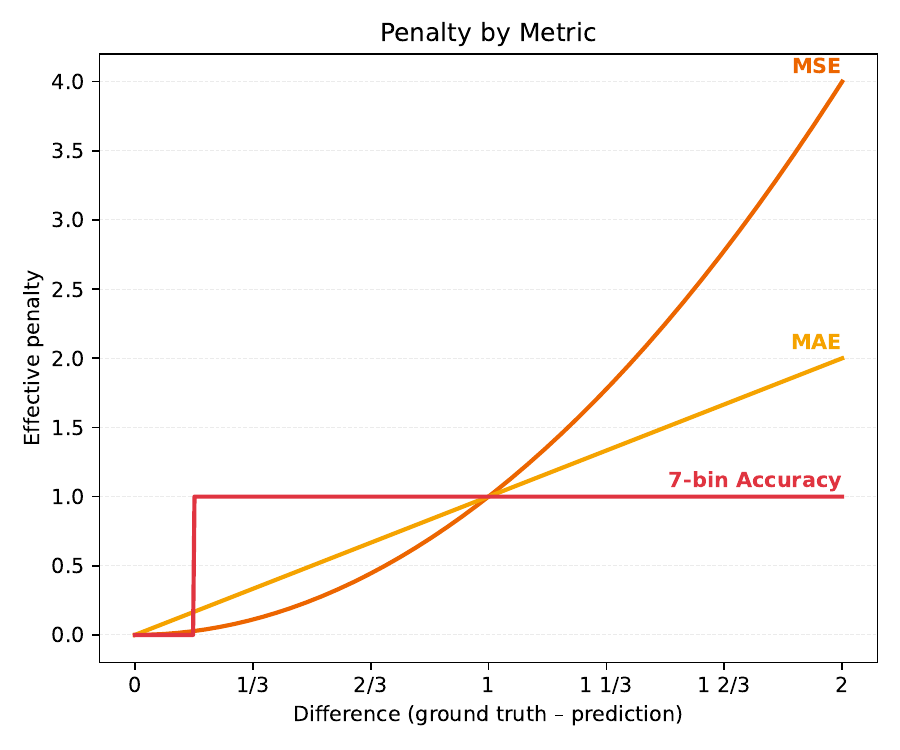}
    \caption{Comparison of metrics used for \method. Lines indicate the effective penalty on the final evaluation score. \highlight{MSE} = Mean Squared Error (main metric of \method), \textbf{MAE} = Mean Absolute Error.}
    \label{fig:metric_comparison}
\end{figure}

\section{\method Release and Subsampling}
\label{app:subsampling}
To compose the final release data, we subsample the available claims so that the number of \intact claims is equal to the number of \compromised claims for each quarter (holding true also for the longitudinal split). Moreover, text-only claims are rather overrepresented compared to media-based claims as the latter are dismissed more frequently for missing media. Therefore, we prioritize media-based claims over text-only claims during subsampling, setting $80\%$ as the maximum share of media-based claims as determined by \citet{dufour2024AMMeBaLargeScaleSurvey}. To improve media quality, claims with media from original appearances are prioritized over claims with media from archives (typically saving media in reduced resolution) and claims with media from fact-checking articles (as the latter often contain annotations/edits). Additionally, original \intact claims take precedence over rectified claims.

\section{Analysis of Potential Stylistic Shortcuts in Rectified Claims}
\label{app:stylistic_shortcuts}
A potential concern is that rectified claims may contain stylistic artifacts that models could exploit as shortcuts, rather than relying on factual reasoning. To assess this empirically, we conducted an additional experiment evaluating \gpt and \textsc{Gemini 3 Pro} on three subsets: (i) all $147$ \intact, original claims, (ii) $300$ \intact, rectified claims, and (iii) $300$ \compromised, original claims.

\begin{table}[t]
\centering
\small
\begin{tabular}{lccc}
\toprule
& \textbf{\makecell{\intact \\ (original)}} & \textbf{\makecell{\intact \\ (rectified)}} & \textbf{\makecell{\compr \\ (original)}} \\
\midrule
\multicolumn{4}{c}{\textbf{\gpt}} \\
\midrule
$N$ Claims & 147 & 300 & 300 \\
MSE $\downarrow$ & 0.577 & 0.460 & 0.189 \\
MAE $\downarrow$ & 0.450 & 0.423 & 0.233 \\
$3$-bin Acc. $\uparrow$ & 71.4 & 66.3 & 82.0 \\
\midrule
\multicolumn{4}{c}{\textbf{\textsc{Gemini 3 Pro}}} \\
\midrule
$N$ Claims & 147 & 300 & 300 \\
MSE $\downarrow$ & 0.264 & 0.270 & 0.349 \\
MAE $\downarrow$ & 0.264 & 0.264 & 0.336 \\
$3$-bin Acc. $\uparrow$ & 88.4 & 83.7 & 74.7 \\
\bottomrule
\end{tabular}
\caption{Performance comparison on original vs.\ rectified claims for intact and compromised subsets, Accuracy in \%.}
\label{tab:stylistic_shortcuts}
\end{table}

The results (depicted in Tab.~\ref{tab:stylistic_shortcuts}) provide no strong evidence for a systematic shortcut effect. While \textsc{GPT-5.2} exhibits a moderate improvement on rectified intact claims over original intact ones (MSE difference of approximately 0.12), performance remains far from trivial, with substantial residual errors. In contrast, \textsc{Gemini 3 Pro} shows no consistent advantage for rectified claims; performance is comparable or slightly worse than on original claims.

Overall, these findings suggest that, even if minor stylistic signals are present, they do not translate into a reliable or substantial shortcut across models. However, this analysis does not preclude the possibility that such patterns could be exploited under targeted prompting or finetuning, which we leave for future investigation.

\section{Analysis of Duplicates in \method}
\label{app:duplicate_analysis}

A potential concern is the presence of duplicate or highly similar claims, which could reduce effective diversity. To quantify redundancy in \method, we computed pairwise cosine similarity over all $25$\,K claim text embeddings using OpenAI's \texttt{text-embedding-3-large} model.

\begin{table*}[t]
\centering
\small
\begin{tabular}{l p{4cm} l l}
\toprule
\textbf{Cos-sim} & \textbf{Typical Observation} & \textbf{\# Claim Pairs} & \textbf{Worst-case Share} \\
\midrule
$\geq 0.90$ & Same topic, but different assertions & $280$ & $1.12\%$ \\
$\geq 0.95$ & Mostly same assertions, minor wording differences & $112$ & $0.45\%$ \\
$\geq 0.99$ & Nearly identical (1--2 word or punctuation differences) & $24$ & $0.10\%$ \\
\bottomrule
\end{tabular}
\caption{Redundancy analysis based on cosine similarity of claim text embeddings.}
\label{tab:duplicates}
\end{table*}

The results in Tab.~\ref{tab:duplicates} indicate that redundancy is limited. At most $0.45\%$ of claims exhibit substantial overlap in their textual assertions (cosine similarity $\geq 0.95$), while near-identical claims (cosine similarity $\geq 0.99$) account for only $0.1\%$. Importantly, these estimates represent a worst-case scenario based solely on textual similarity. When additionally considering differences in associated media or claim publication dates, the effective redundancy is further reduced.

Overall, this analysis suggests that \method maintains a high degree of diversity, with only minimal duplication.

\section{Sensitivity to Proprietary LLM Backbones}
\label{app:llm_dependence}

Dependence on proprietary models raises questions regarding reproducibility. To empirically assess the sensitivity of \method to the choice of backbone, we partially regenerated Q4 2025 using a substantially different LLM family.

Specifically, we replaced \textsc{GPT-5.2} with \textsc{Gemini 3.1 Pro} and \textsc{GPT-5.2 mini} with \textsc{Gemini 3 Flash} for all non-video claims. For claims involving videos, we substituted \textsc{Gemini 2.5 Pro} with \textsc{Gemini 3.1 Pro} and \textsc{Gemini 2.5 Flash} with \textsc{Gemini 3 Flash}. These changes affect stages 5 (Claim Normalization), 6 (Verdict Standardization), and 7 (Claim Rectification), while the ensemble used for verdict prediction in stage 6 remained unchanged.

In total, we regenerated $689$ claims and compared them against the original pipeline outputs. Manual inspection shows that regenerated claims preserve the same core assertions, with only minor differences in wording or level of detail. Quantitatively, comparing ensemble-generated verdicts on original versus regenerated claims yields an MSE of $0.0217$ across all low-level properties and an MSE of $0.0069$ for Integrity.

These results indicate extremely high agreement and suggest that the pipeline is largely invariant to the specific proprietary backbone employed. While this does not yet establish full reproducibility with open-weight models, it provides evidence that \method is not tightly coupled to a single model family. As open-weight multimodal models continue to improve, future iterations of the pipeline can replace individual components accordingly.

\section{\method Statistics}
\subsection{ClaimReview}
\label{app:cr_stats}
We obtained a total of $398,327\,K$ ClaimReviews starting from January 1, 2016. Three sources yielded the data: Google Fact-Check Explorer, Data Commons, and the fact-checking organizations themselves who added ClaimReview metadata to their published article webpages. The same ClaimReview can be obtained from multiple sources, where the ClaimReview directly downloaded from the publisher takes precedence over the others, as we expect it to be most up-to-date. Fig.~\ref{fig:cr_origin} shows the number of reviews returned by the three different sources. Fig.~\ref{fig:quarter_stats/reviews} displays the number of reviews per quarter since the introduction of ClaimReview in 2016. Fig.~\ref{fig:raw_ratings} shows the most common verdict labels as provided by the fact-checkers.
\begin{figure}
    \centering
    \includegraphics[width=\linewidth]{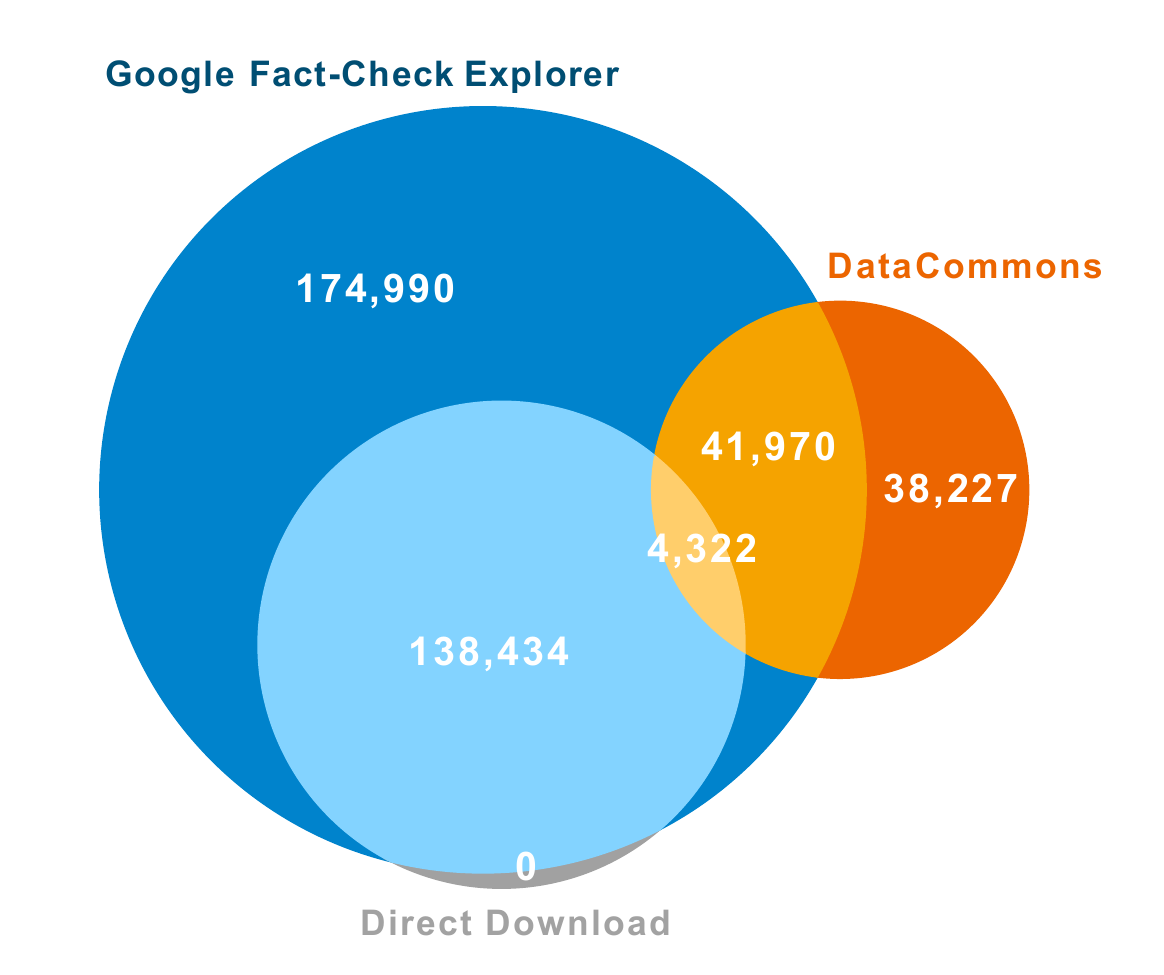}
    \caption{Origins of ClaimReviews as obtained by stage~1.}
    \label{fig:cr_origin}
\end{figure}

\begin{figure*}
    \centering
    \begin{subfigure}[t]{\textwidth}
        \centering
        \includegraphics[width=\linewidth]{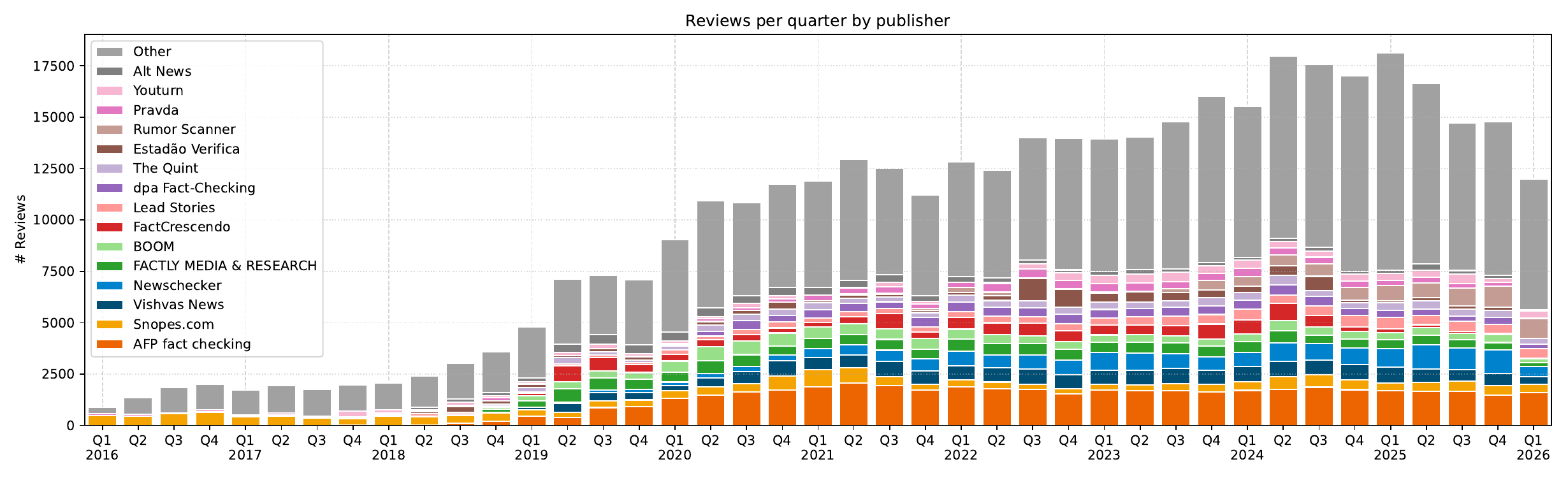}
        \caption{ClaimReviews by publishing organization.}
    \end{subfigure}
    
    \begin{subfigure}[t]{\textwidth}
        \centering
        \includegraphics[width=\linewidth]{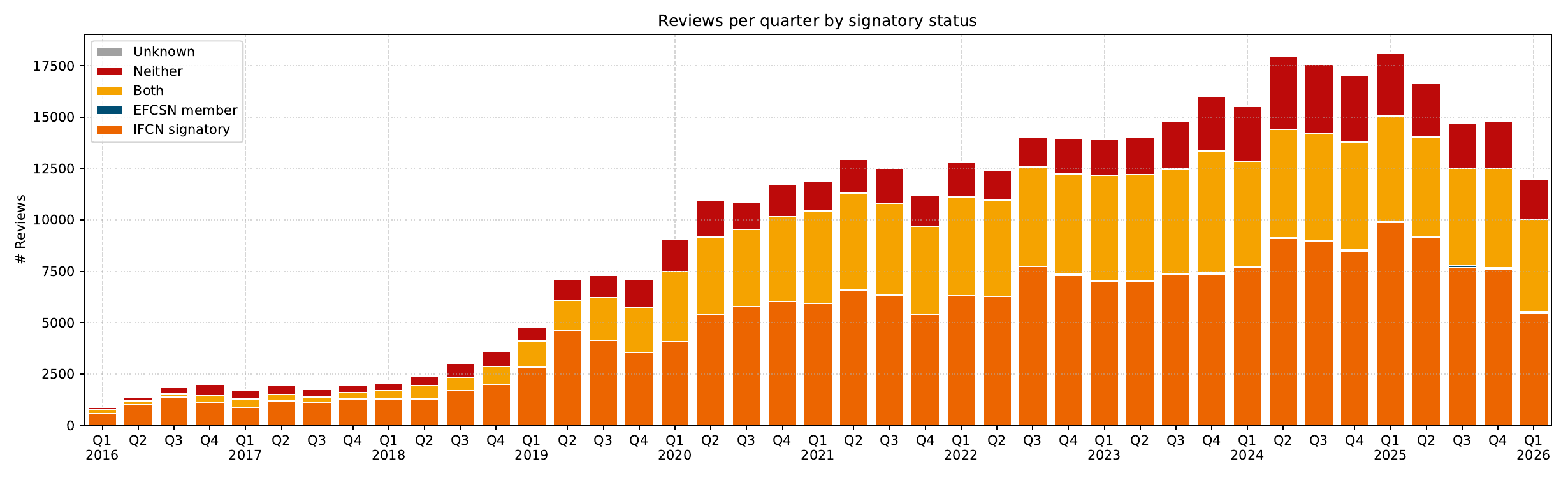}
        \caption{ClaimReviews by publisher signatory status.}
    \end{subfigure}
    
    \caption{ClaimReview statistics, as obtained by stages 1 and 2, showing \textit{all} ClaimReviews exposed by Google and Data Commons retrieved by Dec 31, 2025.}
    \label{fig:quarter_stats/reviews}
\end{figure*}

\begin{figure}
    \centering
    \includegraphics[width=\linewidth]{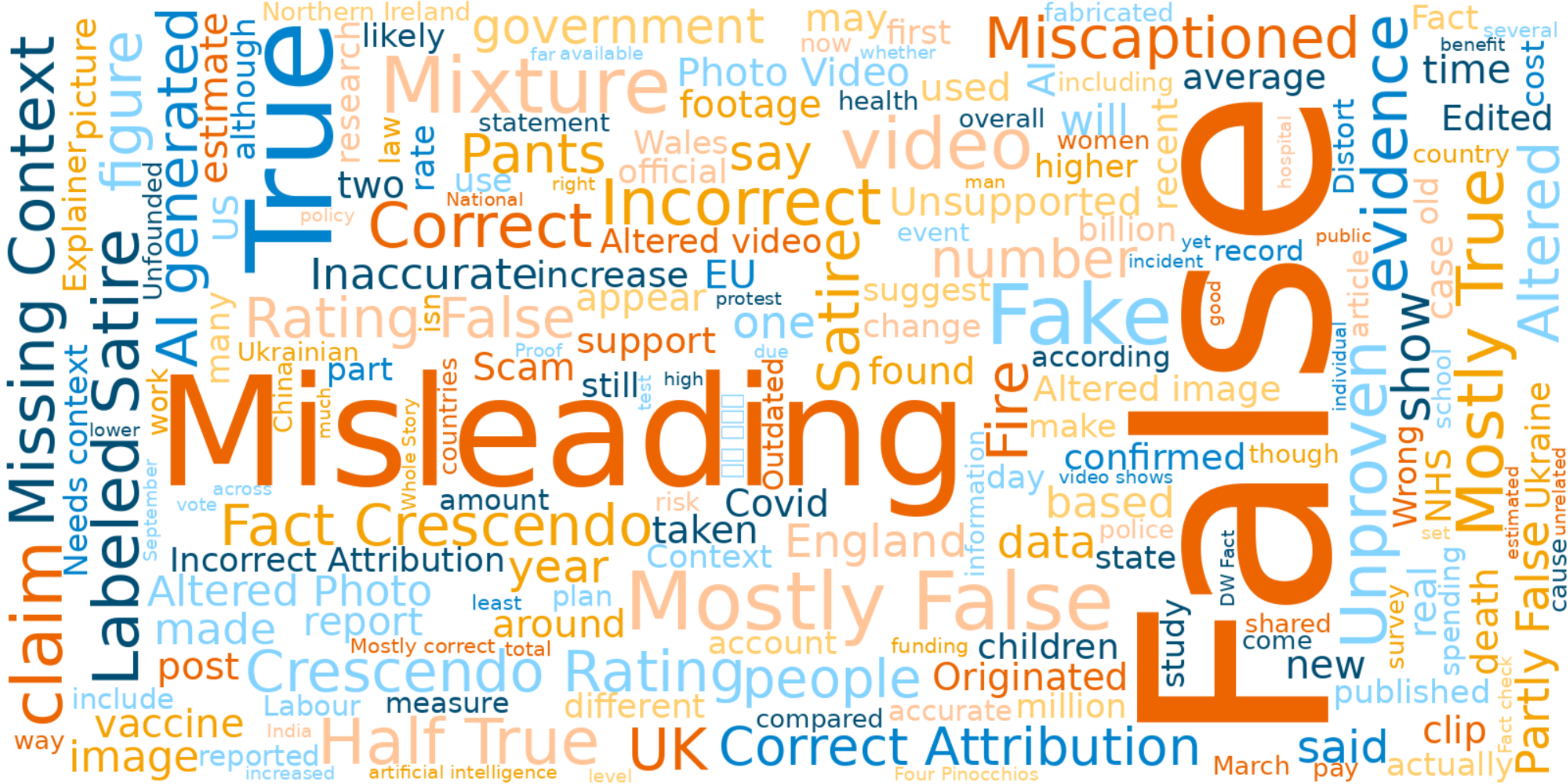}
    \caption{Raw ratings as provided by ClaimReview for English reviews.}
    \label{fig:raw_ratings}
\end{figure}

\subsection{Appearances}
\label{app:app_stats}

We identified $495,642$ appearances. Fig.~\ref{fig:appearance_url_vs_archive} depicts the shares of appearances, containing the URL to the original source and/or to the archived record. In about $92\%$ of the cases, we were able to identify the original appearance URL, only for $8\%$ of the appearances we did not. Note that claims often have multiple appearances. Almost $40\%$ of appearances also have a corresponding URL to an archived record. The top $10$ platforms where appearances occur are listed in Fig.~\ref{fig:appearance_domains}. The distribution of archiving services is shown in Fig.\ \ref{fig:appearance_archivers}. The most prominent are Perma.cc ($41.8\%$) and Archive.today ($38.2\%$). The quarterly shares of appearances by platform are shown in Fig.~\ref{fig:quarter_stats/claims/natural/platforms}.

\begin{figure}[t]
    \centering
    \includegraphics[width=0.7\linewidth]{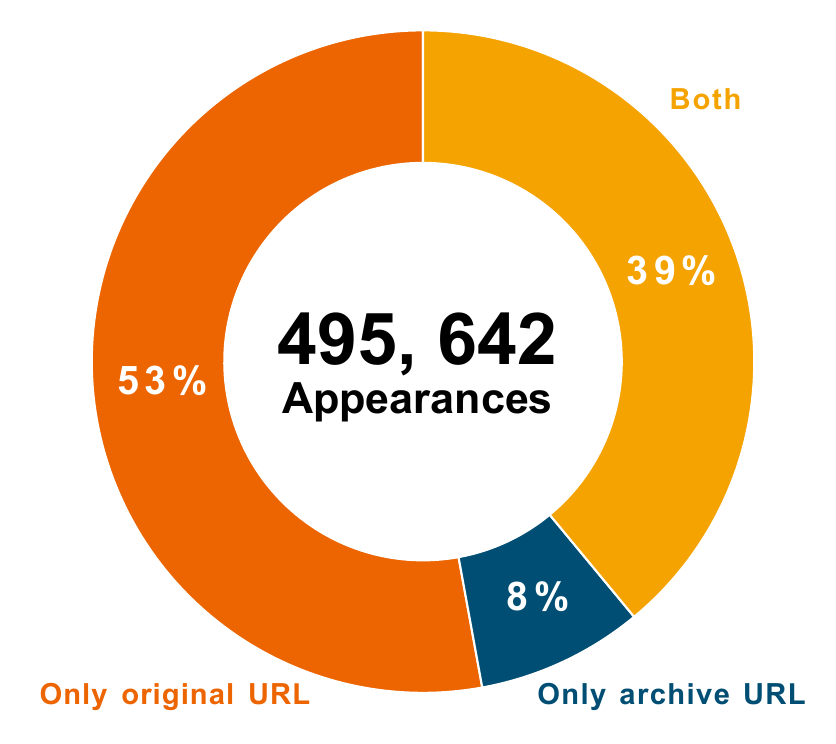}
    \caption{Distribution of appearances that have the original source URL, an archive URL, or both.}
    \label{fig:appearance_url_vs_archive}
\end{figure}

\begin{figure}[t]
    \centering
    \includegraphics[width=\linewidth]{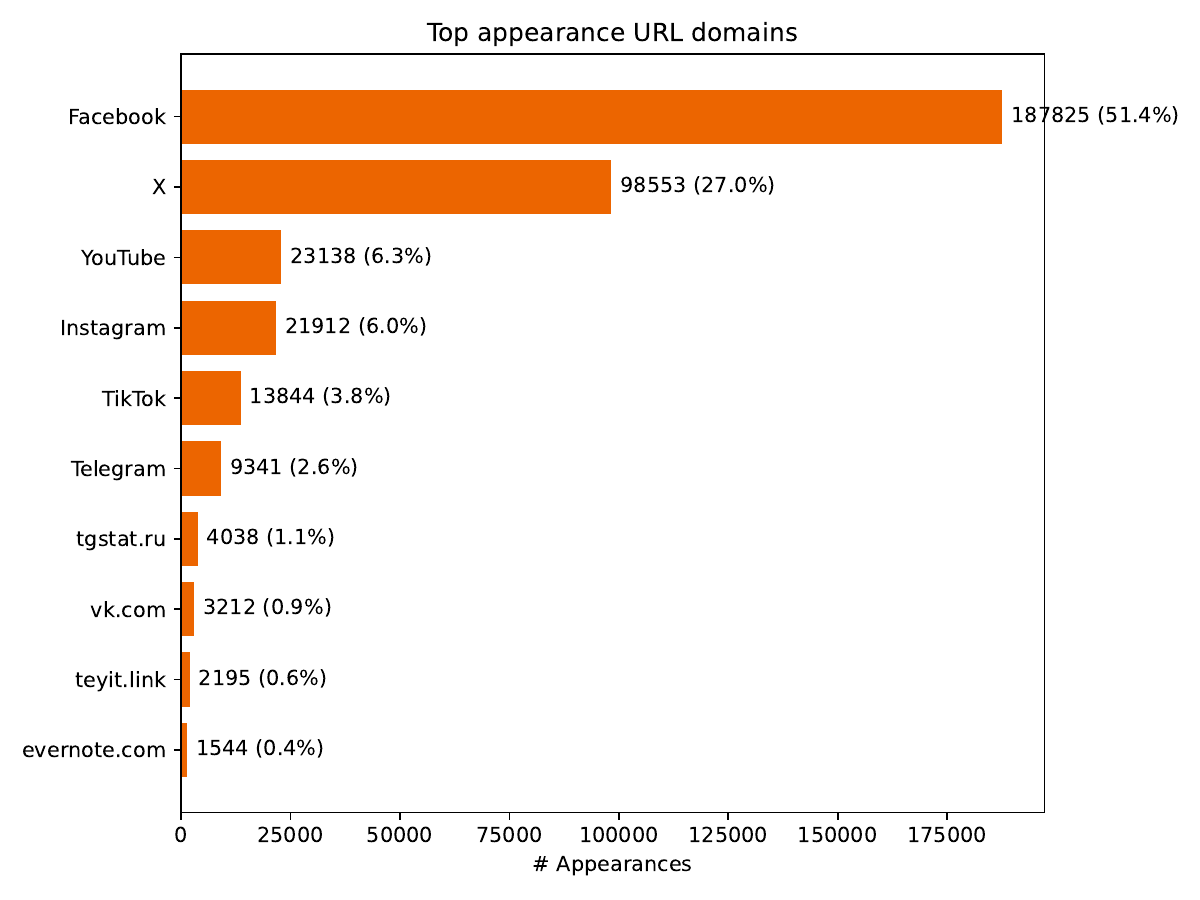}
    \caption{Appearances per platform, showing the top 10.}
    \label{fig:appearance_domains}
\end{figure}

\begin{figure}[t]
    \centering
    \includegraphics[width=\linewidth]{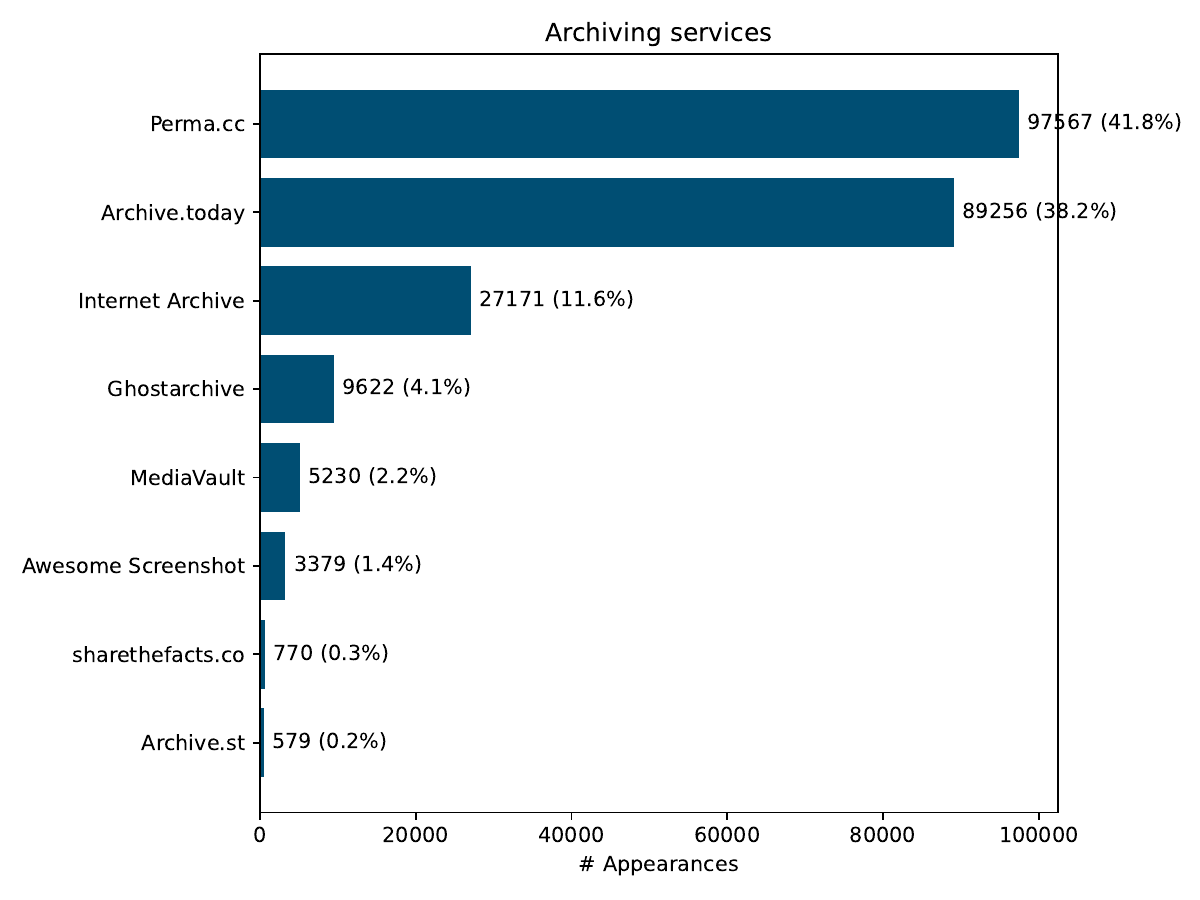}
    \caption{Appearances per archiving service.}
    \label{fig:appearance_archivers}
\end{figure}

\subsection{Claims}
\label{app:claim_stats}
A total of $86,372\,K$ claims (original, rectified, and dismissed) were obtained by stage $5$. The Figures~\ref{fig:claim_stats_1} and~\ref{fig:claim_stats_2} show the quarterly claim distributions for several different properties for the \textbf{released} $24\,K$ claims. The statistics for original (i.e., non-rectified and dismissed) claims are depicted in Figures~\ref{fig:natural_claim_stats_1} and~\ref{fig:natural_claim_stats_2}.

\begin{figure*}
    \centering
    
    \begin{subfigure}[t]{\textwidth}
        \centering
        \includegraphics[width=\linewidth]{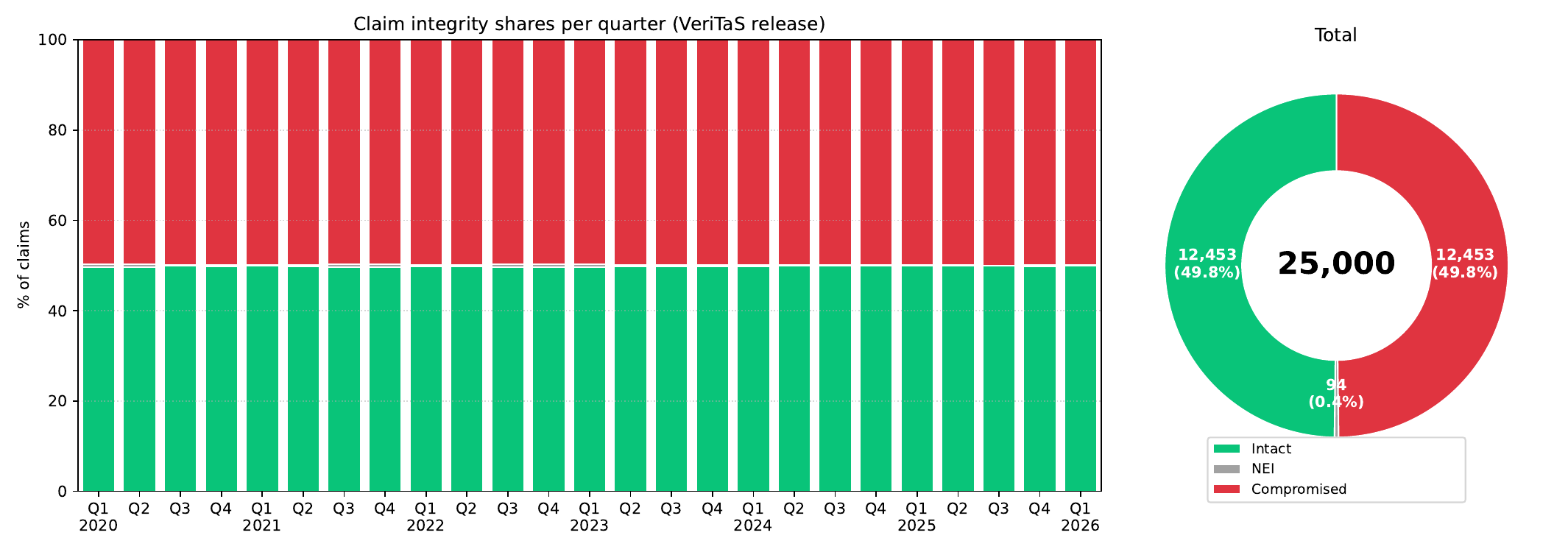}
        \caption{Integrity shares.}
    \end{subfigure}\vspace{1em}
    
    \begin{subfigure}[t]{\textwidth}
        \centering
        \includegraphics[width=\linewidth]{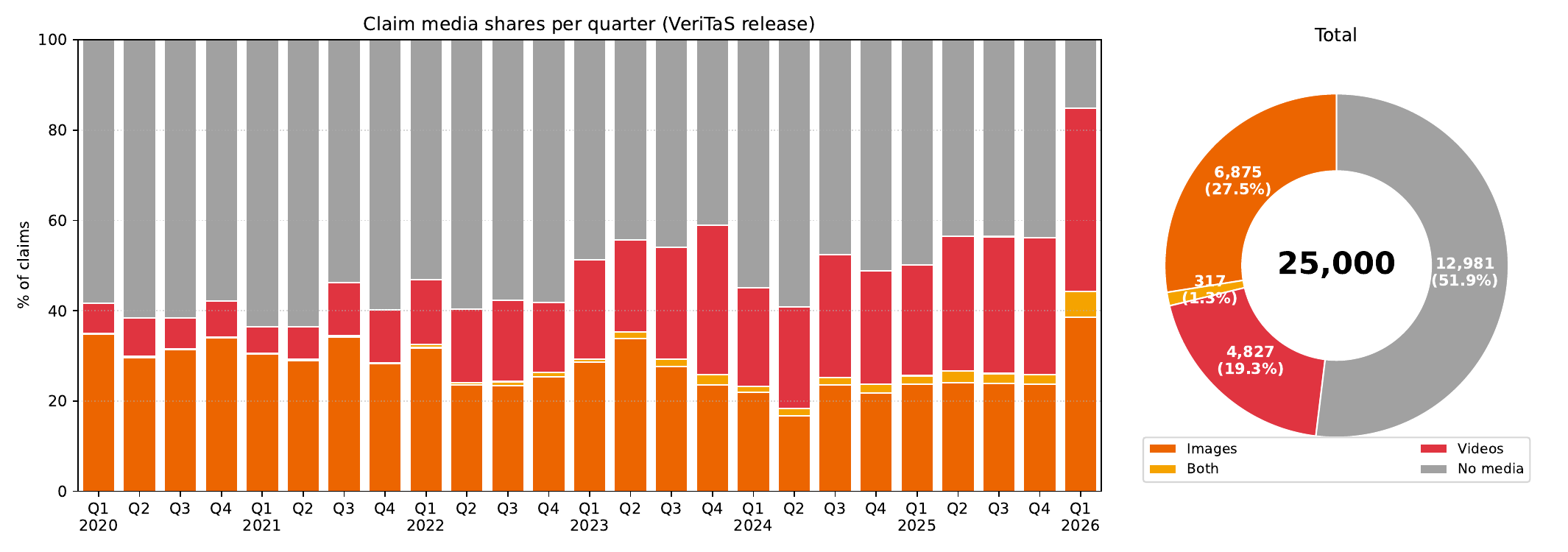}
        \caption{Shares of claims featuring media. Note that in Q1 2026 the \texttt{scrapeMM} web scraping package was improved resulting in significantly higher media download success rates.}
        \label{fig:release_claim_media_share}
    \end{subfigure}\vspace{1em}
    
    \begin{subfigure}[t]{\textwidth}
        \centering
        \includegraphics[width=\linewidth]{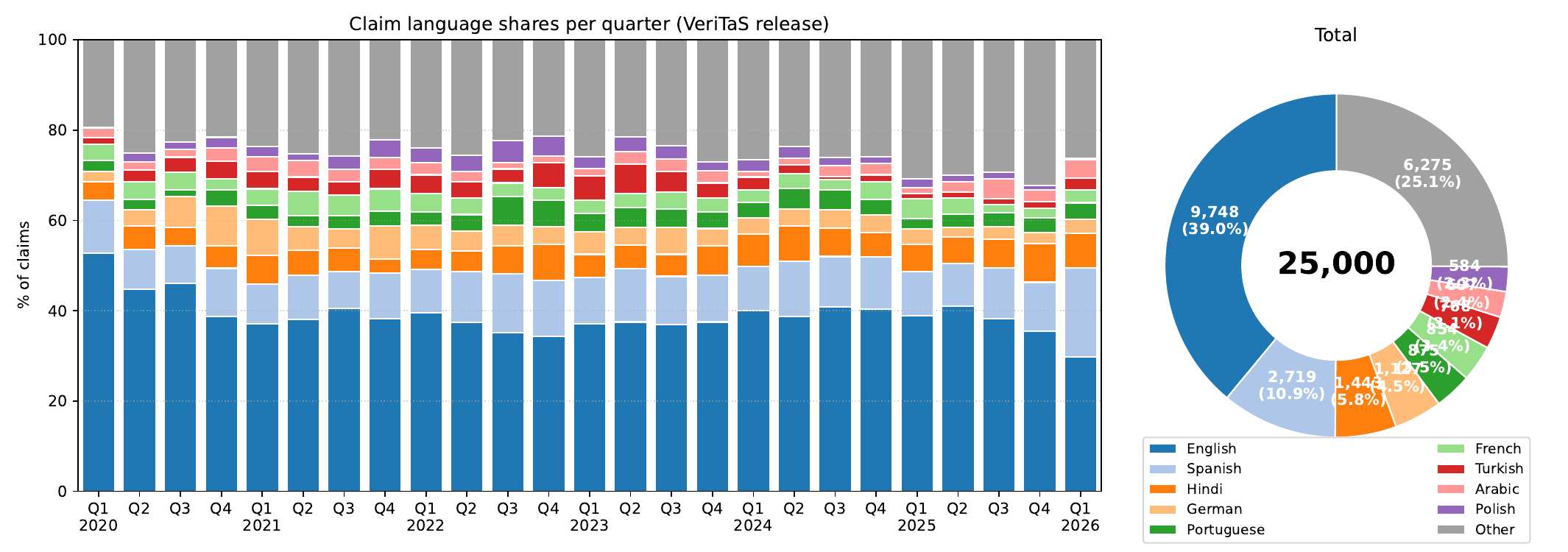}
        \caption{Language shares.}
        \label{fig:quarter_stats/release/claims/claim_languages}
    \end{subfigure}
    
    \caption{Statistics in the \textbf{released} \method benchmark for all quarter splits.}
    \label{fig:claim_stats_1}
\end{figure*}

\begin{figure*}
    \centering
    
    \begin{subfigure}[t]{\textwidth}
        \centering
        \includegraphics[width=\linewidth]{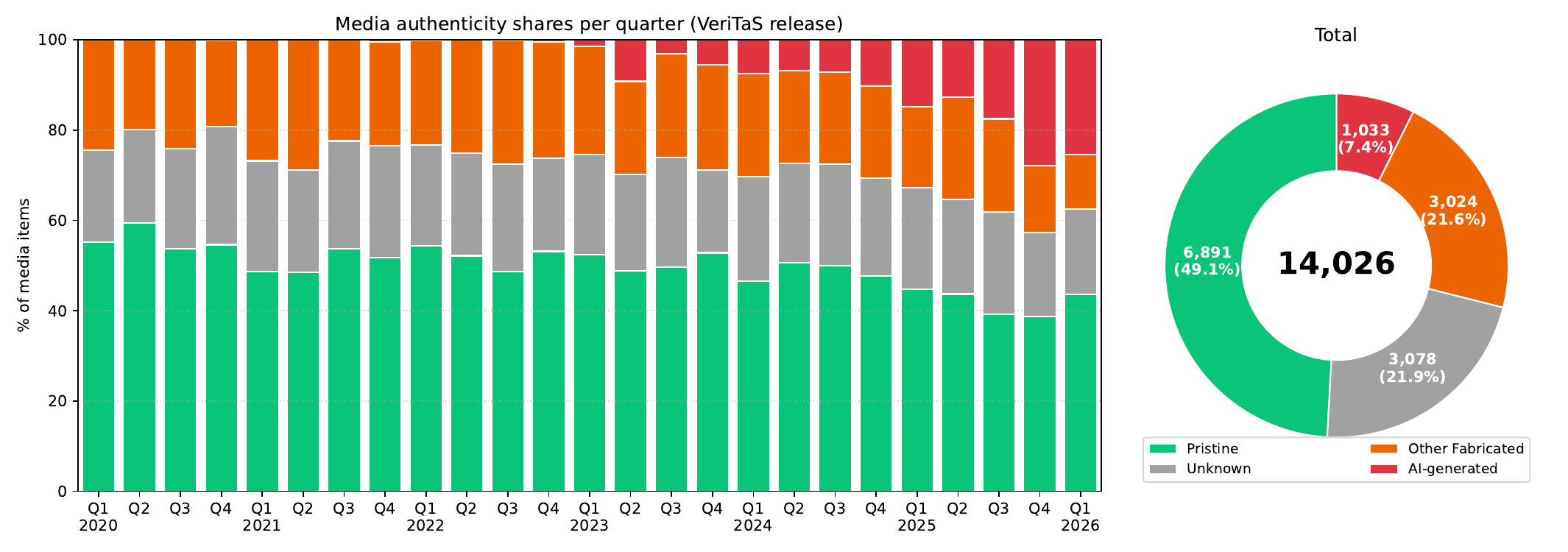}
        \caption{Media authenticity shares.}
    \end{subfigure}\vspace{1em}
    
    \begin{subfigure}[t]{\textwidth}
        \centering
        \includegraphics[width=\linewidth]{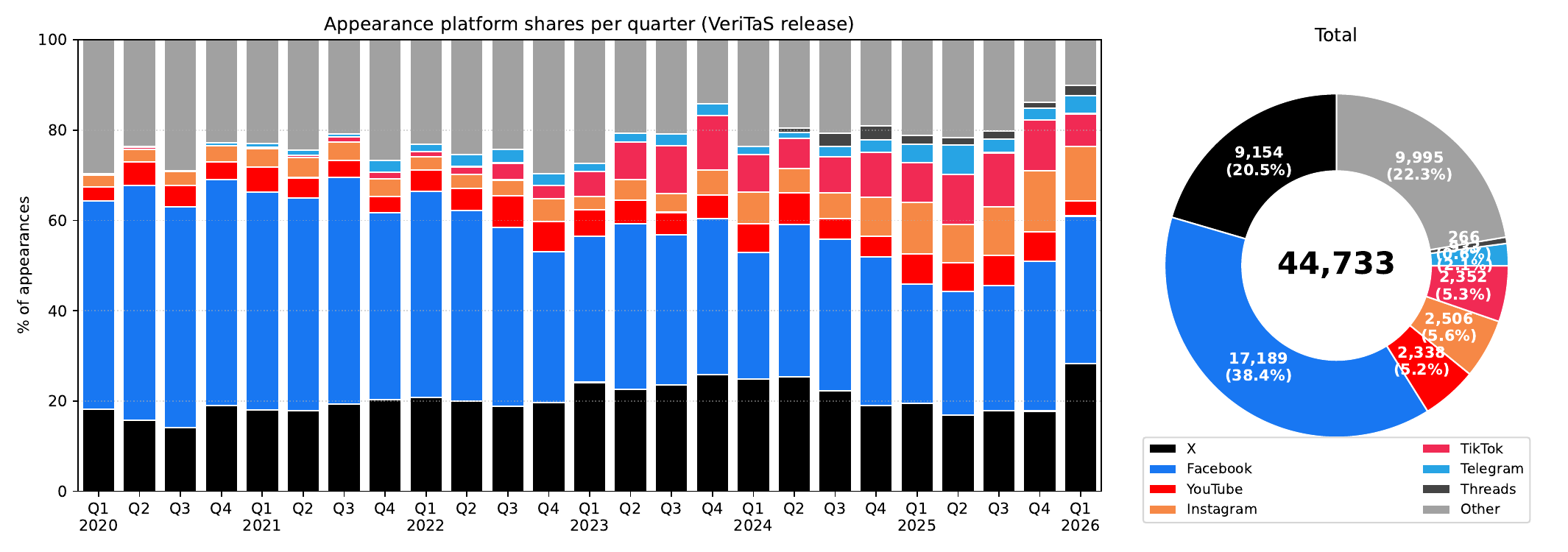}
        \caption{Platform shares for all claim appearances.}
    \end{subfigure}\vspace{1em}
    
    \begin{subfigure}[t]{\textwidth}
        \centering
        \includegraphics[width=\linewidth]{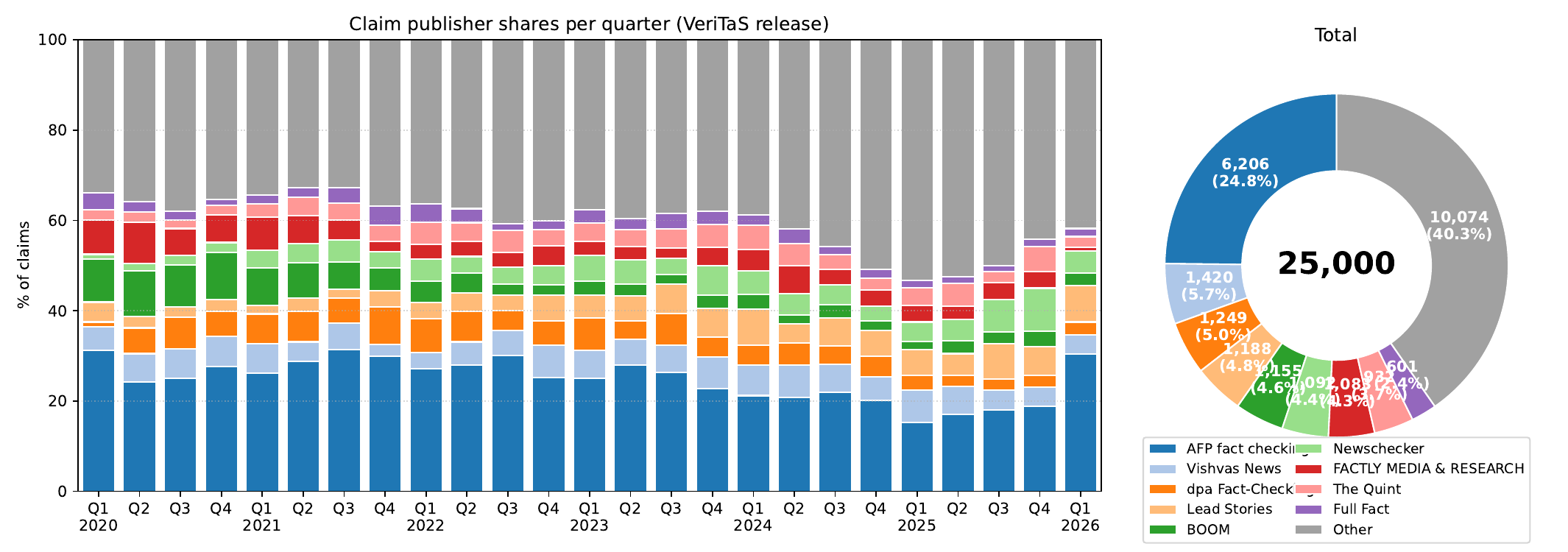}
        \caption{Publisher shares.}
    \end{subfigure}
    
    \caption{Statistics in the \textbf{released} \method benchmark for all quarter splits.}
    \label{fig:claim_stats_2}
\end{figure*}

\begin{figure*}
    \centering
    
    \begin{subfigure}[t]{0.48\textwidth}
        \centering
        \includegraphics[width=\textwidth]{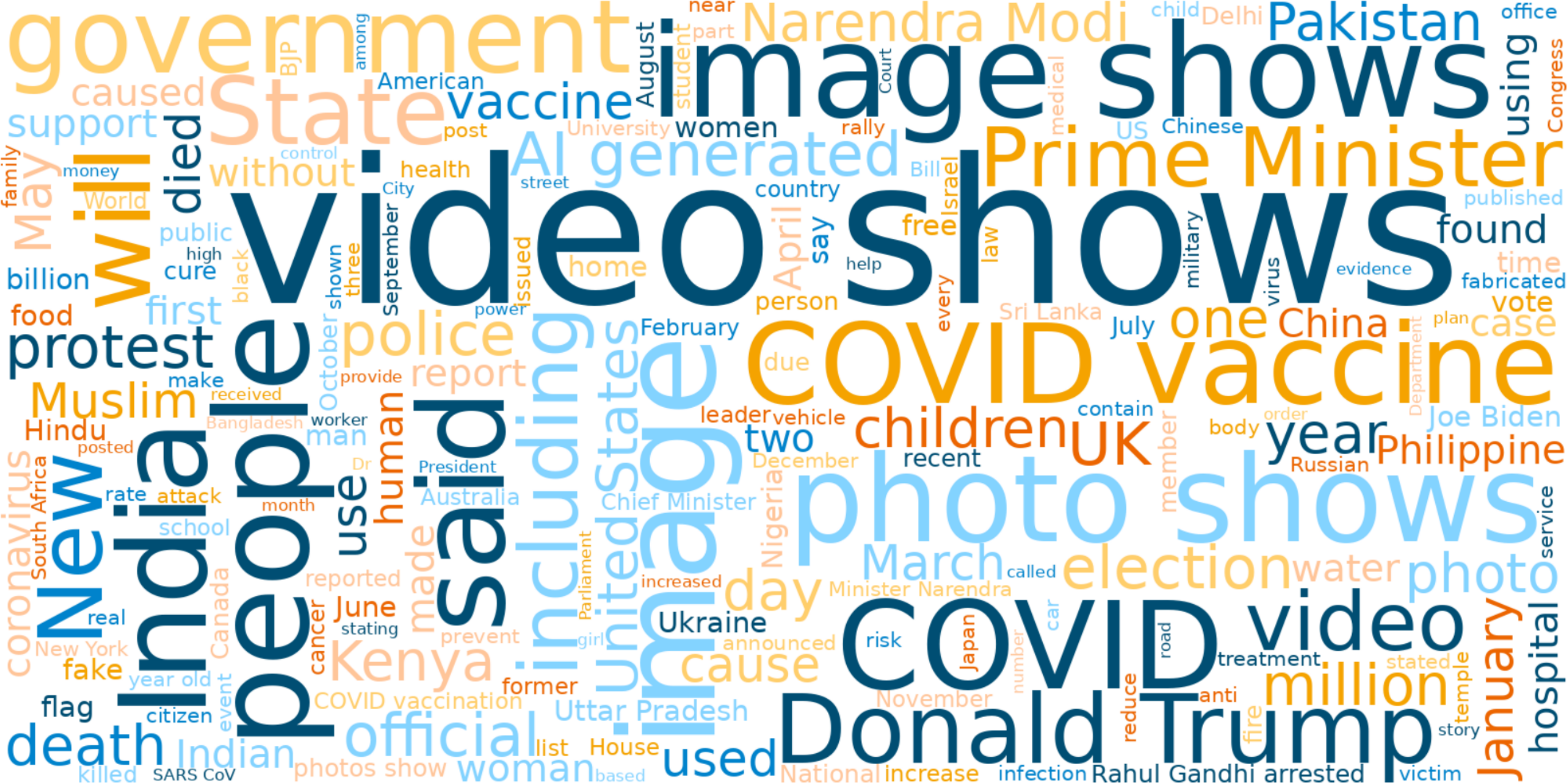}
        \caption{English}
    \end{subfigure}
    \hfill
    \begin{subfigure}[t]{0.48\textwidth}
        \centering
        \includegraphics[width=\textwidth]{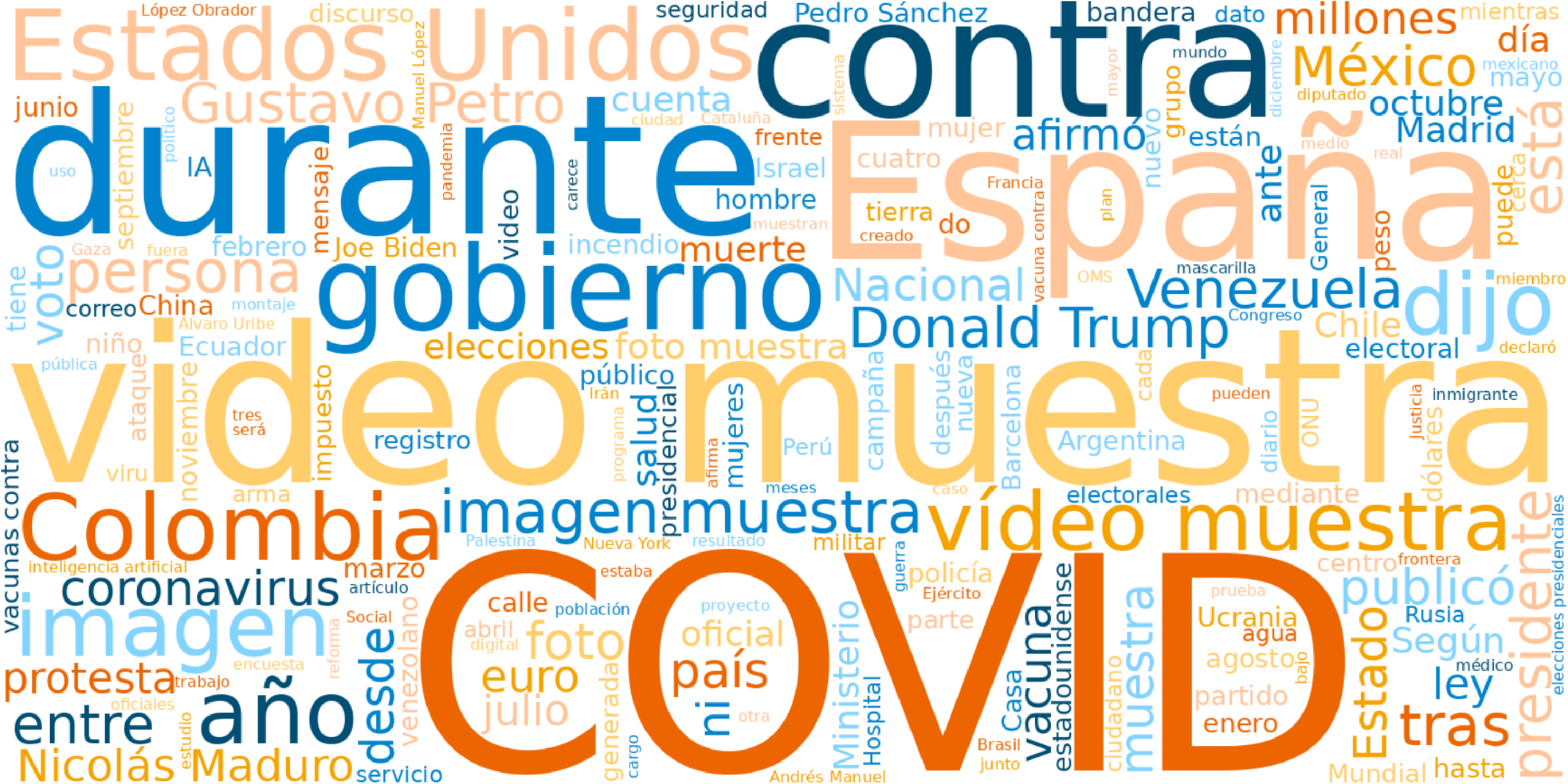}
        \caption{Spanish}
    \end{subfigure}\vspace{0.5em}
    
    \begin{subfigure}[t]{0.48\textwidth}
        \centering
        \includegraphics[width=\textwidth]{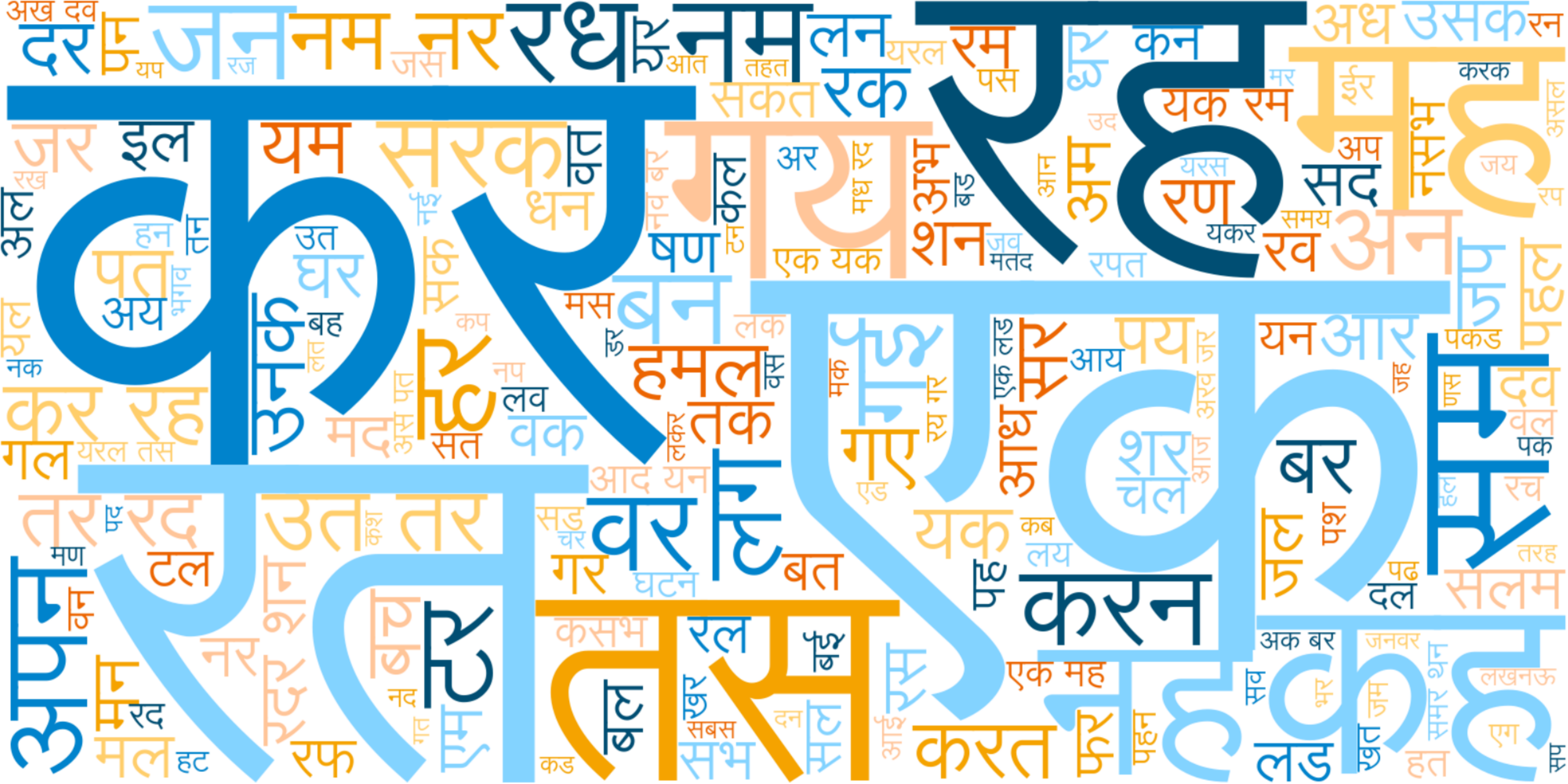}
        \caption{Hindi}
    \end{subfigure}
    \hfill
    \begin{subfigure}[t]{0.48\textwidth}
        \centering
        \includegraphics[width=\textwidth]{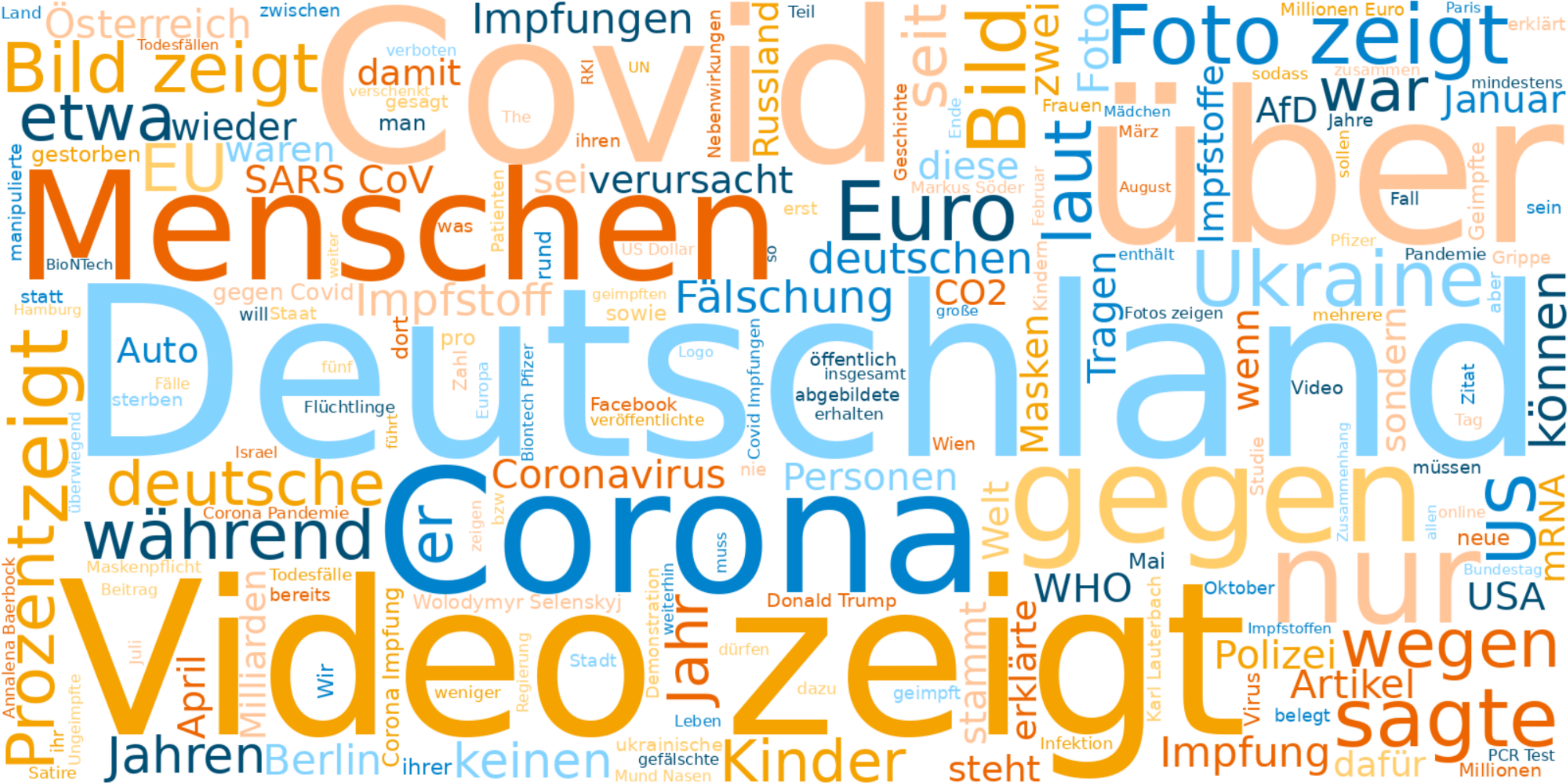}
        \caption{German}
    \end{subfigure}\vspace{0.5em}
    
    \begin{subfigure}[t]{0.48\textwidth}
        \centering
        \includegraphics[width=\textwidth]{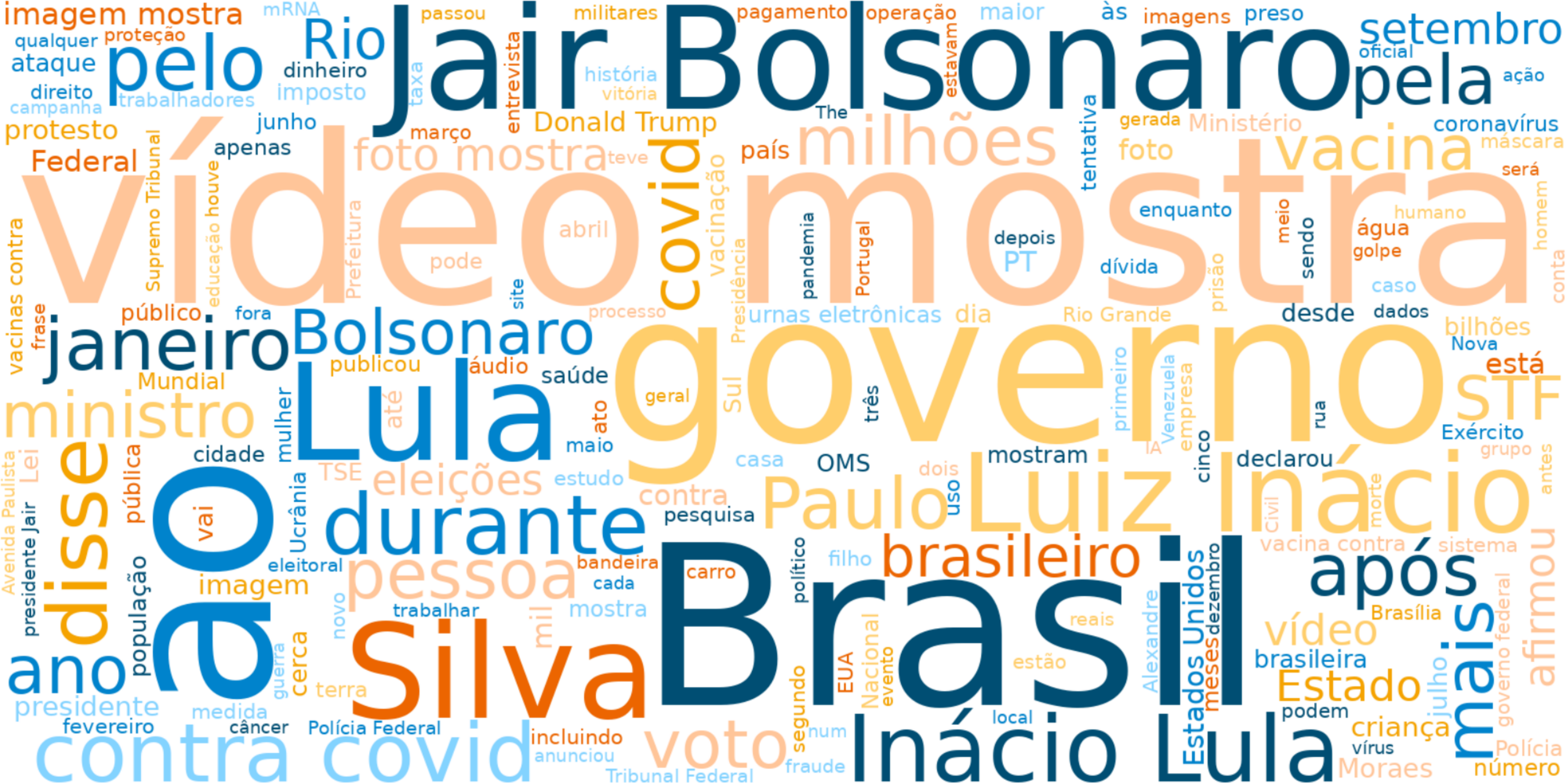}
        \caption{Portuguese}
    \end{subfigure}
    \hfill
    \begin{subfigure}[t]{0.48\textwidth}
        \centering
        \includegraphics[width=\textwidth]{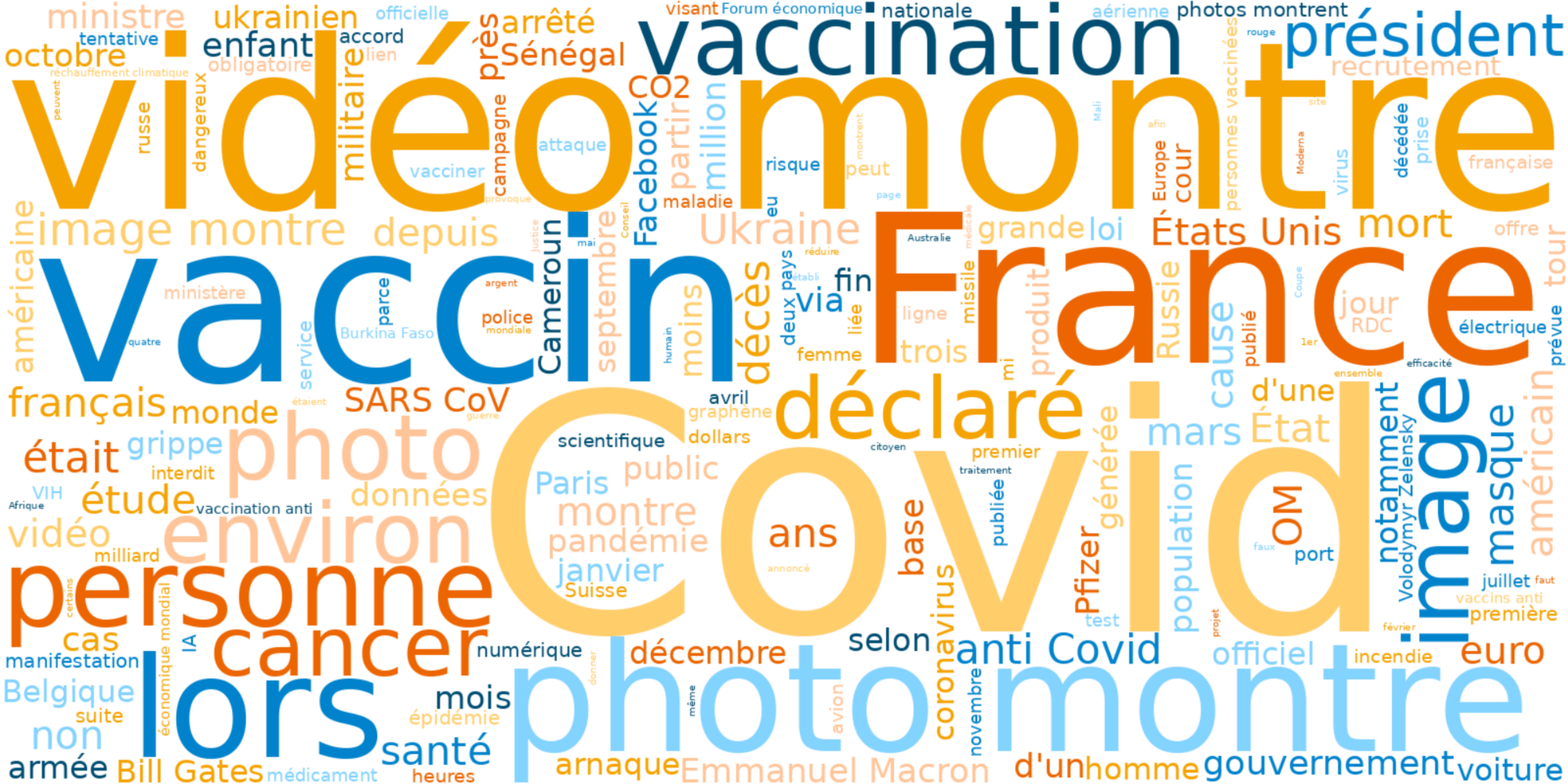}
        \caption{French}
    \end{subfigure}\vspace{0.5em}
    
    \begin{subfigure}[t]{0.48\textwidth}
        \centering
        \includegraphics[width=\textwidth]{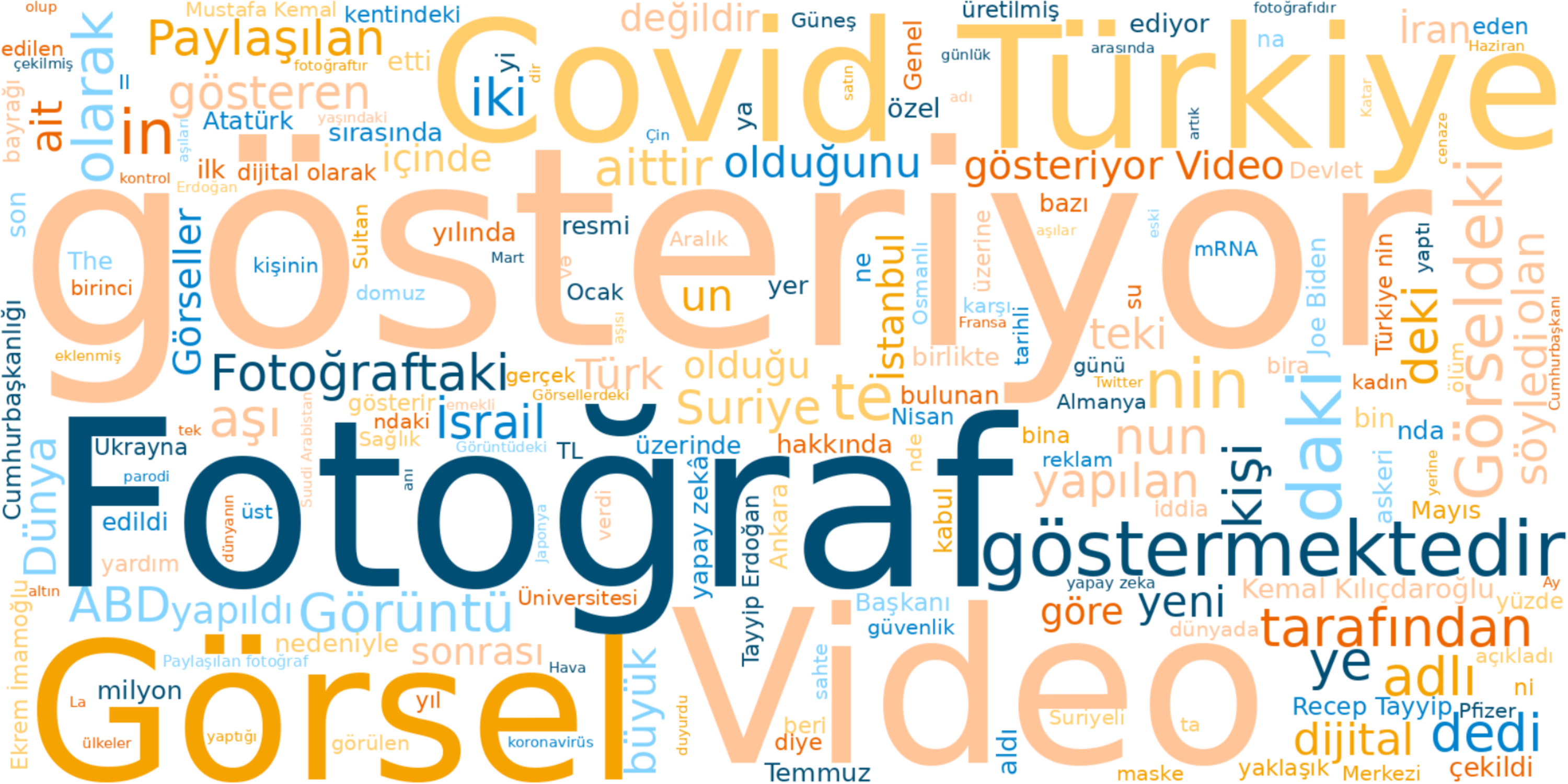}
        \caption{Turkish}
    \end{subfigure}
    \hfill
    \begin{subfigure}[t]{0.48\textwidth}
        \centering
        \includegraphics[width=\textwidth]{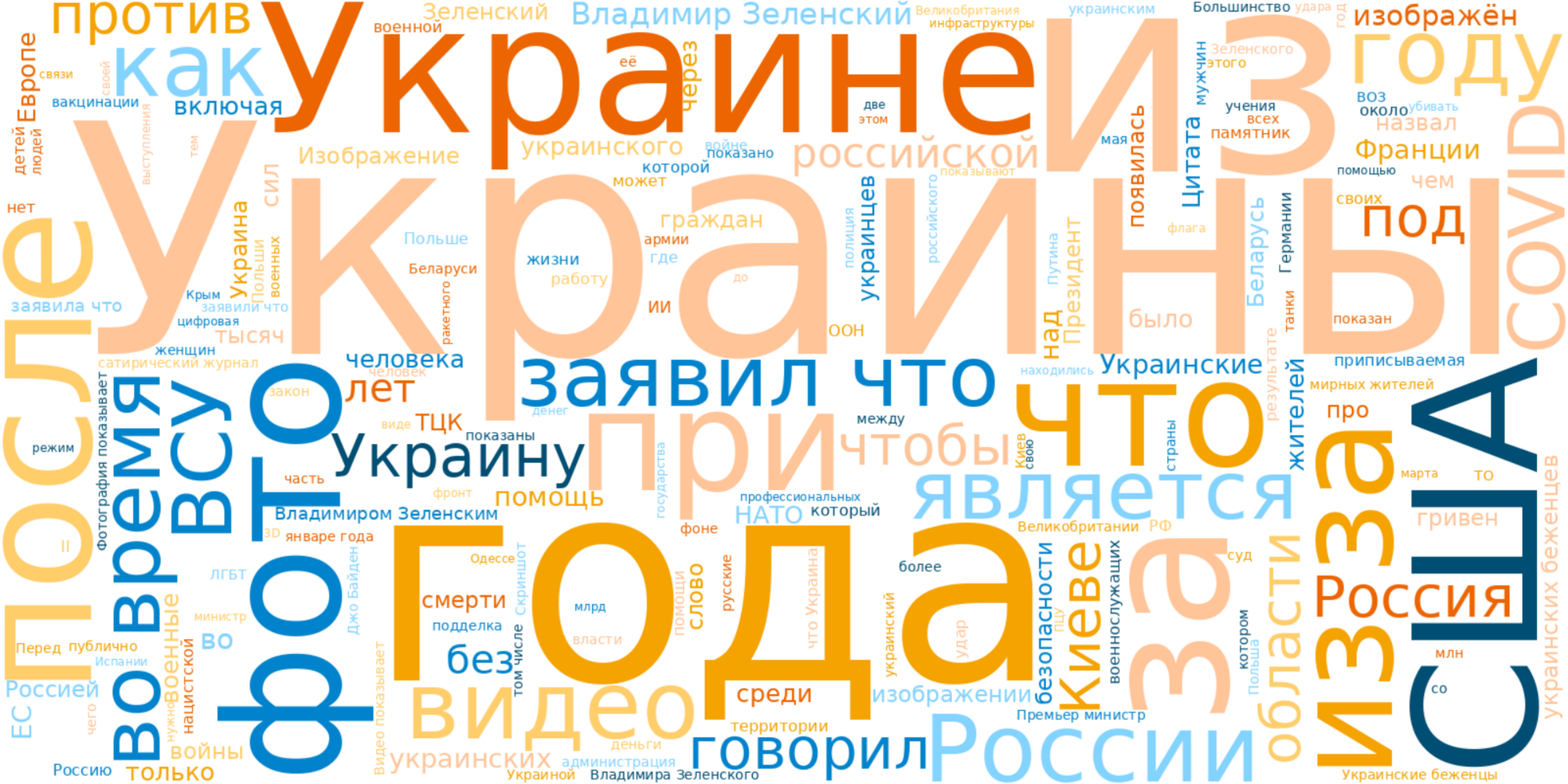}
        \caption{Russian}
    \end{subfigure}\vspace{0.5em}
    
    \begin{subfigure}[t]{0.48\textwidth}
        \centering
        \includegraphics[width=\textwidth]{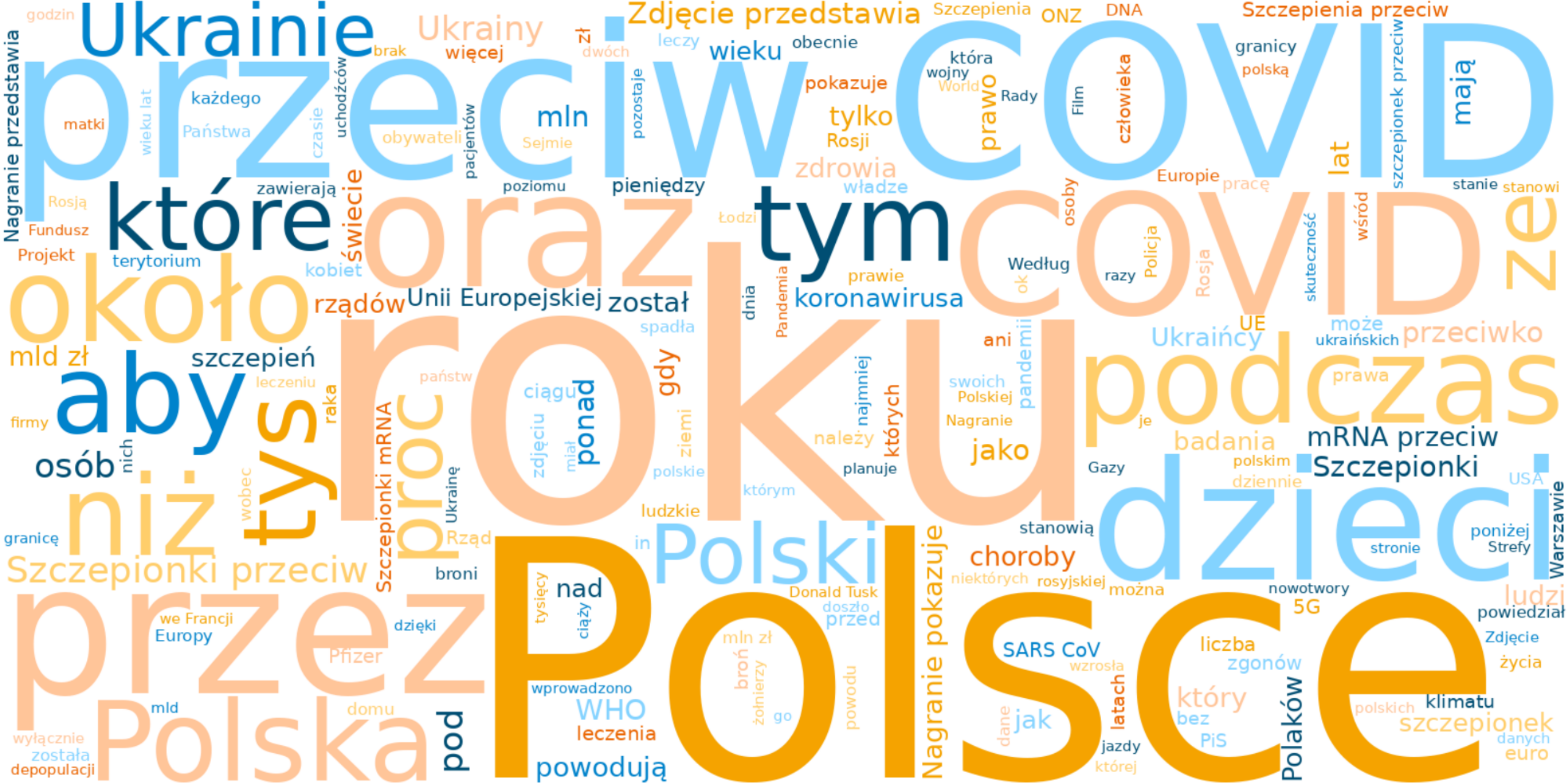}
        \caption{Polish}
    \end{subfigure}
    \hfill
    \begin{subfigure}[t]{0.48\textwidth}
        \centering
        \includegraphics[width=\textwidth]{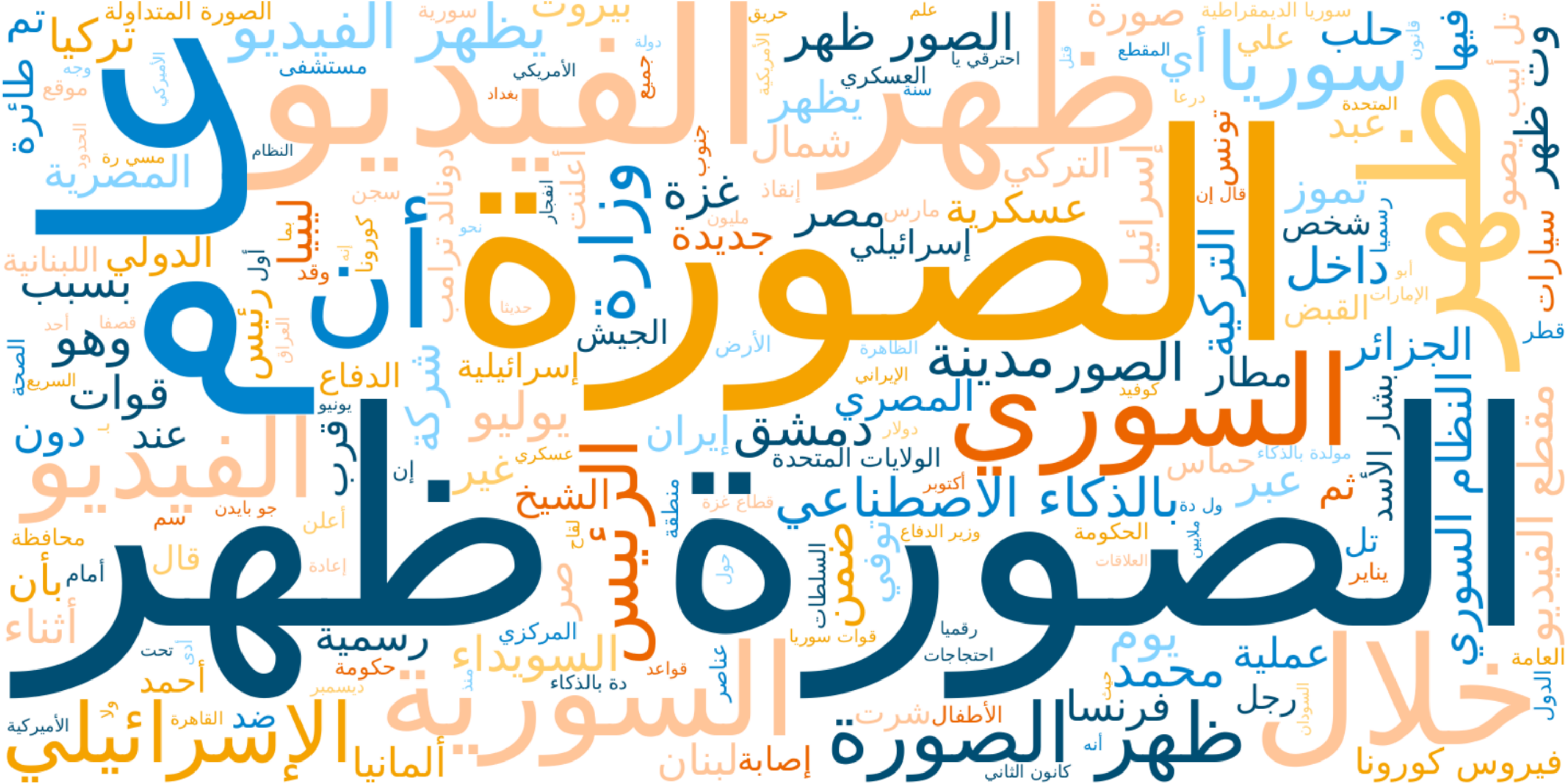}
        \caption{Arabic}
    \end{subfigure}
    
    \caption{Common words occurring in \method' claims for the ten most frequent languages.}
    \label{fig:wordclouds}
\end{figure*}

\begin{figure*}
    \centering
    
    \begin{subfigure}[t]{\textwidth}
        \centering
        \includegraphics[width=\linewidth]{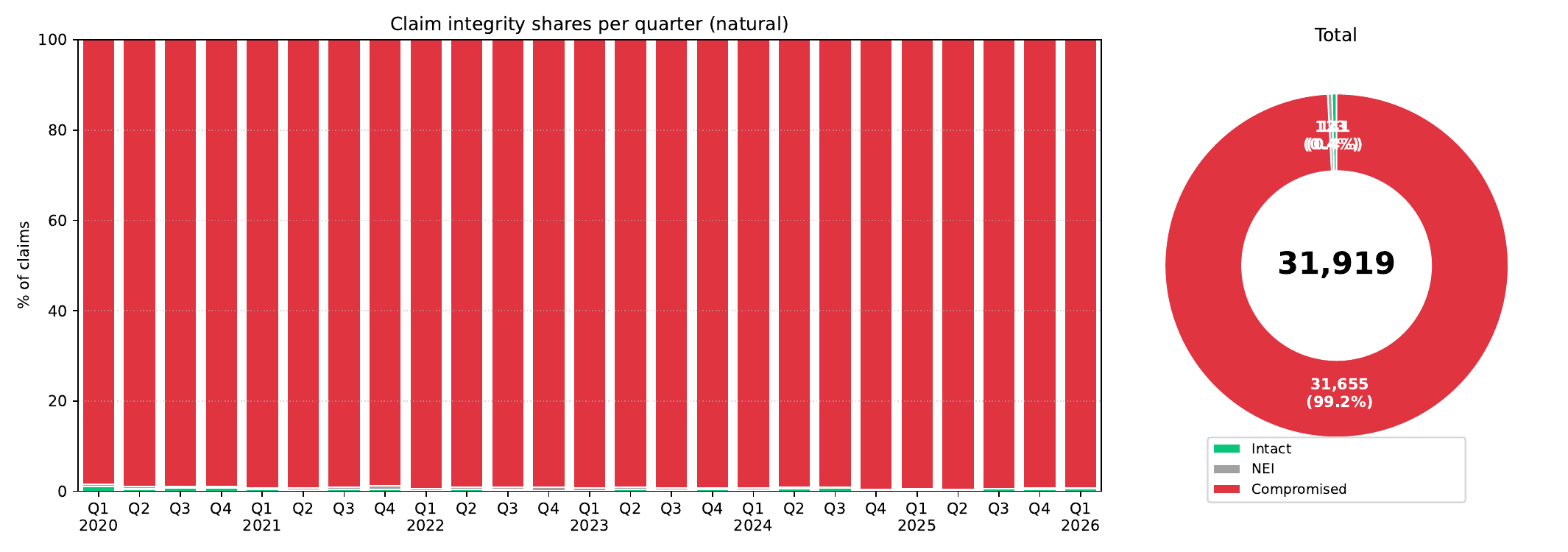}
        \caption{Integrity shares.}
        \label{fig:quarter_stats/natural/claims/integrity}
    \end{subfigure}\vspace{1em}
    
    \begin{subfigure}[t]{\textwidth}
        \centering
        \includegraphics[width=\linewidth]{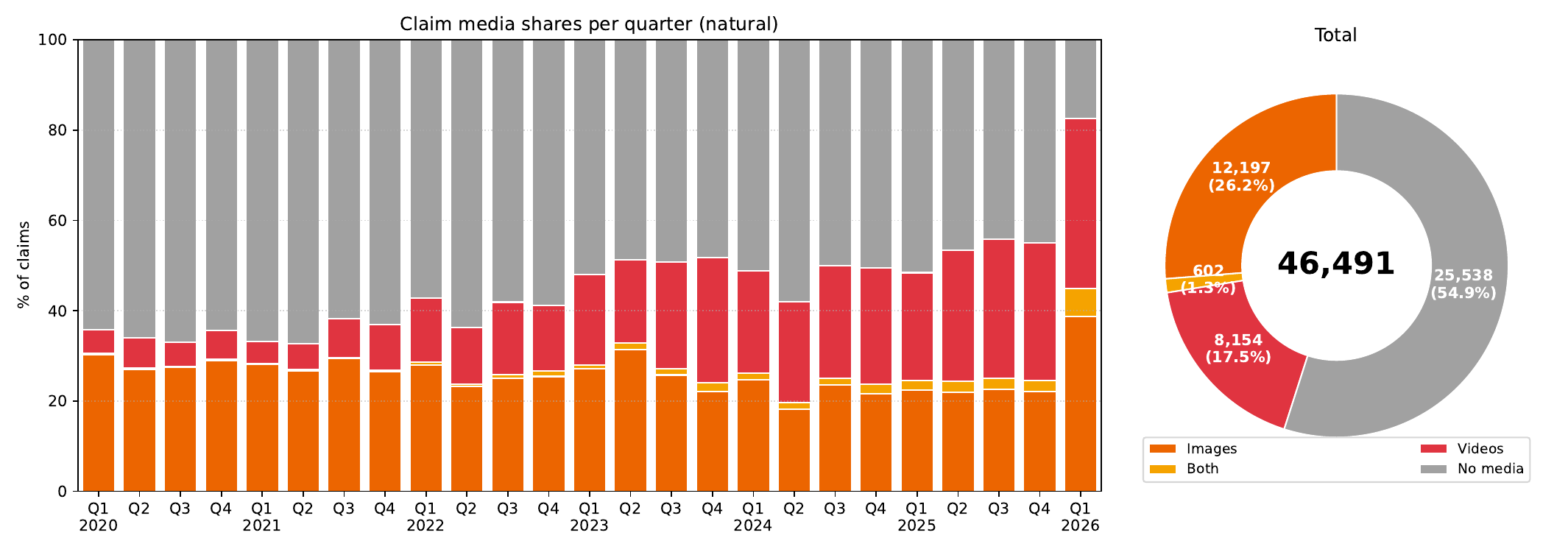}
        \caption{Shares of claims featuring media. Note that in Q1 2026 the \texttt{scrapeMM} web scraping package was improved resulting in significantly higher media download success rates.}
    \end{subfigure}\vspace{1em}
    
    \begin{subfigure}[t]{\textwidth}
        \centering
        \includegraphics[width=\linewidth]{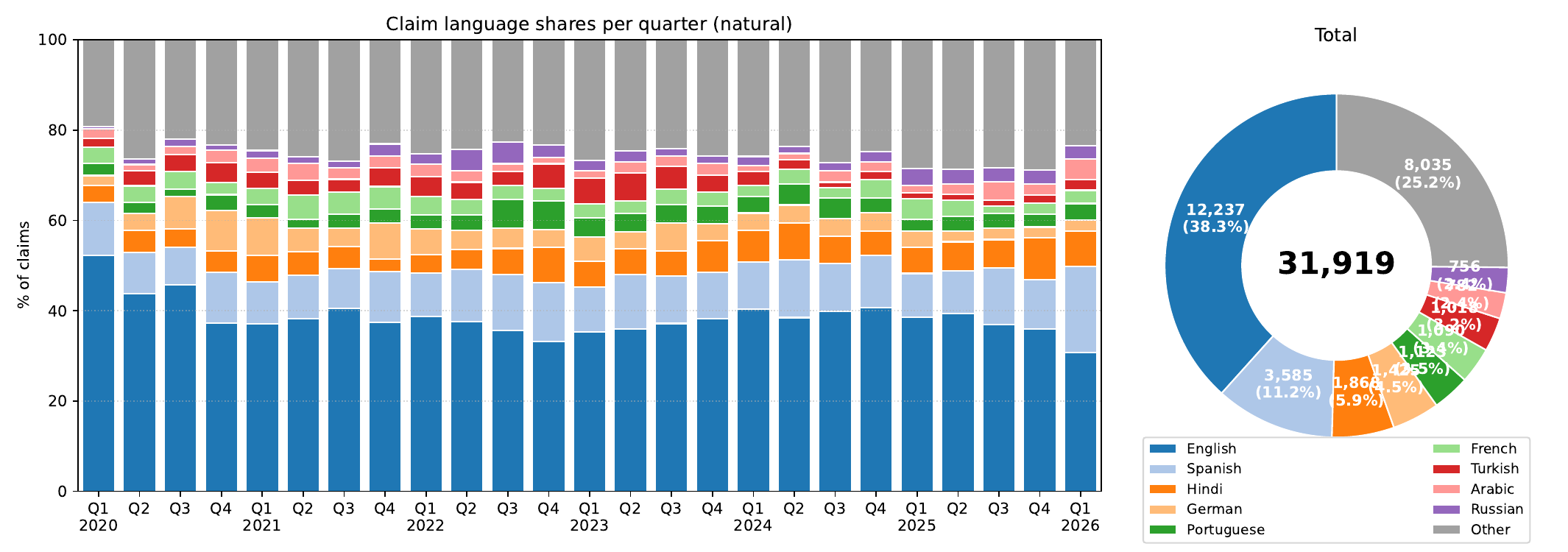}
        \caption{Language shares.}
    \end{subfigure}
    
    \caption{\textbf{Natural} data statistics as obtained before stage 7, i.e., before balancing/rectification and sampling.}
    \label{fig:natural_claim_stats_1}
\end{figure*}

\begin{figure*}
    \centering
    
    \begin{subfigure}[t]{\textwidth}
        \centering
        \includegraphics[width=\linewidth]{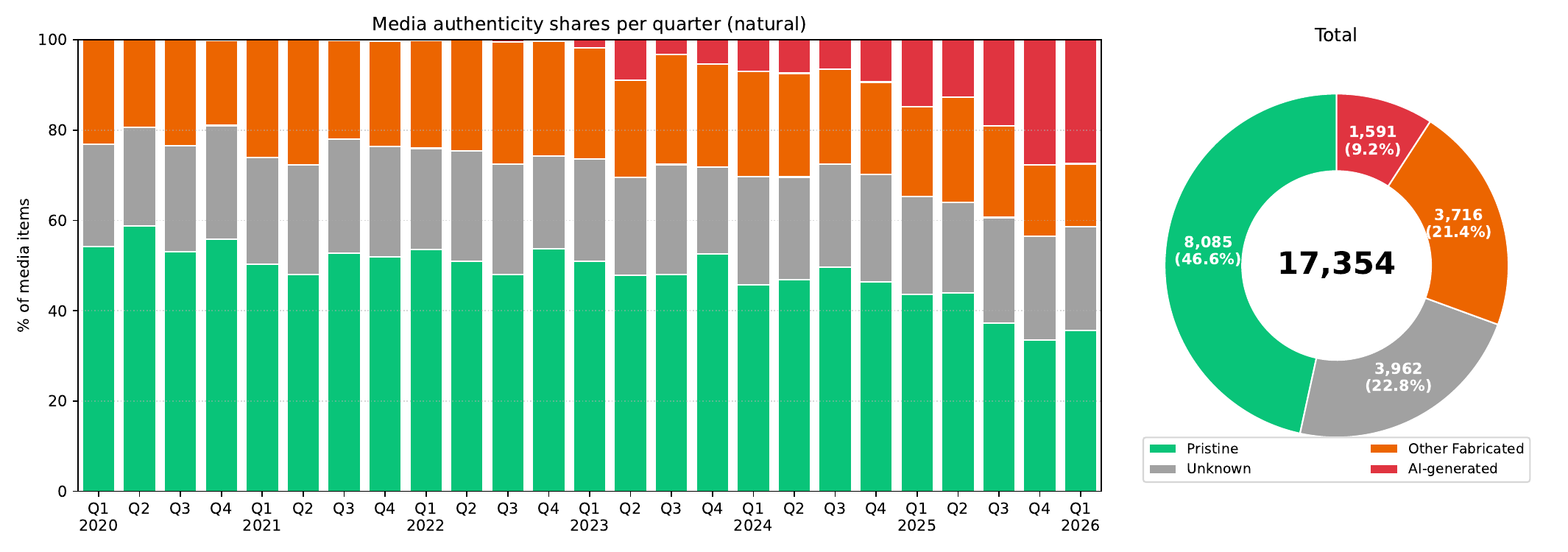}
        \caption{Media authenticity shares.}
    \end{subfigure}\vspace{1em}
    
    \begin{subfigure}[t]{\textwidth}
        \centering
        \includegraphics[width=\linewidth]{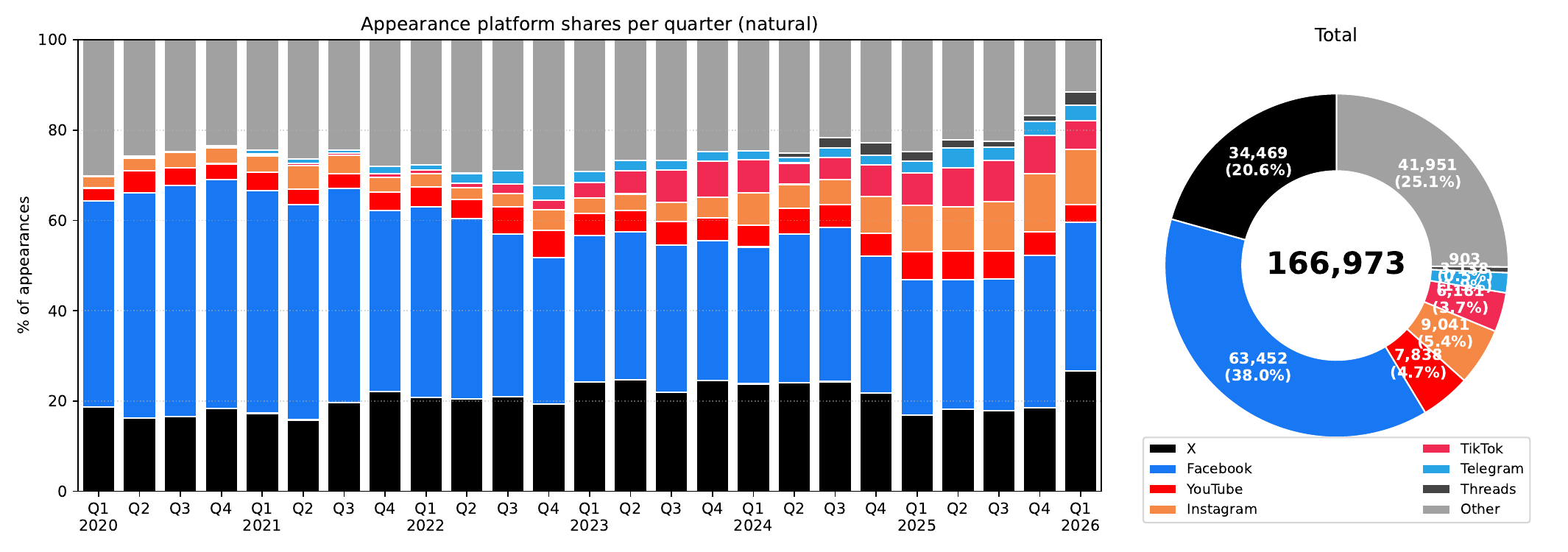}
        \caption{Platform shares for all claim appearances.}
        \label{fig:quarter_stats/claims/natural/platforms}
    \end{subfigure}\vspace{1em}
    
    \begin{subfigure}[t]{\textwidth}
        \centering
        \includegraphics[width=\linewidth]{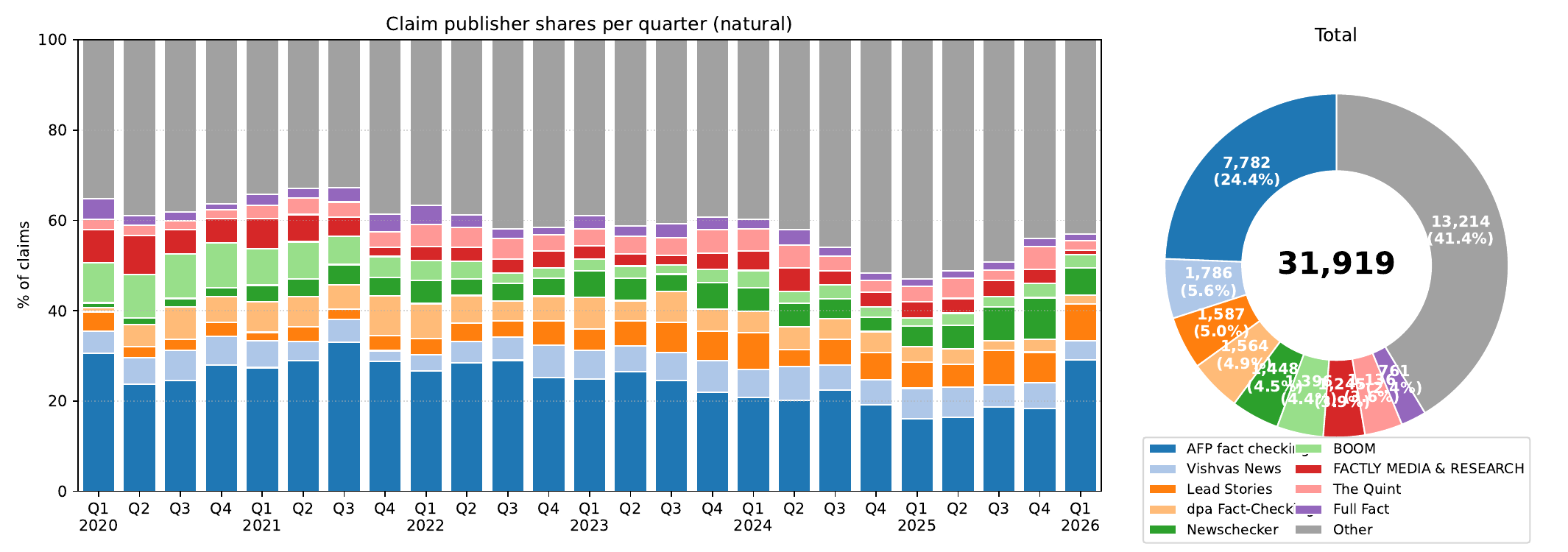}
        \caption{Publisher shares.}
    \end{subfigure}
    
    \caption{\textbf{Natural} data statistics as obtained before stage 7, i.e., before balancing and sampling.}
    \label{fig:natural_claim_stats_2}
\end{figure*}

\subsection{Verdicts}
\label{app:verdict_stats}
Tab.~\ref{tab:verdict_stats} shows the total number of labels after discretizing scores into $3$ bins as described in App.~\ref{app:score_discretization}. Note that the totals across properties differ for two reasons: The number of media is different to the number of claims and the properties \textit{Veracity} and \textit{Context Coverage} are evaluated only if the previous properties did not result in a \negative decision. Table~\ref{tab:verdict_stats} and Figure~\ref{fig:score_distributions} show the distribution of scores for all properties.
\begin{table}[h]
    \centering
    \resizebox{\linewidth}{!}{
    \begin{tabular}{l|rrr}
        \toprule
         \textbf{Property} & \negative & \nei & \positive  \\
         \midrule
         Authenticity      & 3,659  & 2,591 & 5,769 \\
         Contextualization & 5,726  & 462  & 5,831 \\
         Veracity          & 6,539  & 129  & 12,610 \\
         Context Coverage  & 192   & 4    & 12,446 \\
         Integrity         & 12,453 & 94   & 12,453 \\
         \bottomrule
    \end{tabular}
    }
    \caption{Decision counts for all five properties.}
    \label{tab:verdict_stats}
\end{table}

\begin{figure*}
    \centering
    \includegraphics[width=\linewidth]{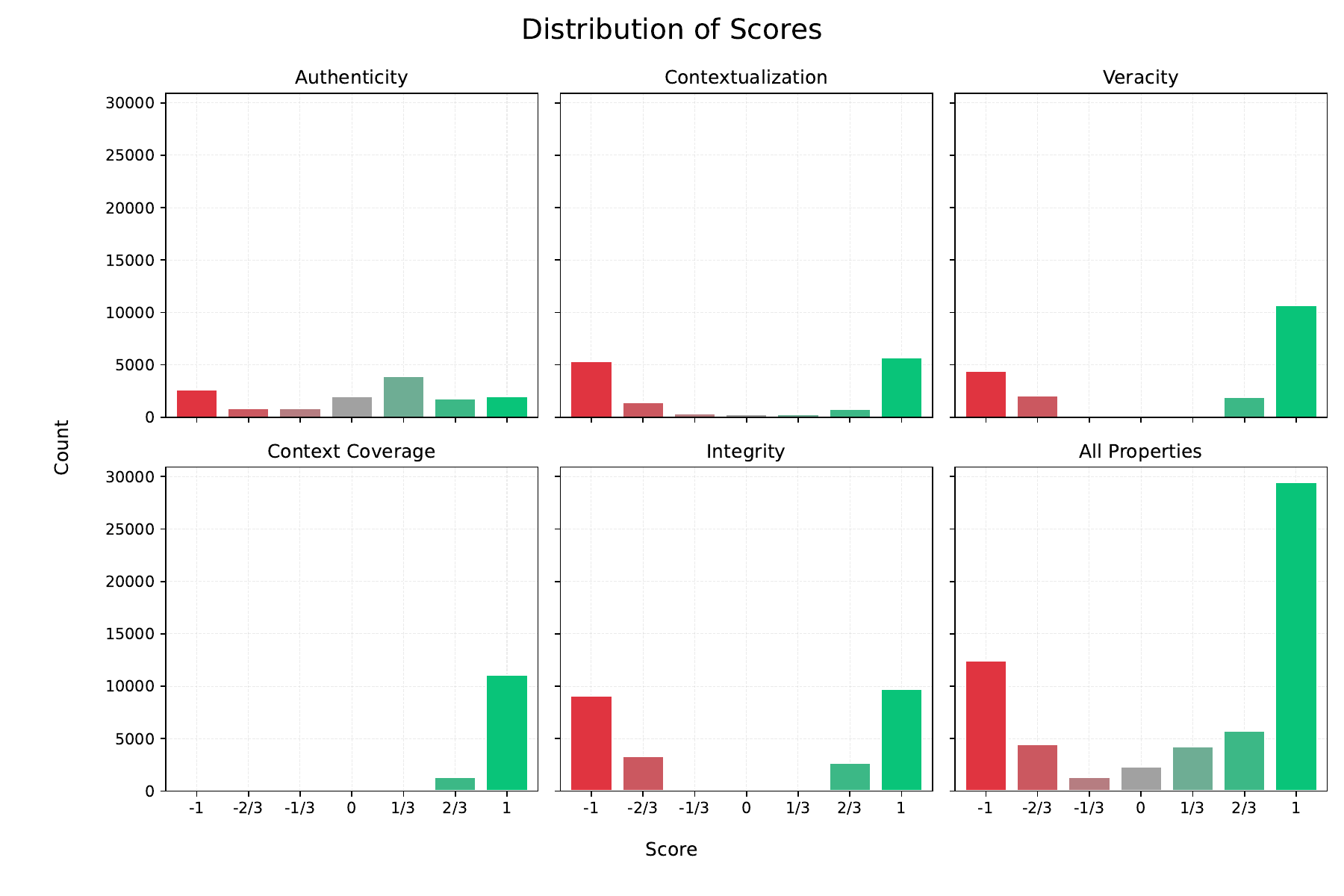}
    \caption{Score distribution of the released \method benchmark for all five properties and their sum.}
    \label{fig:score_distributions}
\end{figure*}

\subsection{Dismissal Reasons}
The \method pipeline can discard reviews and claims for various reasons. Fig.~\ref{fig:dismissal_reasons} depicts the $10$ most common reasons.

\begin{figure}[t]
    \centering
    \includegraphics[width=\linewidth]{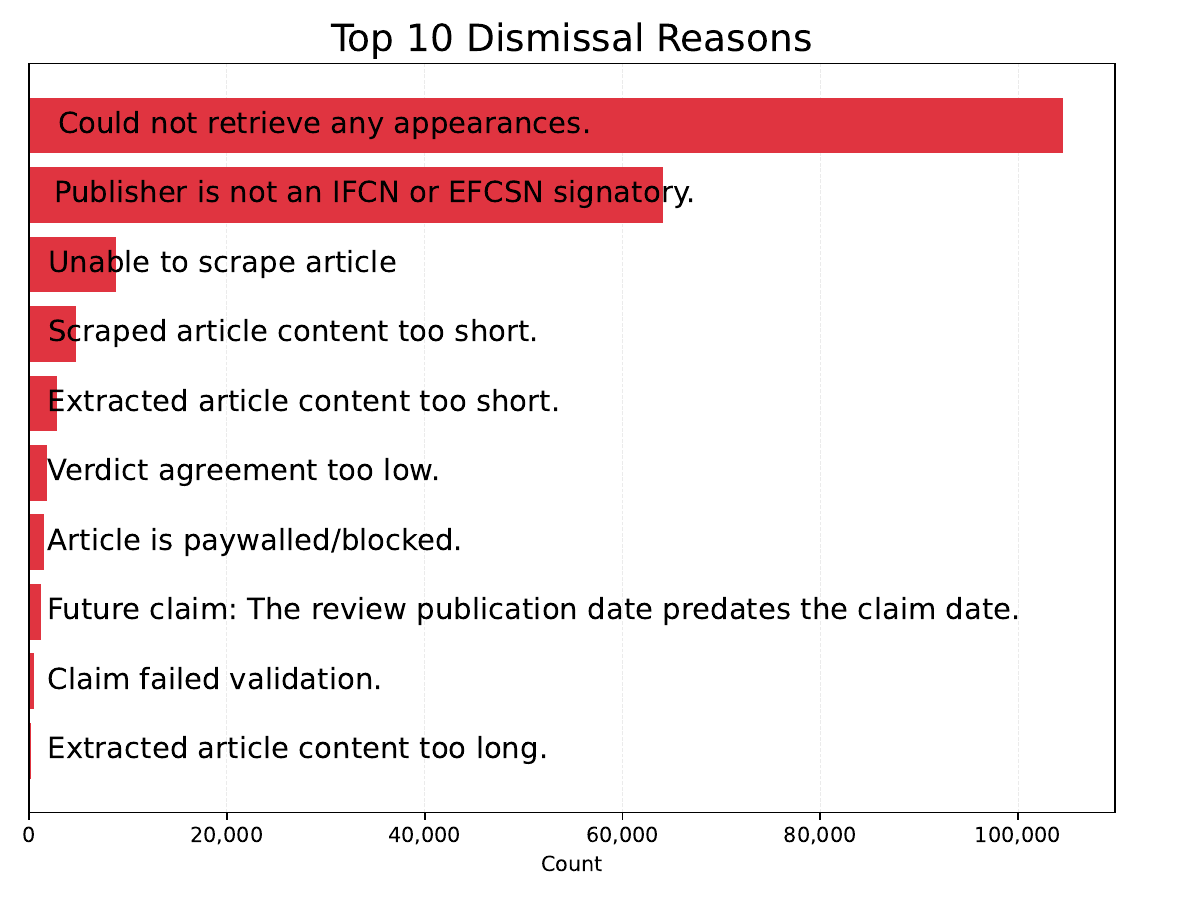}
    \caption{Most common reasons for a claim/review to be dismissed by the \method pipeline}
    \label{fig:dismissal_reasons}
\end{figure}

\section{Baseline Experiments}
\subsection{Web Search Tool}\label{app:web-search-tool}
When provided with search access, models can query the internet via Google's Serper API, which returns the top $10$ results per query. We apply temporal filtering (only content published before the claim date is retained) to avoid temporal leakage. For the top $3$ filtered results, we attempt to scrape the full page content using \texttt{scrapeMM}; models receive both the search snippets from Serper and the complete scraped content when available. For each claim, up to $5$ searches are possible.

The full results for the longitudinal split are shown in Tab.~\ref{tab:results_longitudinal} and Fig.~\ref{fig:results_longitudinal_full}. Figure \ref{fig:results_longitudinal_per_modality} shows the results per modality.

\begin{table}[]
    \centering
    \resizebox{\linewidth}{!}{
    \begin{tabular}{lc|cccc}
        \toprule
        \textbf{Method} & \textbf{Search} & \multicolumn{2}{c}{\textbf{Error Rates (↓)}} & \multicolumn{2}{c}{\textbf{Accuracy (↑)}} \\
        & & \highlight{MSE} & MAE & $7$-bin & $3$-bin \\
        \midrule
        \textsc{Gemini~3~Flash}  & - & 0.334 & 0.248 & \textbf{70.4} & 89.7 \\
        \textsc{Gemini~3.1~Pro}  & - & \textbf{0.309} & \textbf{0.235} & \textbf{70.3} & \textbf{90.3} \\
        \textsc{Claude~Opus~4.6} & - & 0.363 & 0.375 & 41.0 & 76.0 \\
        \gptfour                 & - & 0.565 & 0.504 & 37.0 & 64.0 \\
        \gpt                     & - & 0.520 & 0.552 & 21.4 & 56.3 \\
        \llama                   & - & 0.735 & 0.622 & 24.6 & 58.1 \\
        \textsc{Gemma~4~(31B)}   & - & 0.525 & 0.374 & 59.0 & 82.2 \\
        \textsc{Qwen~3.5~(397B)} & - & 0.600 & 0.440 & 50.9 & 74.4 \\
        \midrule
        \textsc{Gemini~3~Flash}  & \checkmark & \textbf{0.158} & \textbf{0.168} & \textbf{72.5} & \textbf{94.1} \\
        \textsc{Gemini~3.1~Pro}  & \checkmark & 0.234 & 0.222 & 68.0 & 88.5 \\
        \textsc{Claude~Opus~4.6} & \checkmark & 0.305 & 0.358 & 42.3 & 72.6 \\
        \gptfour                 & \checkmark & 0.609 & 0.502 & 42.5 & 65.7 \\
        \gpt                     & \checkmark & 0.325 & 0.412 & 25.2 & 76.1 \\
        \llama                   & \checkmark & 0.641 & 0.585 & 26.7 & 59.2 \\
        \textsc{Gemma~4~(31B)}   & \checkmark & 0.265 & 0.243 & 65.0 & 89.2 \\
        \textsc{Qwen~3.5~(397B)} & \checkmark & 0.267 & 0.278 & 56.4 & 85.9 \\
        \bottomrule
    \end{tabular}
    }
    \caption{Results on the \textbf{longitudinal data split}. Baselines have been tested once using parametric knowledge only (without search tool) and once using web search (with search tool), single runs.}
    \label{tab:results_longitudinal}
\end{table}

\begin{figure*}
    \centering
    \begin{subfigure}[t]{0.48\textwidth}
        \centering
        \includegraphics[width=\linewidth]{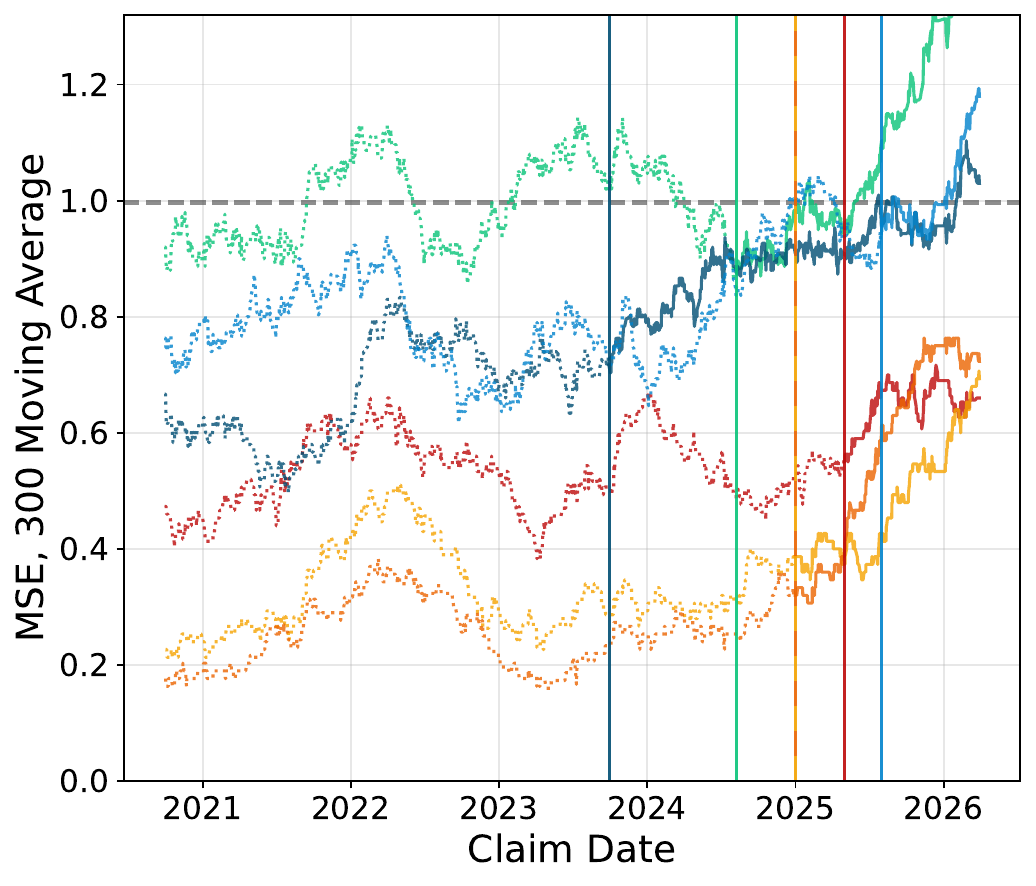}
        \caption{MSE \textbf{without} web search.}
        \label{fig:avg_300_claims_MSE_no-search}
    \end{subfigure}
    \hfill
    \begin{subfigure}[t]{0.48\textwidth}
        \centering
        \includegraphics[width=\linewidth]{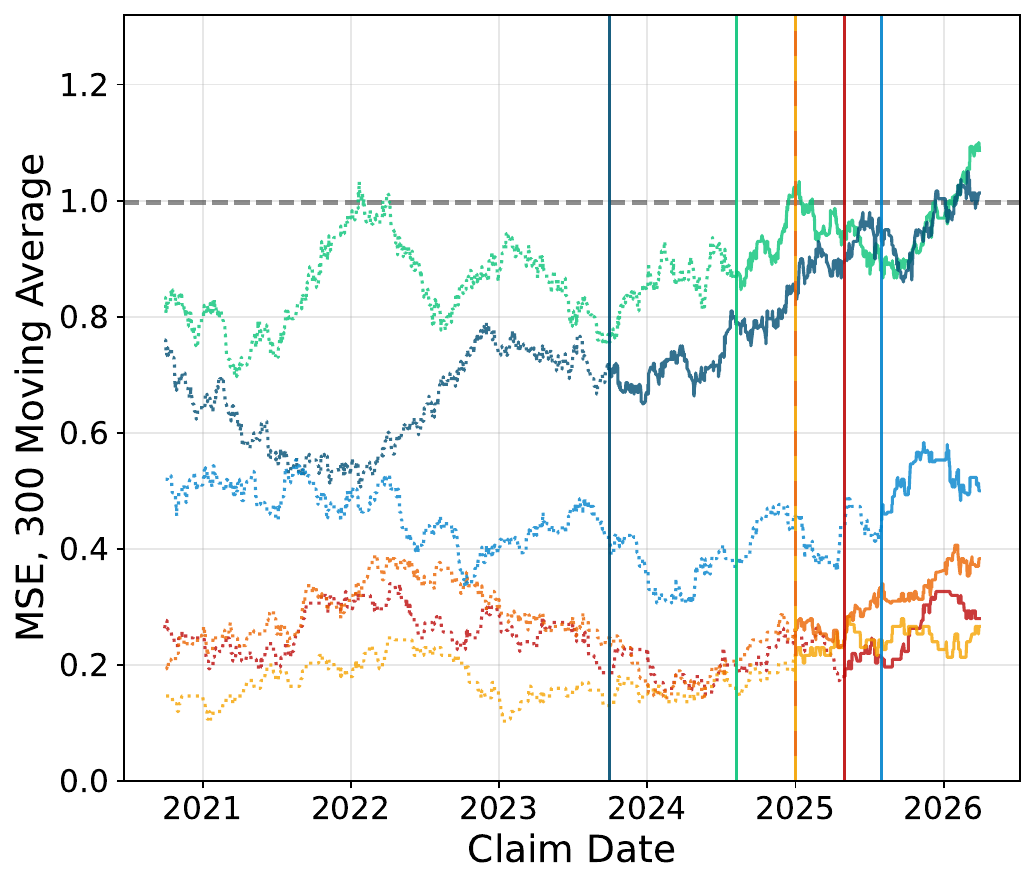}
        \caption{MSE \textbf{with} web search.}
        \label{fig:avg_300_claims_MSE_custom-search}
    \end{subfigure}\vspace{1em}
    
    \begin{subfigure}[t]{0.48\textwidth}
        \centering
        \includegraphics[width=\linewidth]{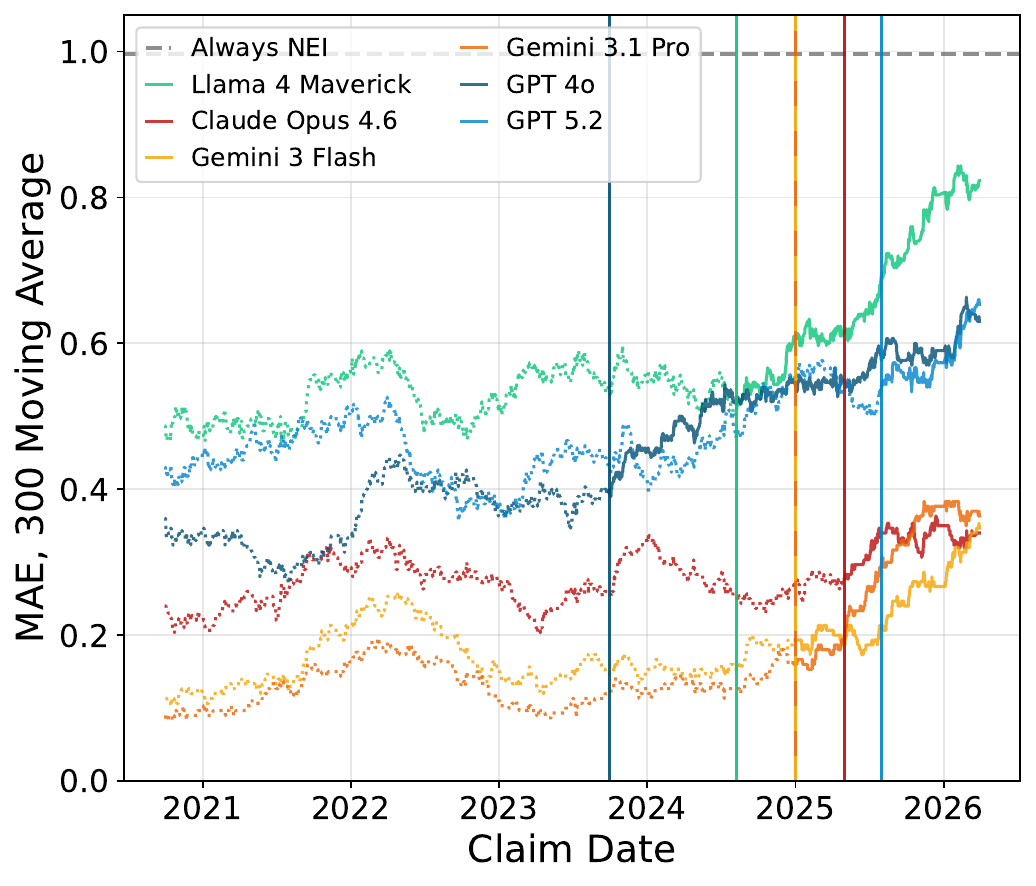}
        \caption{MAE \textbf{without} web search.}
        \label{fig:avg_300_claims_MAE_no-search}
    \end{subfigure}
    \hfill
    \begin{subfigure}[t]{0.48\textwidth}
        \centering
        \includegraphics[width=\linewidth]{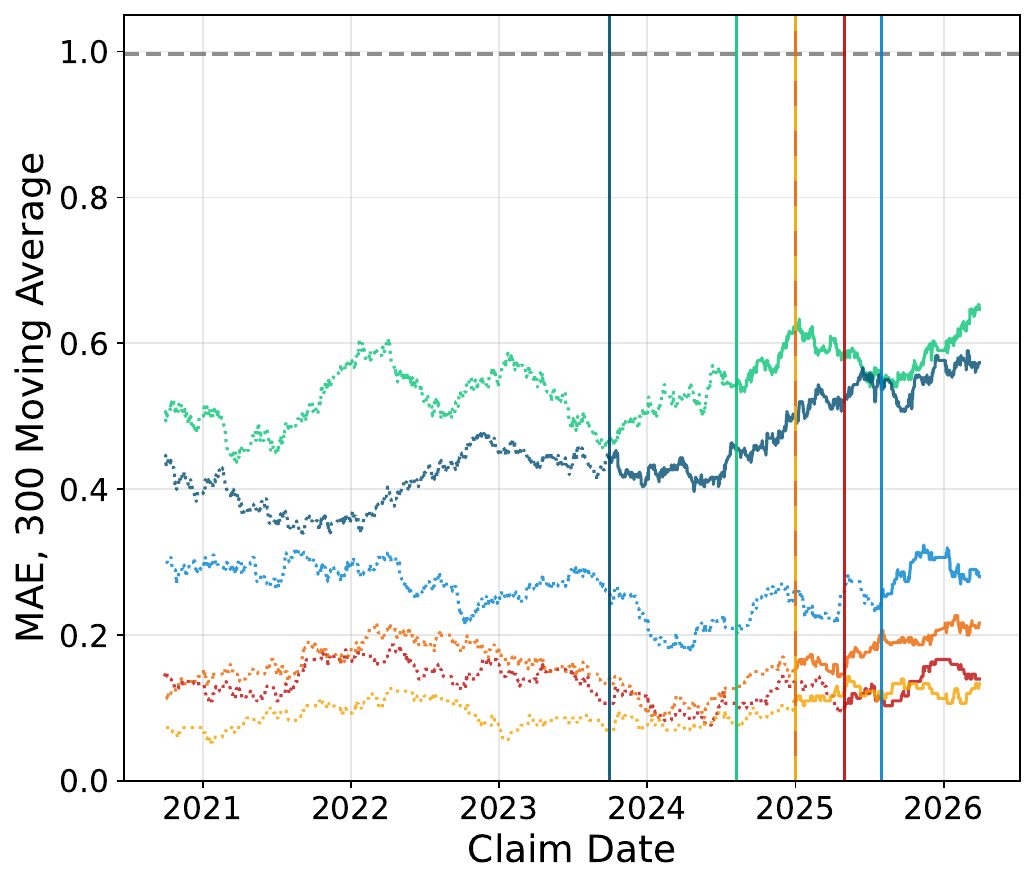}
        \caption{MAE \textbf{with} web search.}
        \label{fig:avg_300_claims_MAE_custom-search}
    \end{subfigure}\vspace{1em}
    
    \begin{subfigure}[t]{0.48\textwidth}
        \centering
        \includegraphics[width=\linewidth]{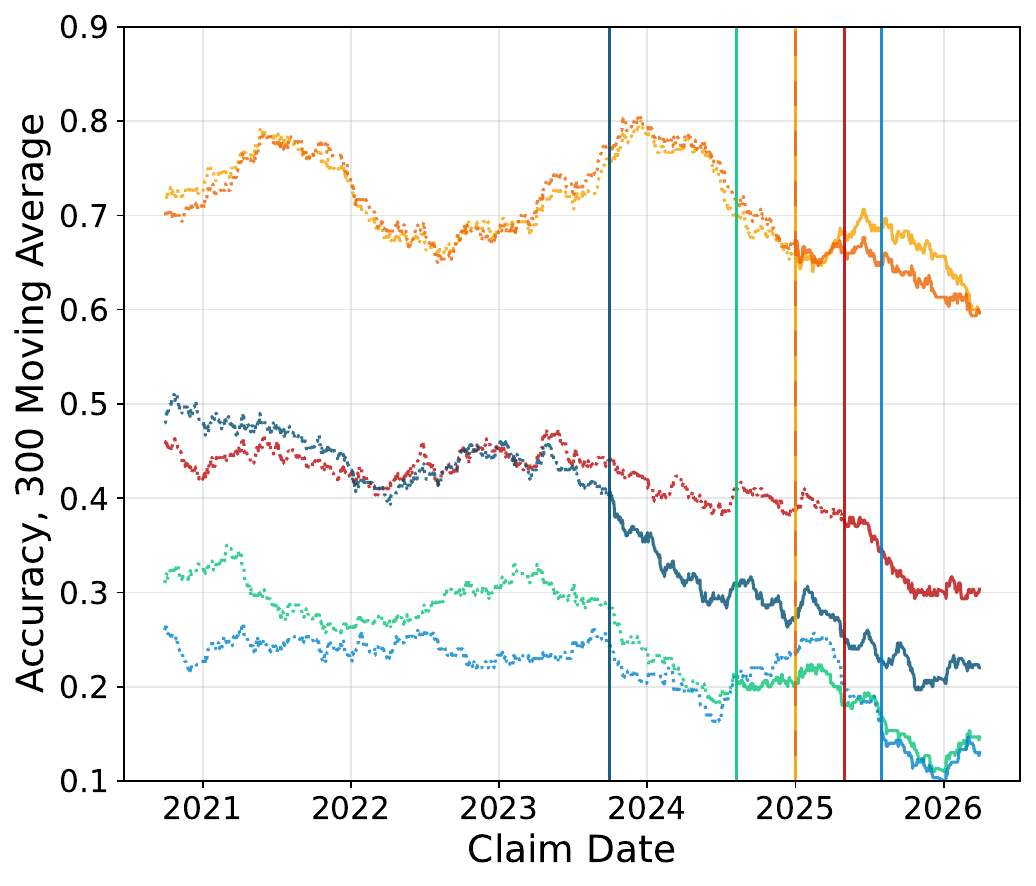}
        \caption{Accuracy \textbf{without} web search.}
        \label{fig:avg_300_claims_Acc_no-search}
    \end{subfigure}
    \hfill
    \begin{subfigure}[t]{0.48\textwidth}
        \centering
        \includegraphics[width=\linewidth]{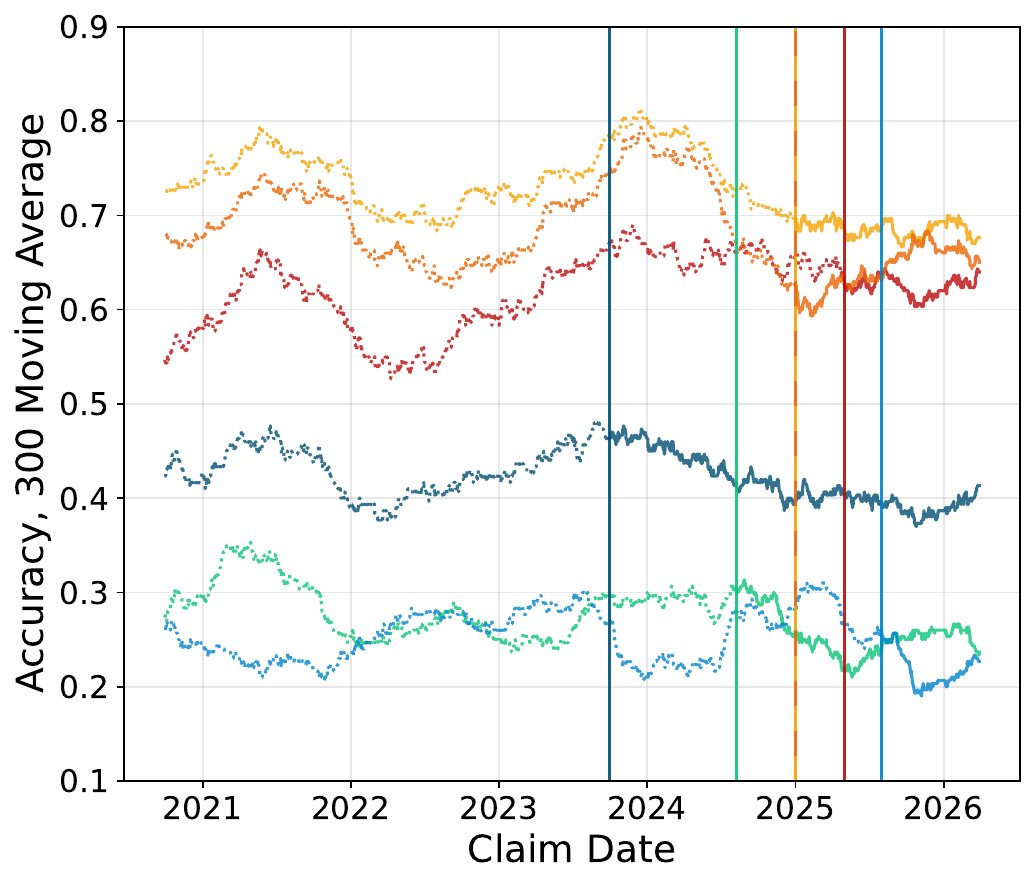}
        \caption{Accuracy \textbf{with} web search.}
        \label{fig:avg_300_claims_Acc_custom-search}
    \end{subfigure}
    
    \caption{Baseline results on the \textbf{longitudinal split} for all three metrics Mean Squared Error (MSE), Mean Absolute Error (MAE), and Accuracy (by 3-bin discretization). All plots use a 200-claim moving average window. Vertical lines indicate knowledge cutoff dates.}
    \label{fig:results_longitudinal_full}
\end{figure*}

\begin{figure*}
    \centering
    \begin{subfigure}[t]{0.48\textwidth}
        \centering
        \includegraphics[width=\linewidth]{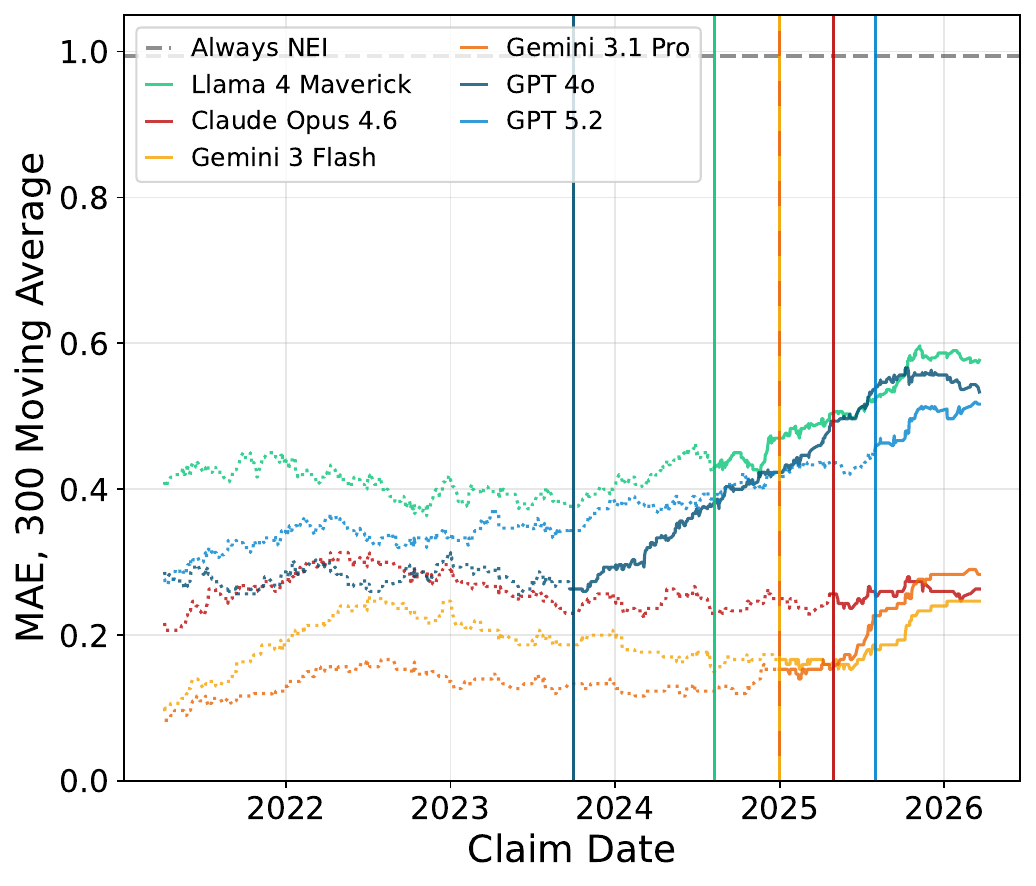}
        \caption{MAE for text-only claims \textbf{without} web search.}
        \label{fig:MA300_claims_MAE_text-only_no-search}
    \end{subfigure}
    \hfill
    \begin{subfigure}[t]{0.48\textwidth}
        \centering
        \includegraphics[width=\linewidth]{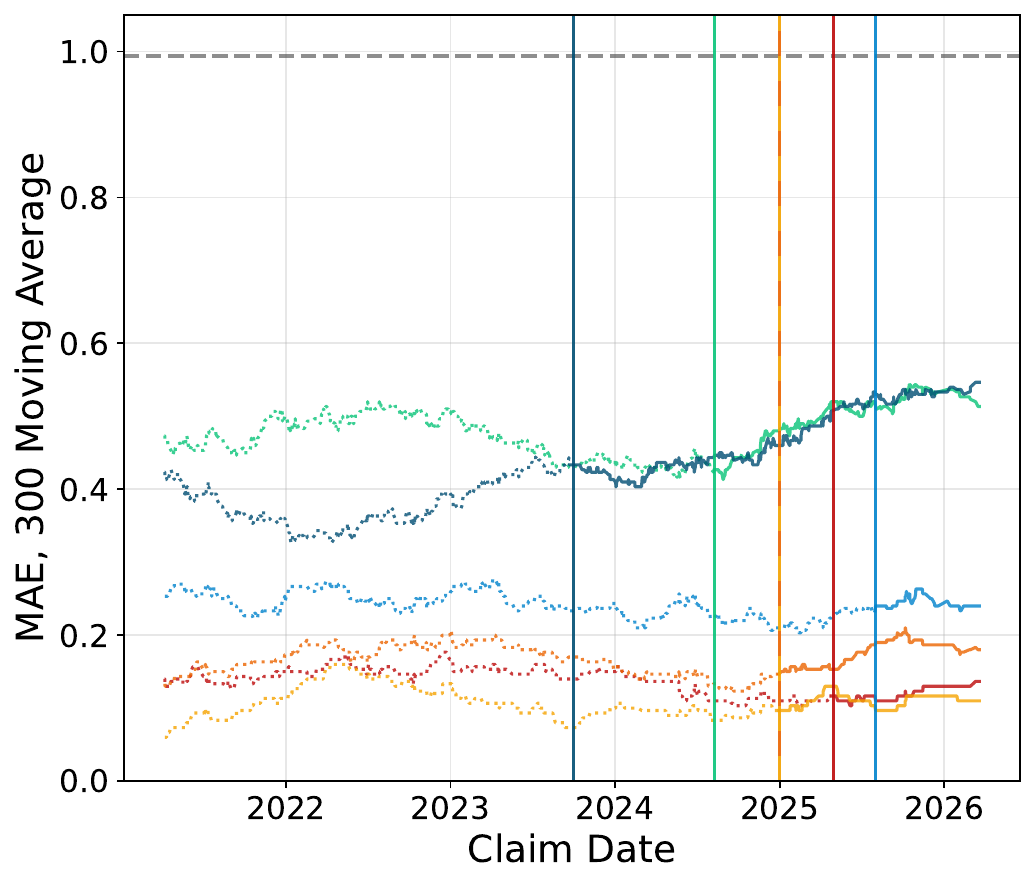}
        \caption{MAE for text-only claims \textbf{with} web search.}
        \label{fig:MA300_claims_MAE_text-only_custom-search}
    \end{subfigure}\vspace{1em}
    
    \begin{subfigure}[t]{0.48\textwidth}
        \centering
        \includegraphics[width=\linewidth]{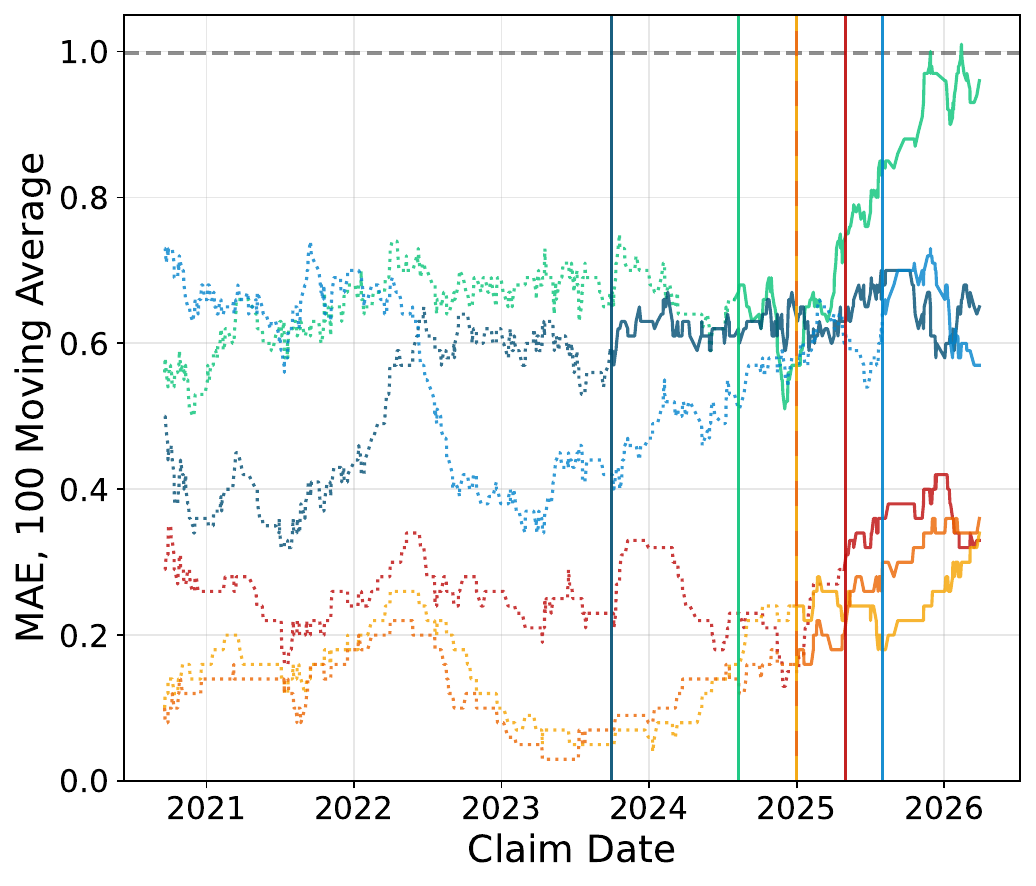}
        \caption{MAE for text+image claims \textbf{without} web search.}
        \label{fig:MA100_claims_MAE_image-only_no-search}
    \end{subfigure}
    \hfill
    \begin{subfigure}[t]{0.48\textwidth}
        \centering
        \includegraphics[width=\linewidth]{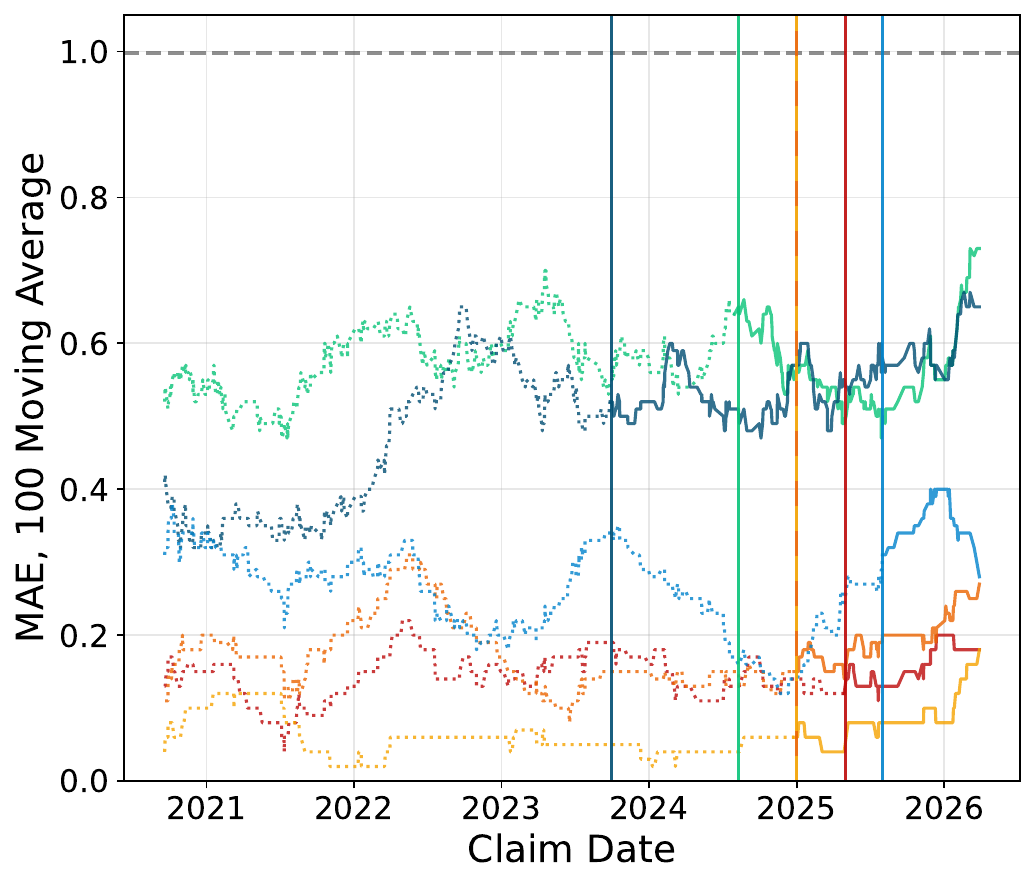}
        \caption{MAE for text+image claims \textbf{with} web search.}
        \label{fig:MA200_claims_MAE_image-only_custom-search}
    \end{subfigure}\vspace{1em}
    
    \begin{subfigure}[t]{0.48\textwidth}
        \centering
        \includegraphics[width=\linewidth]{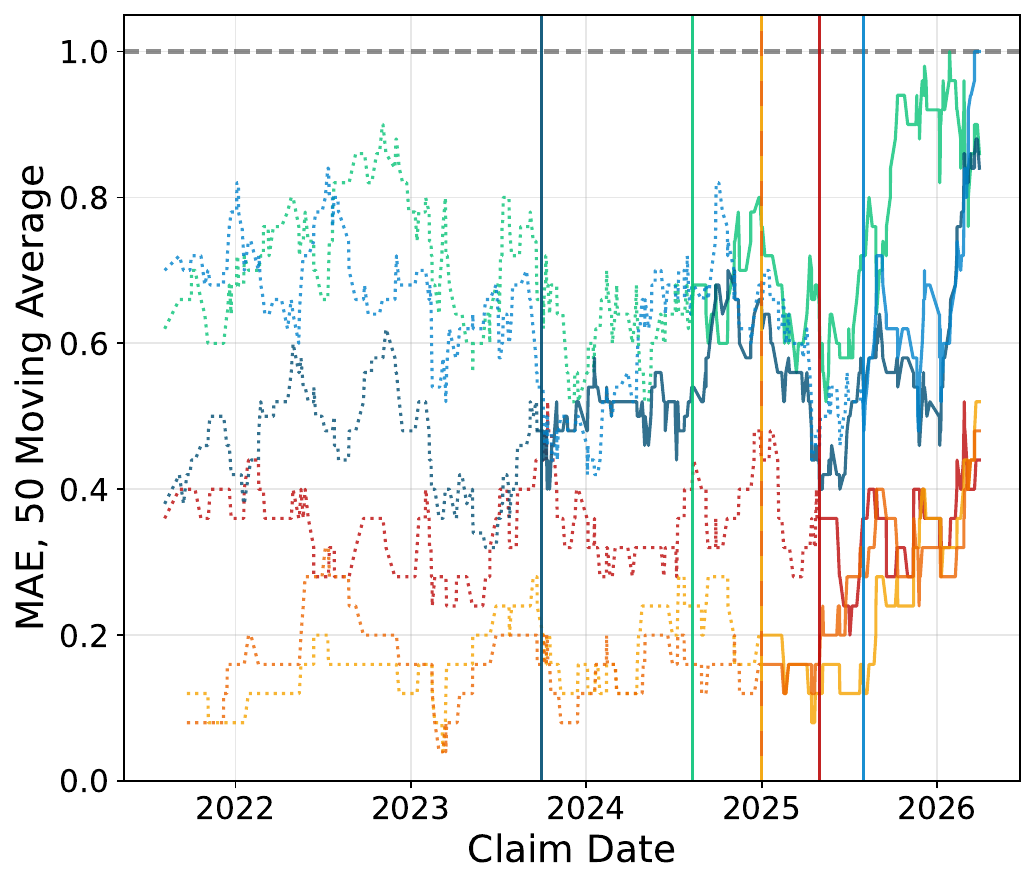}
        \caption{MAE for text+video claims \textbf{without} web search.}
        \label{fig:MA500_claims_MAE_video-only_no-search}
    \end{subfigure}
    \hfill
    \begin{subfigure}[t]{0.48\textwidth}
        \centering
        \includegraphics[width=\linewidth]{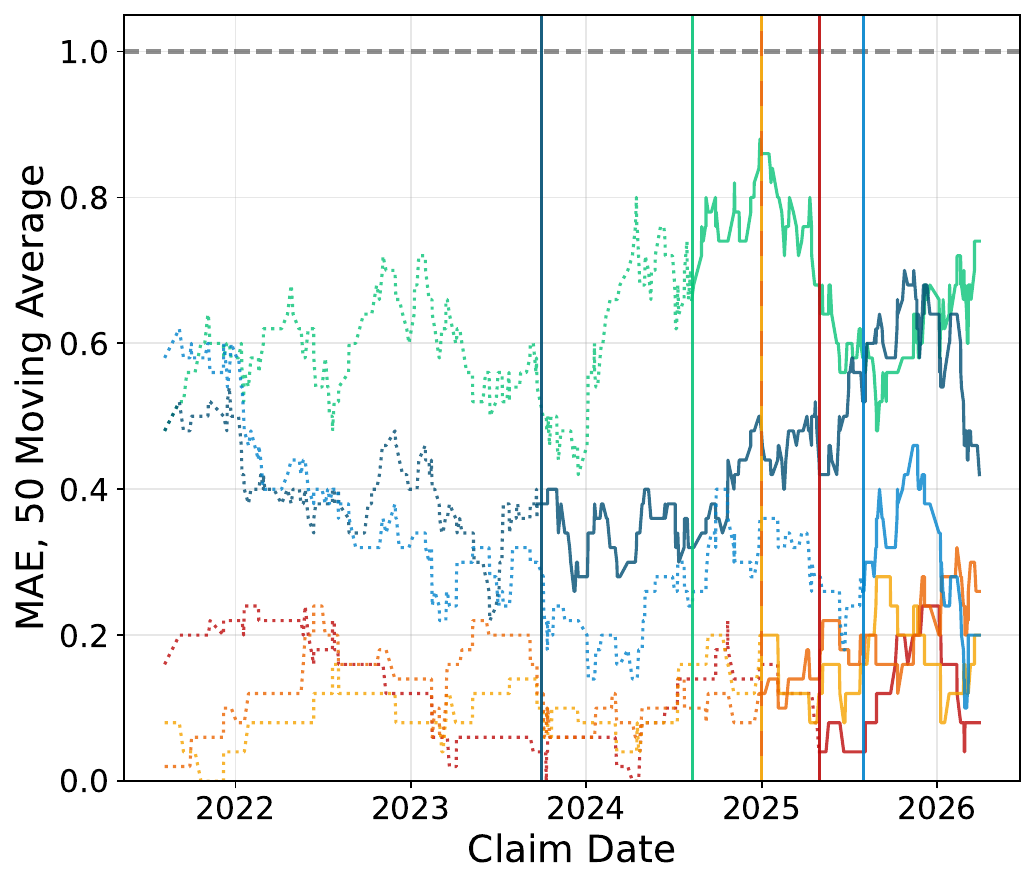}
        \caption{MAE for text+video claims \textbf{with} web search.}
        \label{fig:MA50_claims_MAE_video-only_custom-search}
    \end{subfigure}
    
    \caption{Baseline results on the \textbf{longitudinal split} for modality specific Mean Absolute Error (MAE). All plots use a moving average window. Vertical lines indicate knowledge cutoff dates.}
    \label{fig:results_longitudinal_per_modality}
\end{figure*}

\begin{figure*}
    \centering
    \begin{subfigure}[t]{0.48\textwidth}
        \centering
        \includegraphics[width=\textwidth]{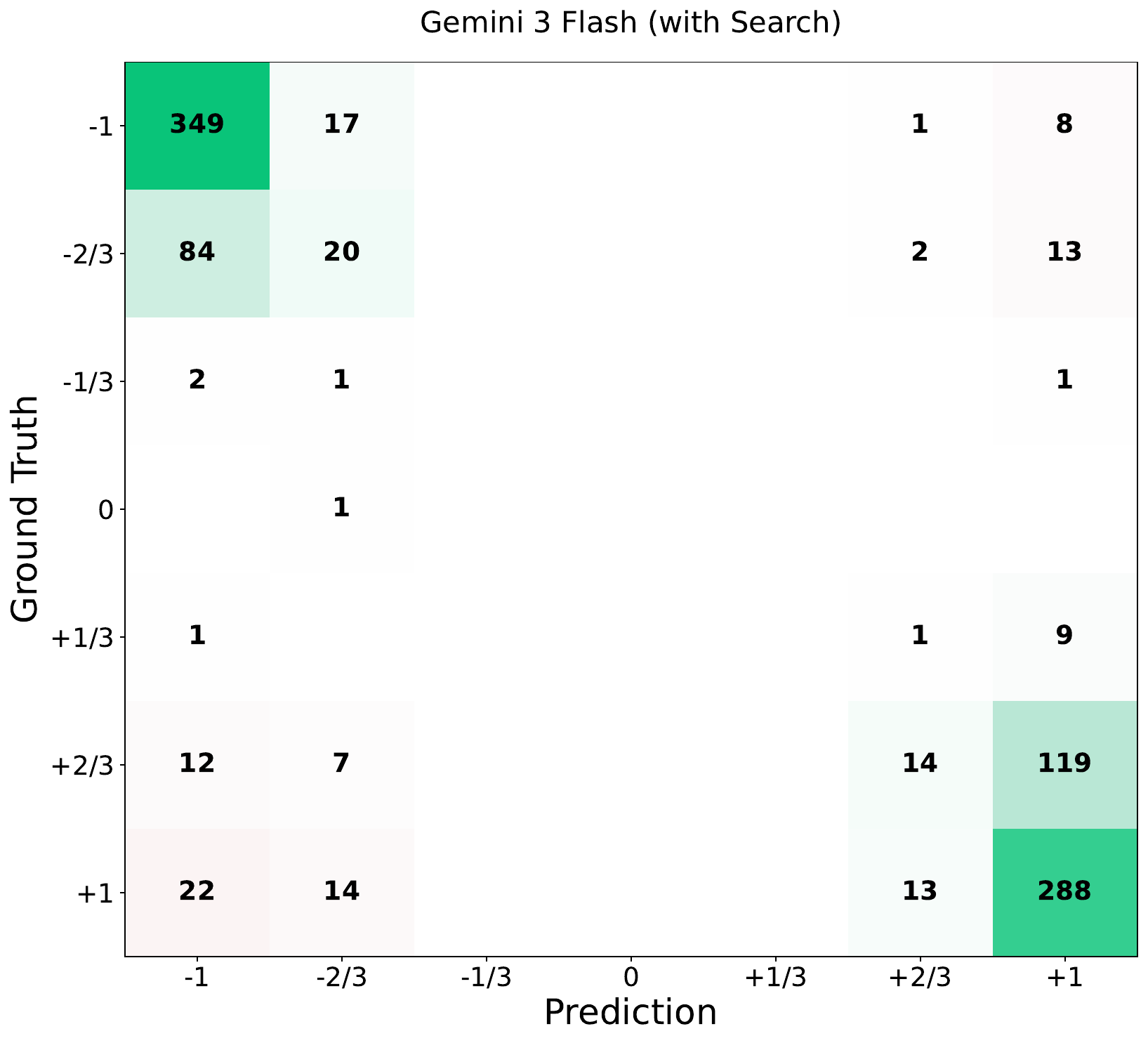}
    \end{subfigure}
    \hfill
    \begin{subfigure}[t]{0.48\textwidth}
        \centering
        \includegraphics[width=\textwidth]{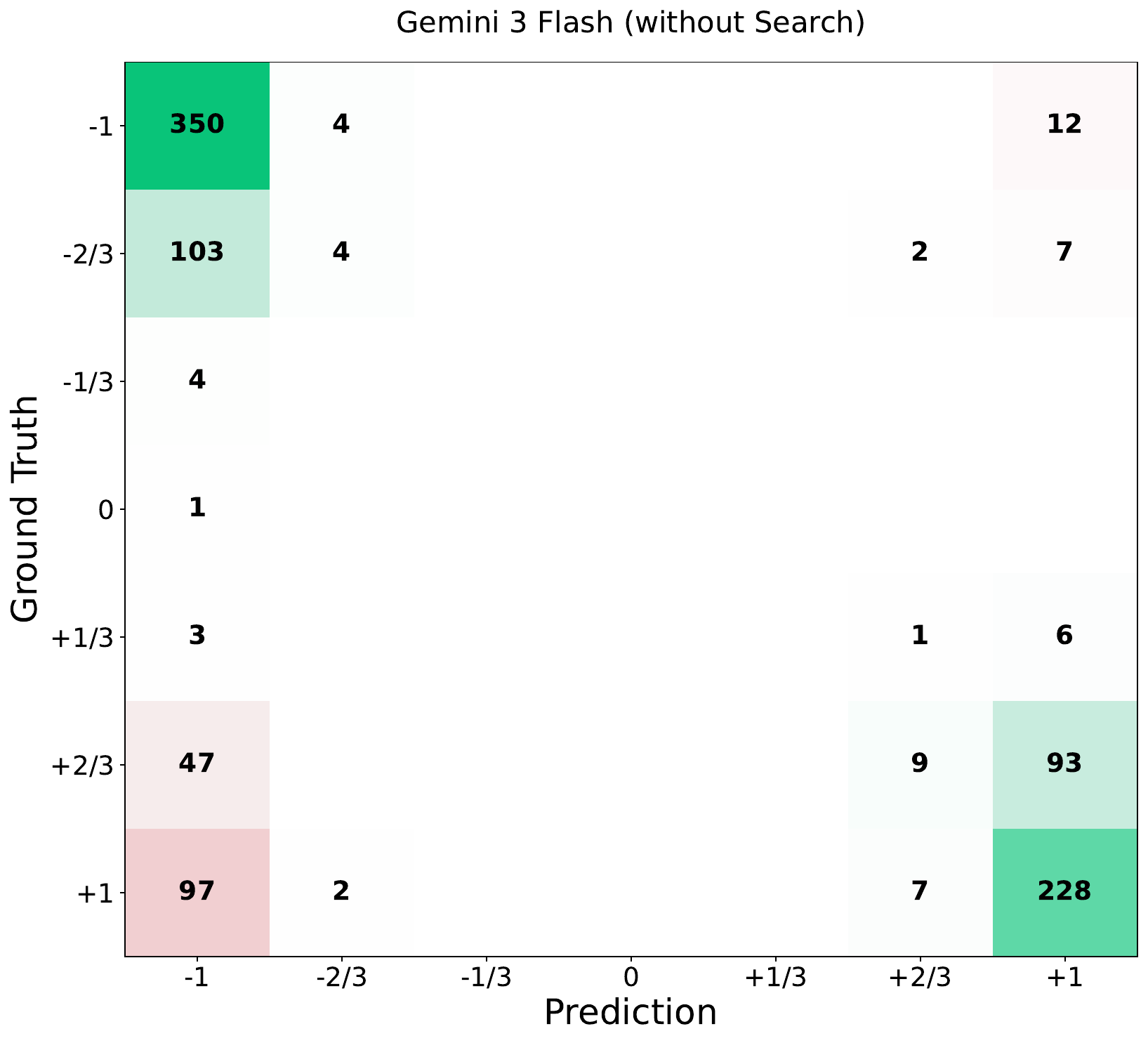}
    \end{subfigure}\vspace{0.5em}

    \begin{subfigure}[t]{0.48\textwidth}
        \centering
        \includegraphics[width=\textwidth]{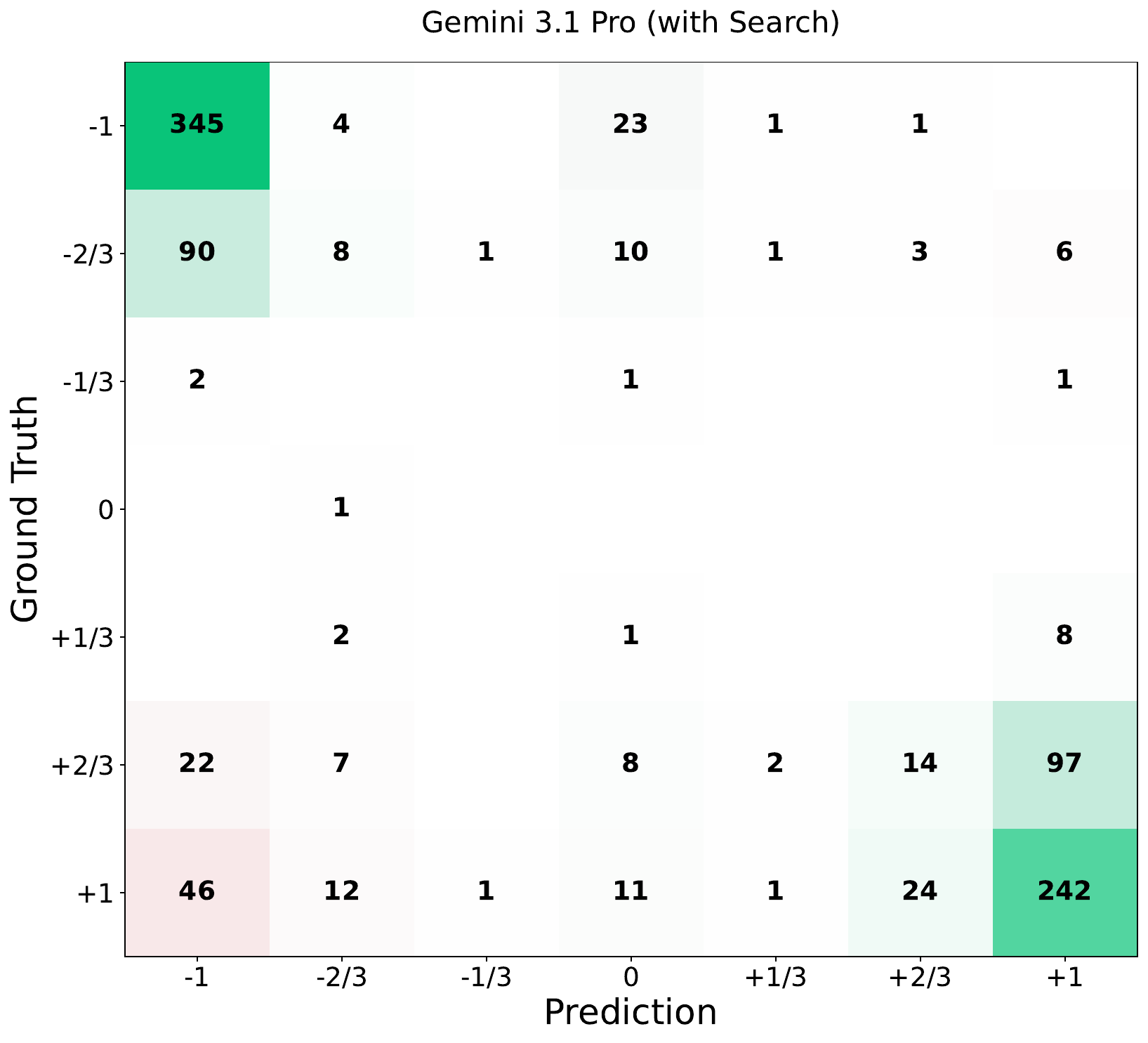}
    \end{subfigure}
    \hfill
    \begin{subfigure}[t]{0.48\textwidth}
        \centering
        \includegraphics[width=\textwidth]{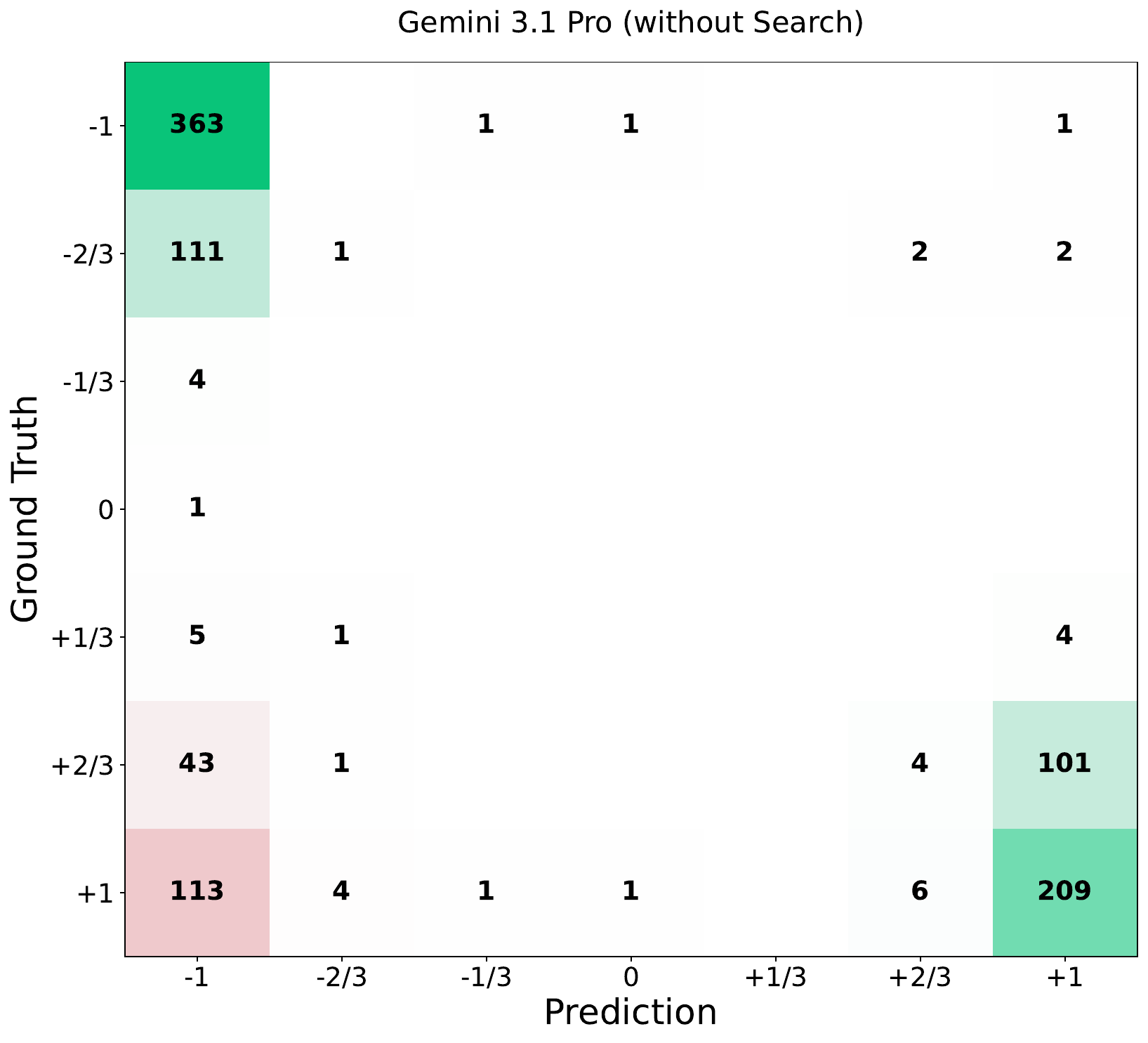}
    \end{subfigure}\vspace{0.5em}
    
    \begin{subfigure}[t]{0.48\textwidth}
        \centering
        \includegraphics[width=\textwidth]{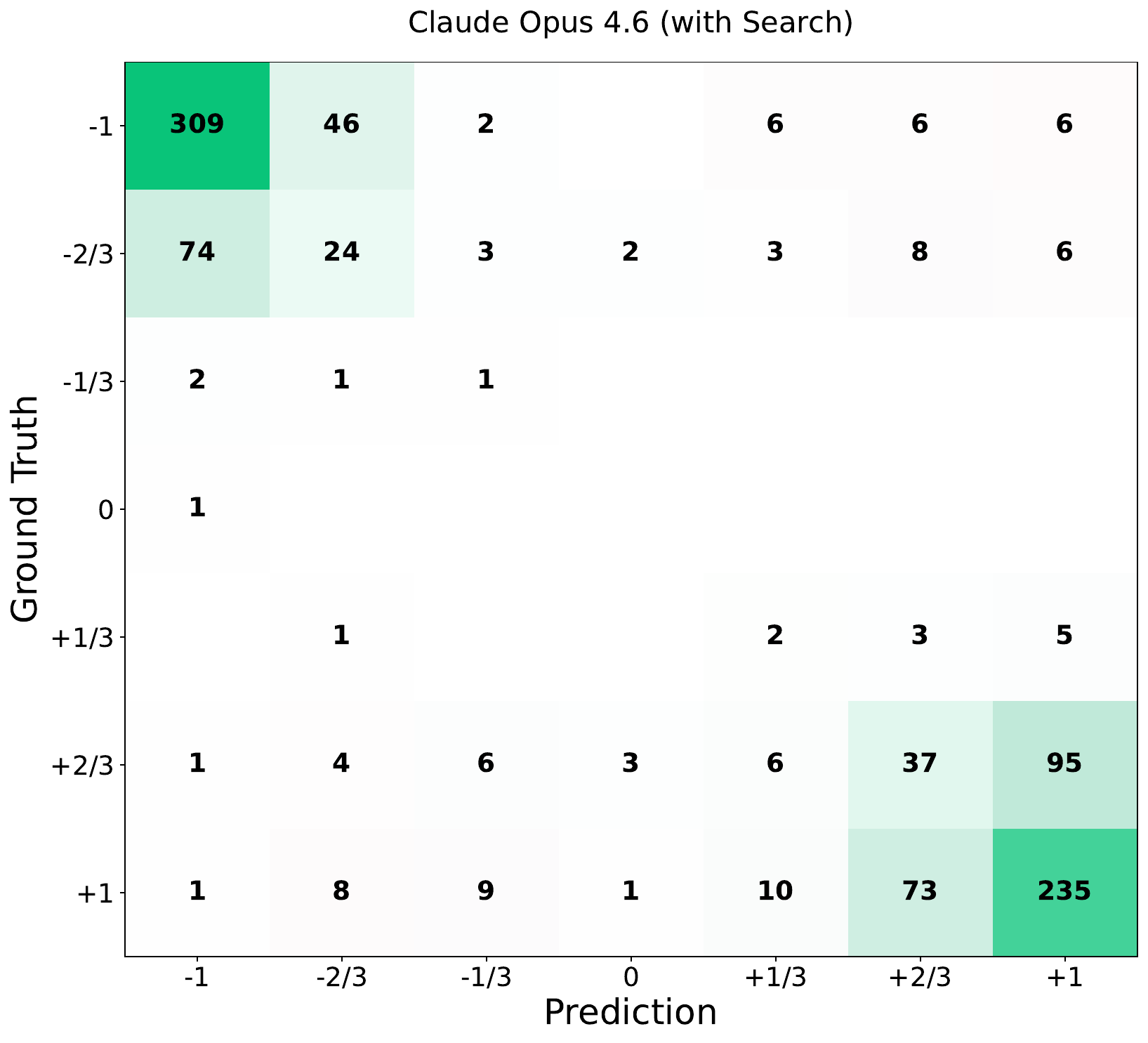}
    \end{subfigure}
    \hfill
    \begin{subfigure}[t]{0.48\textwidth}
        \centering
        \includegraphics[width=\textwidth]{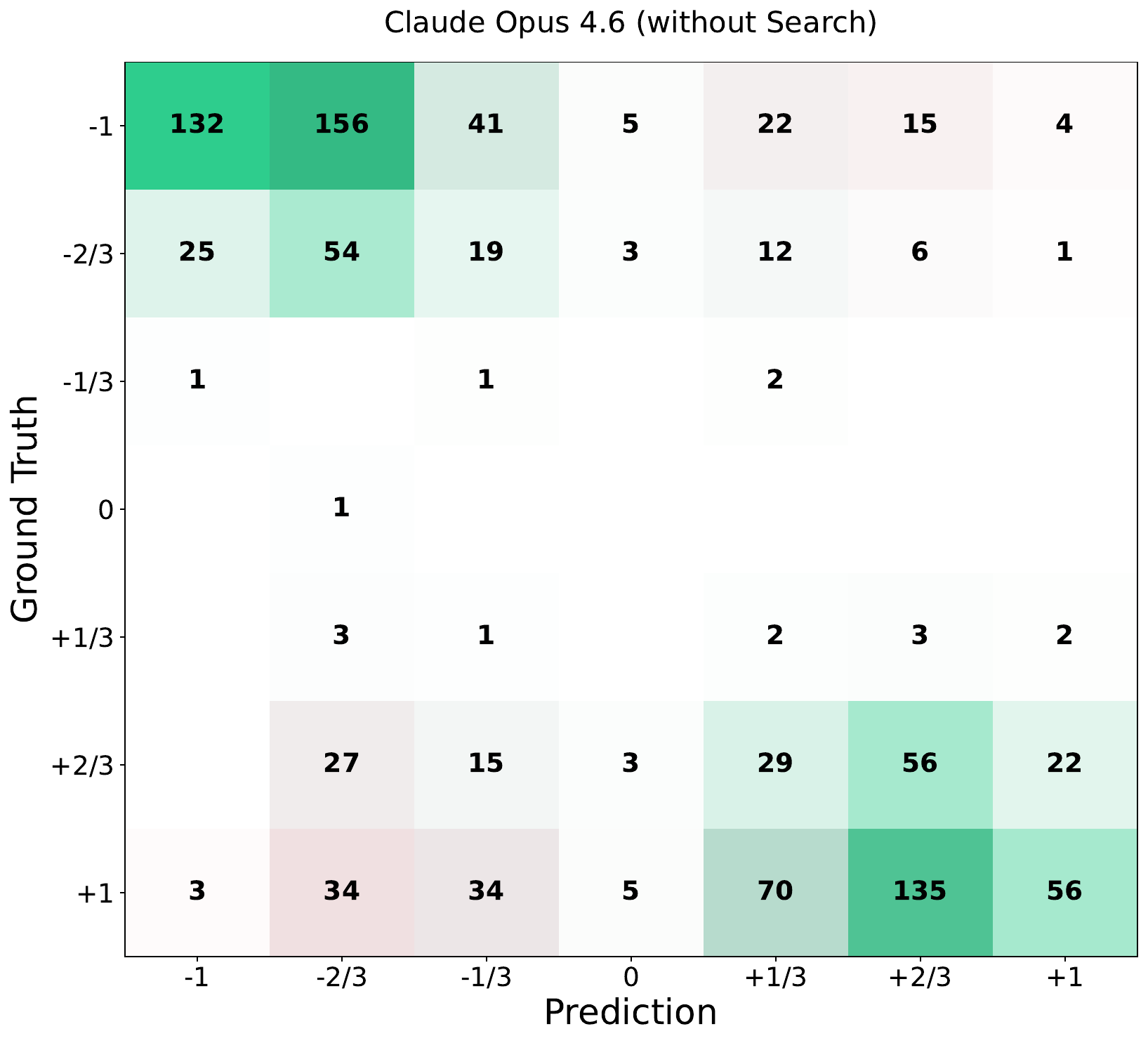}
    \end{subfigure}\vspace{0.5em}
\end{figure*}

\begin{figure*}\ContinuedFloat
    \centering
    \begin{subfigure}[t]{0.48\textwidth}
        \centering
        \includegraphics[width=\textwidth]{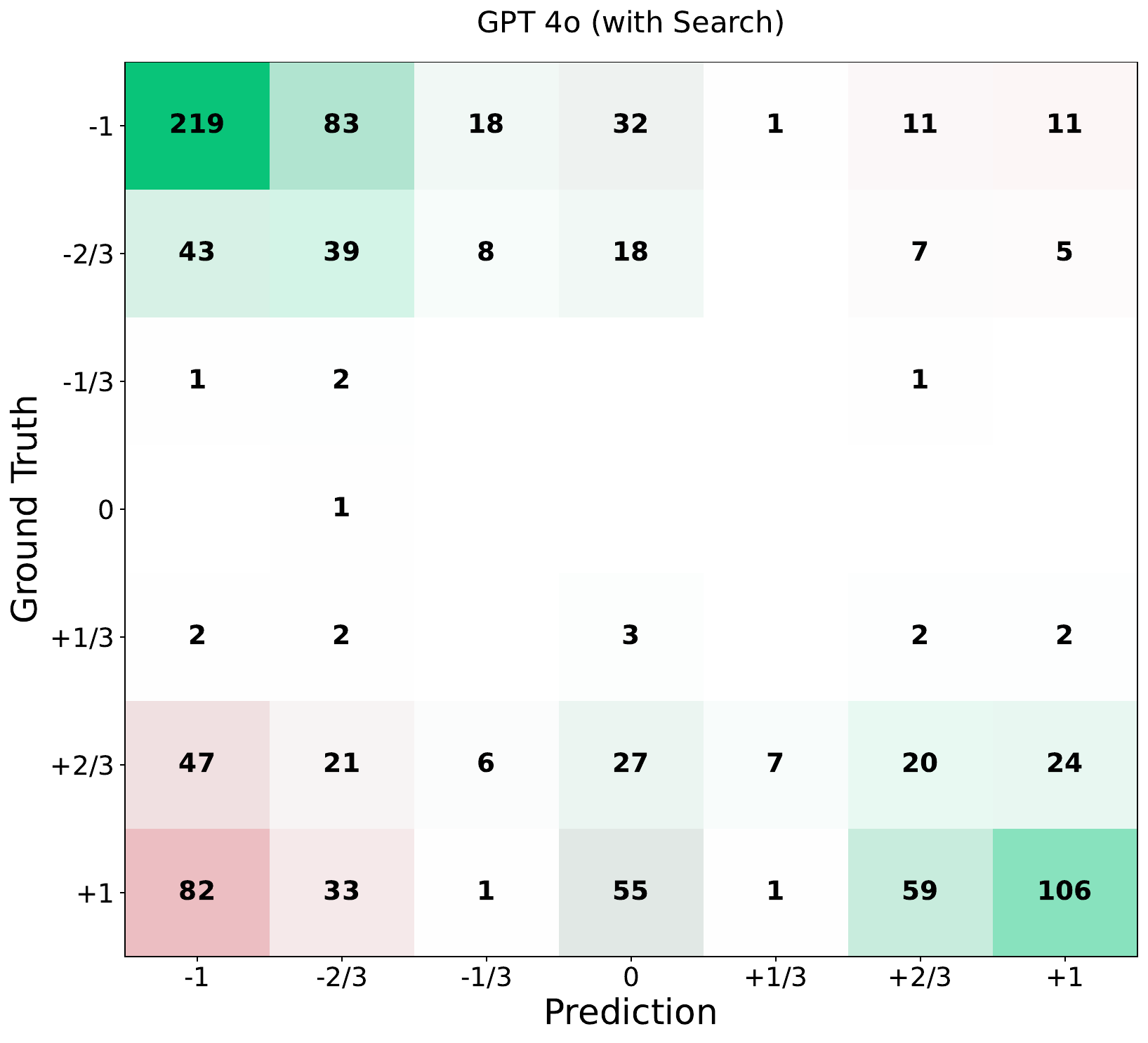}
    \end{subfigure}
    \hfill
    \begin{subfigure}[t]{0.48\textwidth}
        \centering
        \includegraphics[width=\textwidth]{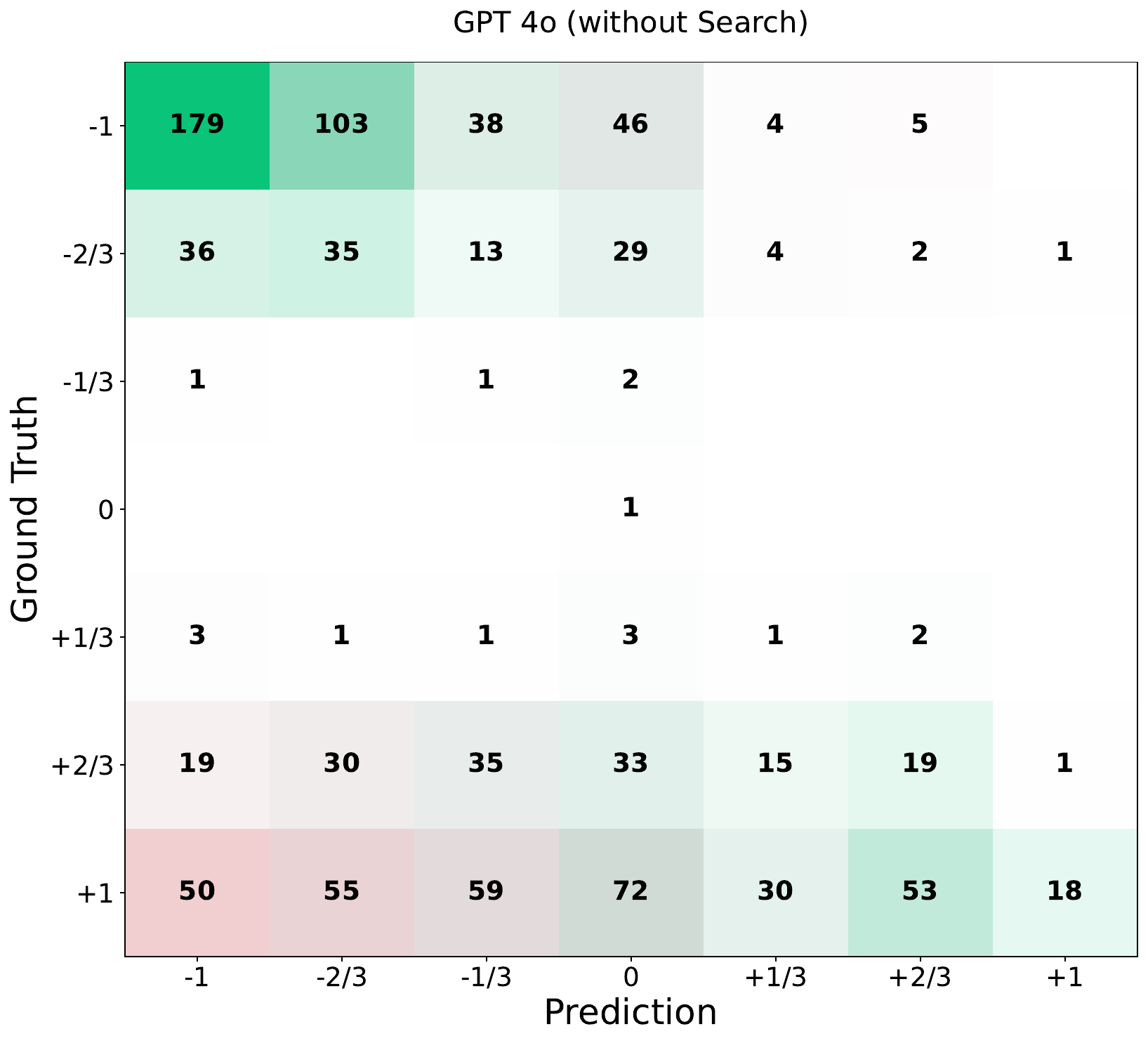}
    \end{subfigure}\vspace{0.5em}

    \begin{subfigure}[t]{0.48\textwidth}
        \centering
        \includegraphics[width=\textwidth]{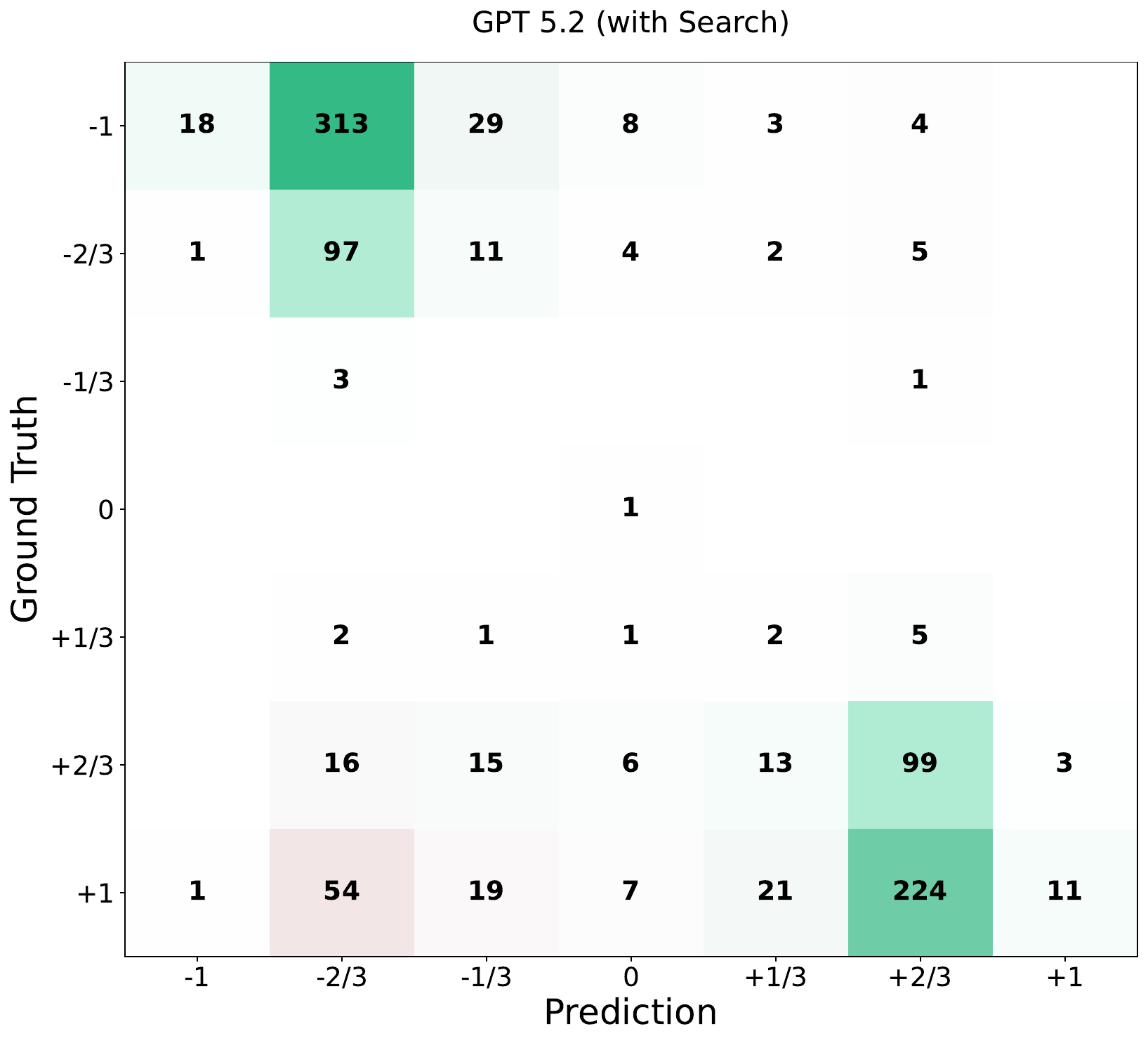}
    \end{subfigure}
    \hfill
    \begin{subfigure}[t]{0.48\textwidth}
        \centering
        \includegraphics[width=\textwidth]{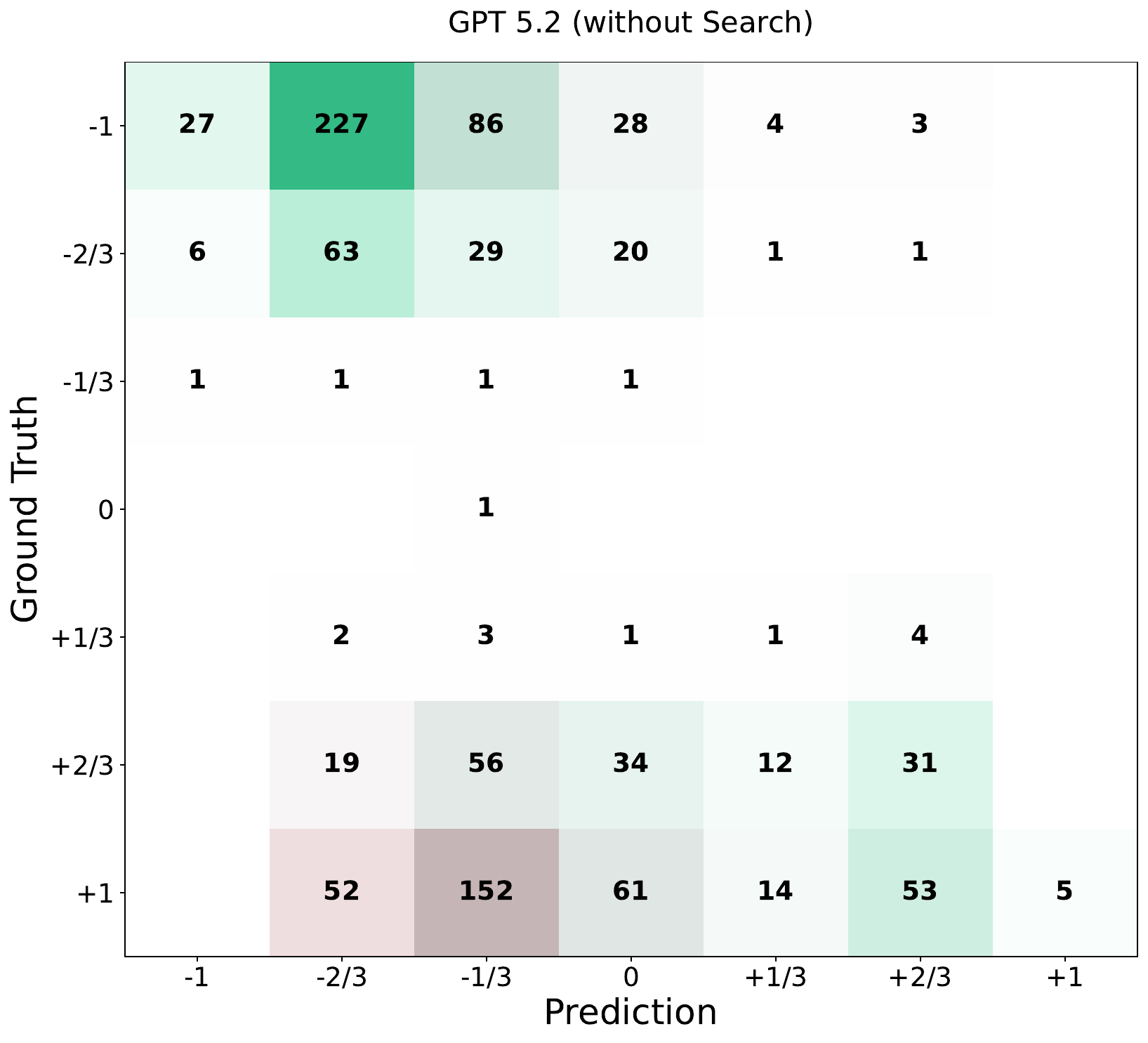}
    \end{subfigure}\vspace{0.5em}

    \begin{subfigure}[t]{0.48\textwidth}
        \centering
        \includegraphics[width=\textwidth]{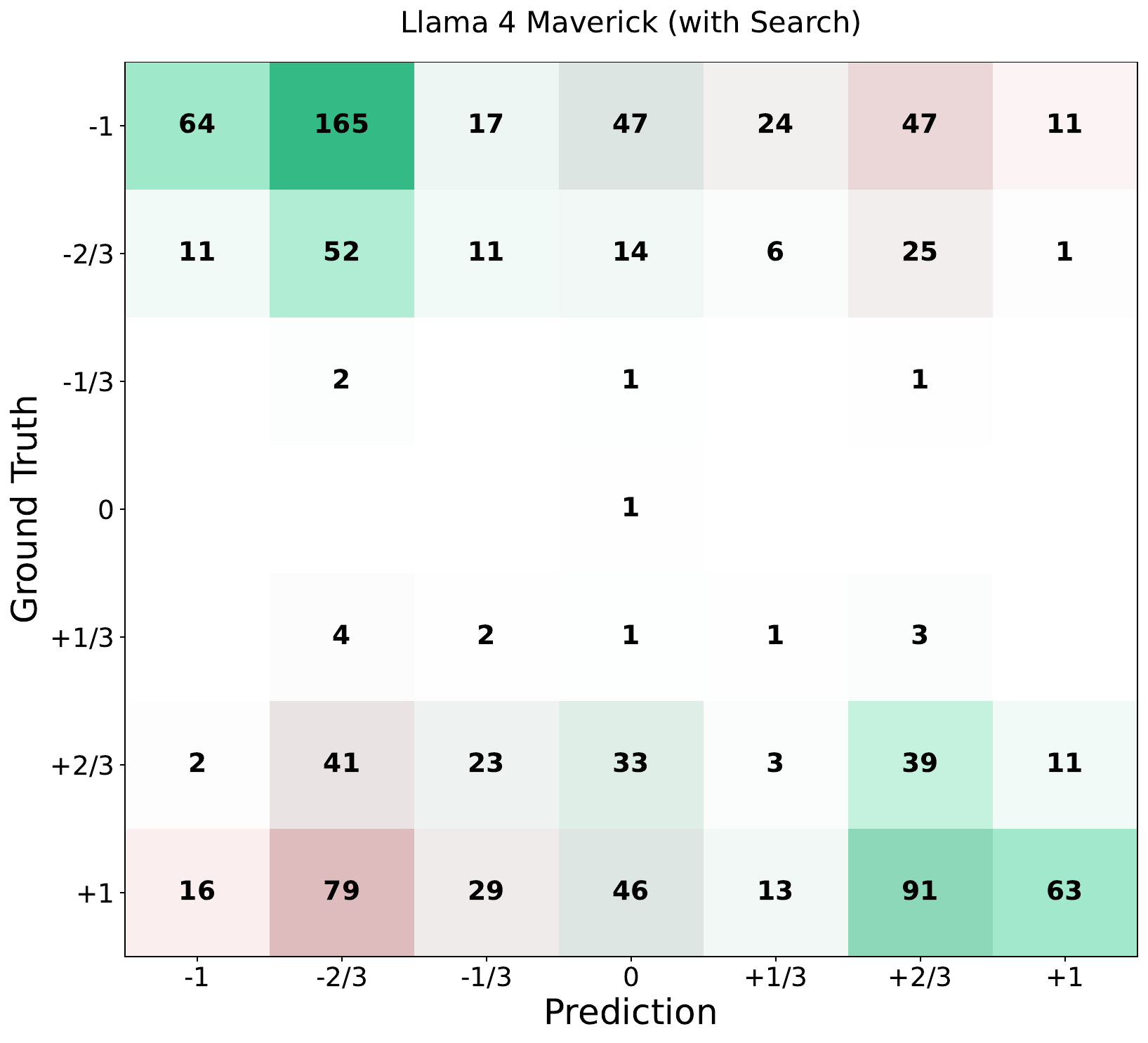}
    \end{subfigure}
    \hfill
    \begin{subfigure}[t]{0.48\textwidth}
        \centering
        \includegraphics[width=\textwidth]{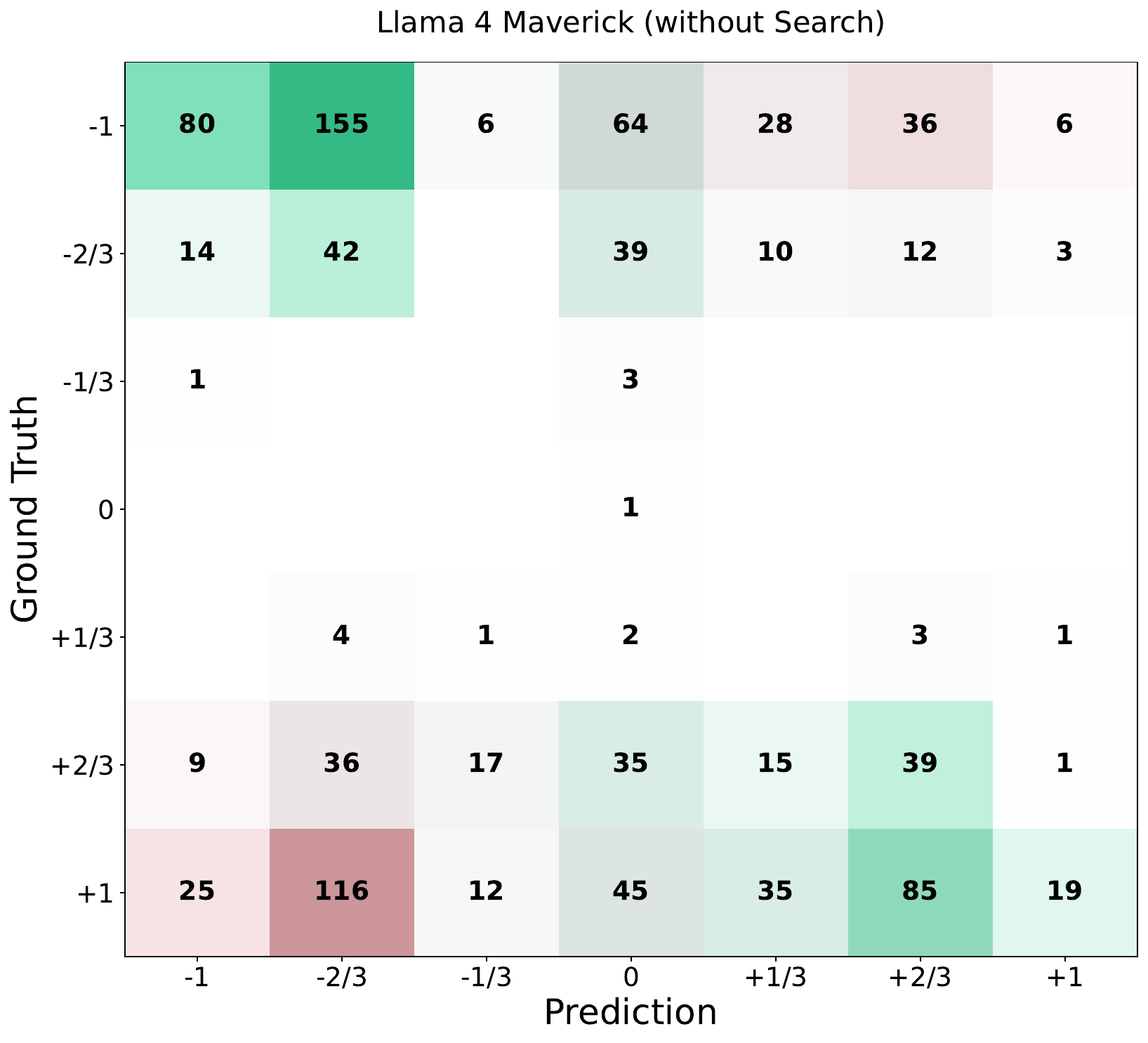}
    \end{subfigure}\vspace{0.5em}
\end{figure*}

\begin{figure*}\ContinuedFloat
    \begin{subfigure}[t]{0.48\textwidth}
        \centering
        \includegraphics[width=\textwidth]{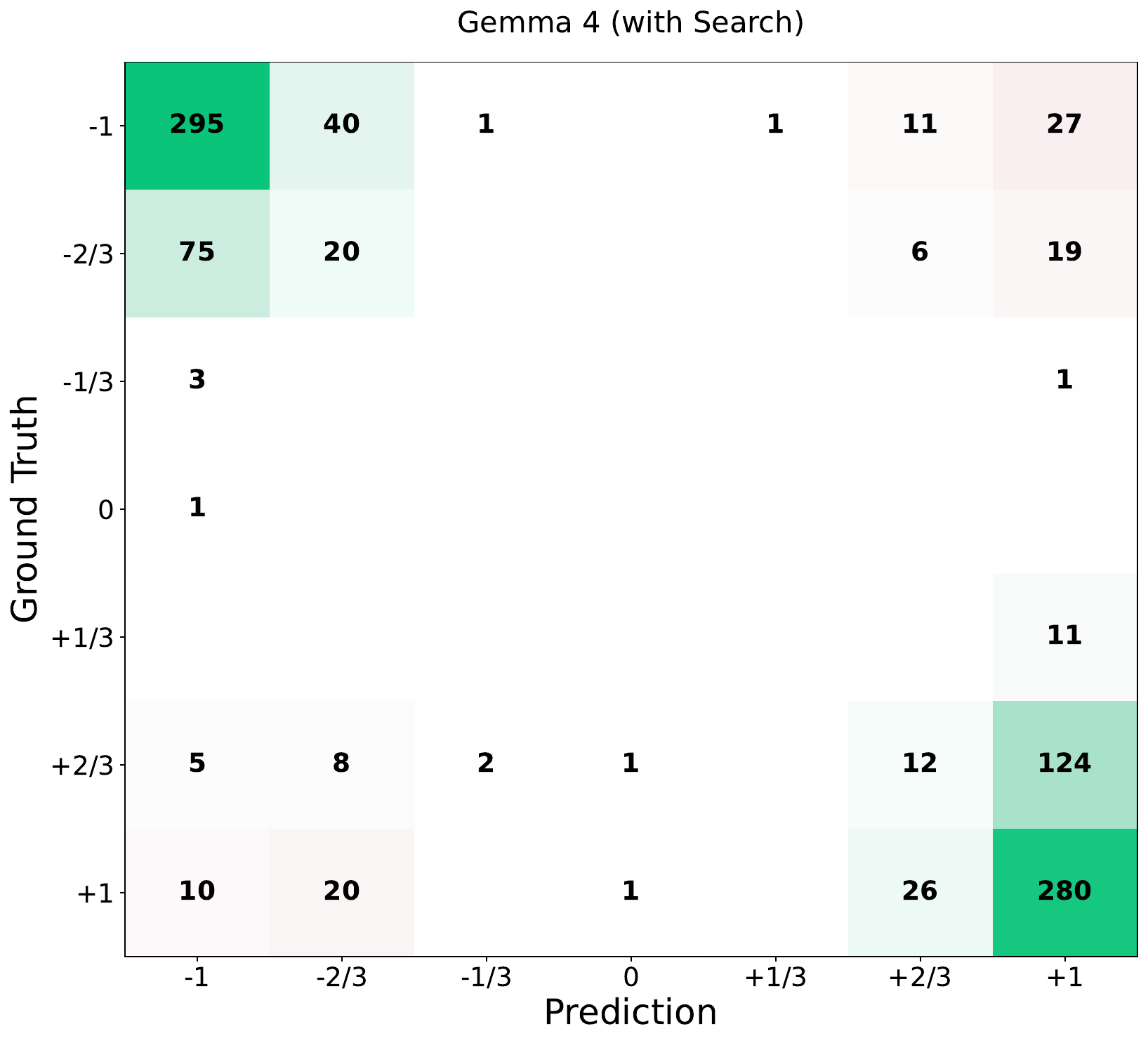}
    \end{subfigure}
    \hfill
    \begin{subfigure}[t]{0.48\textwidth}
        \centering
        \includegraphics[width=\textwidth]{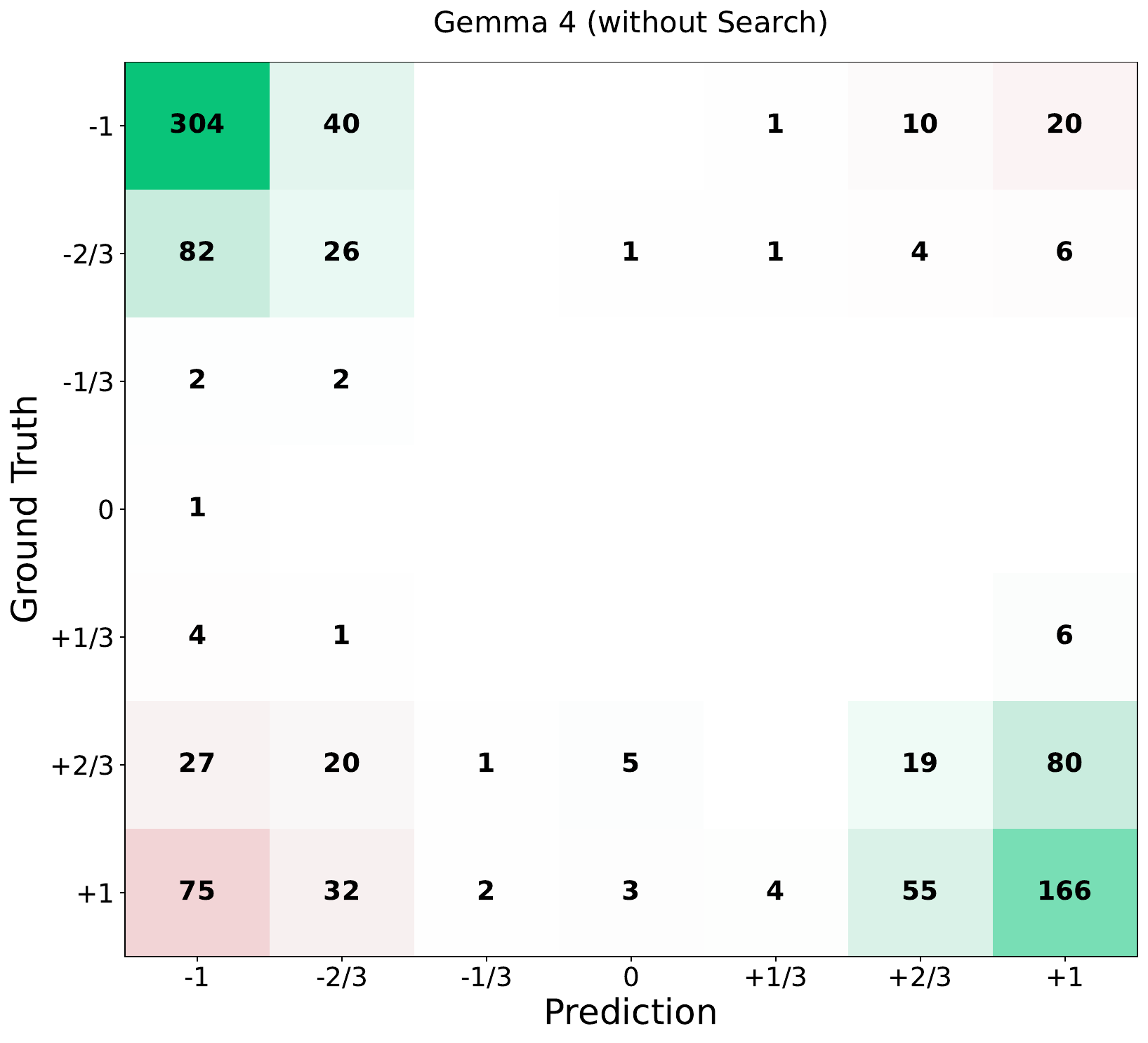}
    \end{subfigure}\vspace{0.5em}

    \begin{subfigure}[t]{0.48\textwidth}
        \centering
        \includegraphics[width=\textwidth]{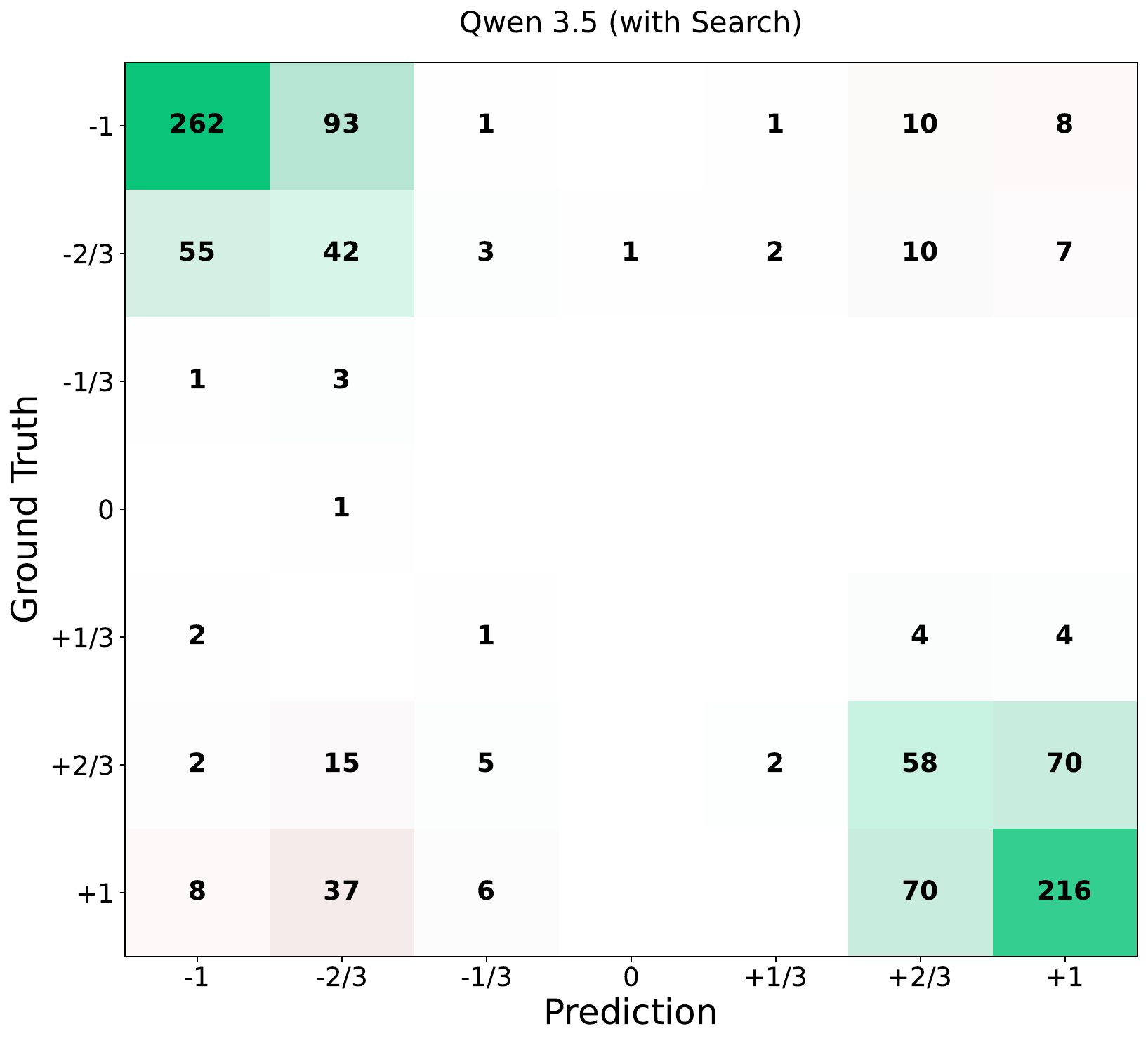}
    \end{subfigure}
    \hfill
    \begin{subfigure}[t]{0.48\textwidth}
        \centering
        \includegraphics[width=\textwidth]{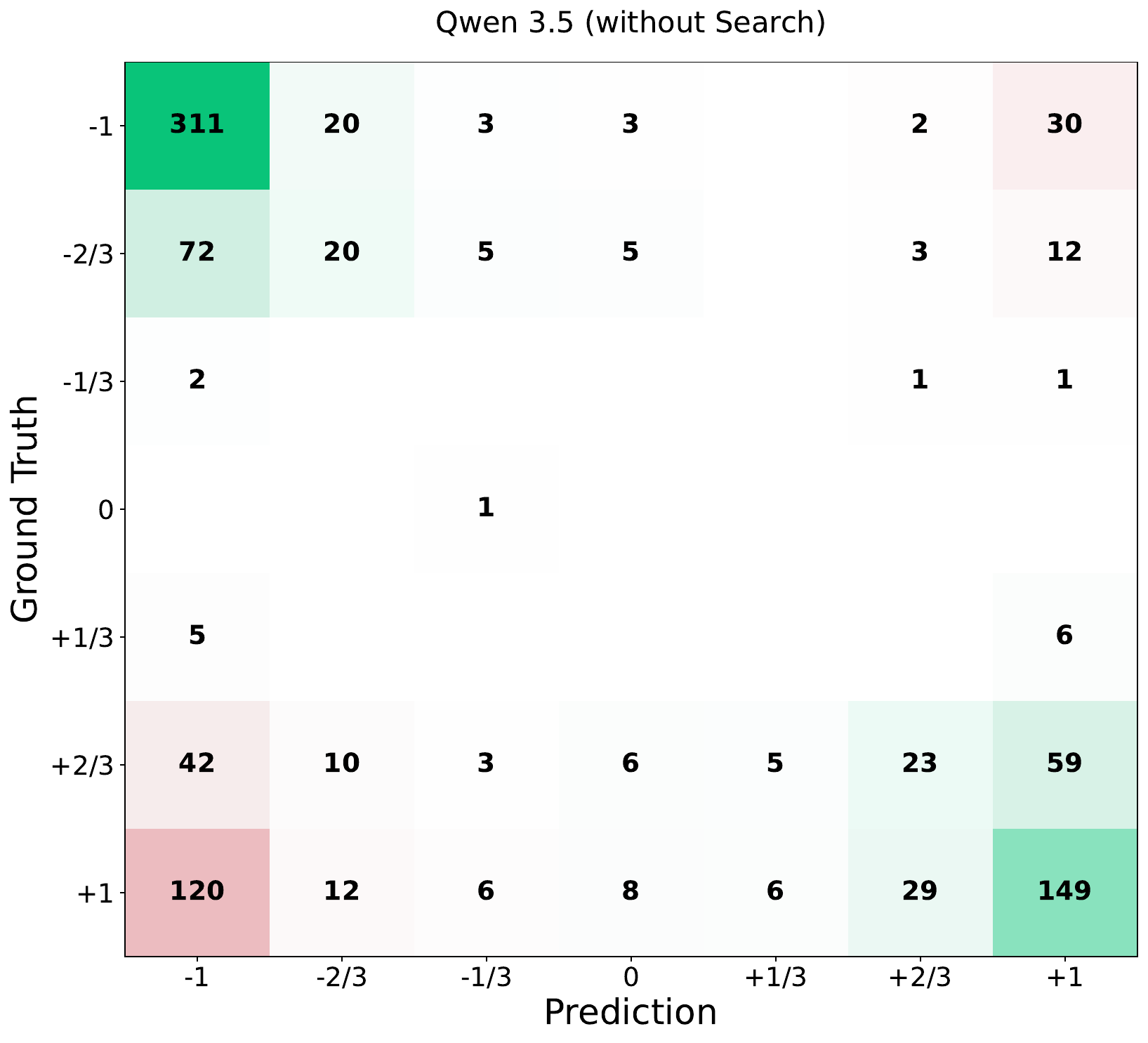}
    \end{subfigure}\vspace{0.5em}
    
    \caption{Confusion matrices for predicting Integrity for Q1 2026.}
    \label{fig:q1_2026_confusion}
\end{figure*}

\section{Human Evaluation}
\label{app:human_eval}

\subsection{Setup Details}

\paragraph{Annotation Organization.}
We conducted two phases of evaluation---both employing the practically same annotation process.

In the first phase, we selected $12$ annotators having at least graduate-level education in Computer Science. $372$ annotations (verdict assessment per claim) were gathered, spanning $9$ different languages. Annotation was done without fees, although two small presents were raffled among the annotators to honor their contribution. The annotation procedure was introduced in a $14$\,min annotation tutorial video\footnote{Phase 1 annotation tutorial: \href{https://youtu.be/ep9ssCf1ets}{youtu.be/ep9ssCf1ets}}. We dismissed annotations with score difference larger than $1$, same for claims that received less than $2$ annotations.

In the second phase, we extended the evaluation to a crowd of $46$ people recruited via Prolific in order strongly to increase the number of evaluated claims. An updated tutorial video was presented to them \footnote{Phase 2 annotation tutorial: \href{https://youtu.be/lYAggVNrynY}{youtu.be/lYAggVNrynY}}, incorporating audience-adequate clarifications and explanation improvements learned from phase one. After watching the tutorial, evaluators were screened: They must correctly answer $10$ out of $10$ non-trivial multiple-choice questions in order to pass the screening. Additionally, we manually reviewed the quality of the annotations of each annotator. $24$ annotators passed this screening. They yielded a total of $444$ additional annotations.

\paragraph{Annotation Process.}Annotators first read the original fact-checking article associated with a claim and perform a manual validation, assessing that the claim (i) is unambiguous, (ii) does not contain text exposing or hinting at the verdict, (iii) any attached media does not contain overlays or labels from the fact-checking article, and (iv) all media referenced in the claim text is attached. Claims failing any criterion are discarded with the failed checks recorded.

The evaluation then proceeds sequentially through the properties, following the same procedure as stage 6 of the \method pipeline, cf.\ Fig.~\ref{fig:stage_6}. Refer to the screenshots in Figure~\ref{fig:human-annotation-process} for a full annotation walkthrough.

For each property, annotators provide a judgment on a seven-point scale incorporating the uncertainty values as defined in the $7$ bin mapping, see Fig.~\ref{fig:binning}. Each judgment requires a written explanation, informing our manual screening. When annotators select a confidence level below ``rather certain,'' they must additionally describe the reason for their uncertainty.

\begin{figure*}
    \centering
    \begin{subfigure}[t]{0.94\textwidth}
        \centering
        \includegraphics[width=\textwidth]{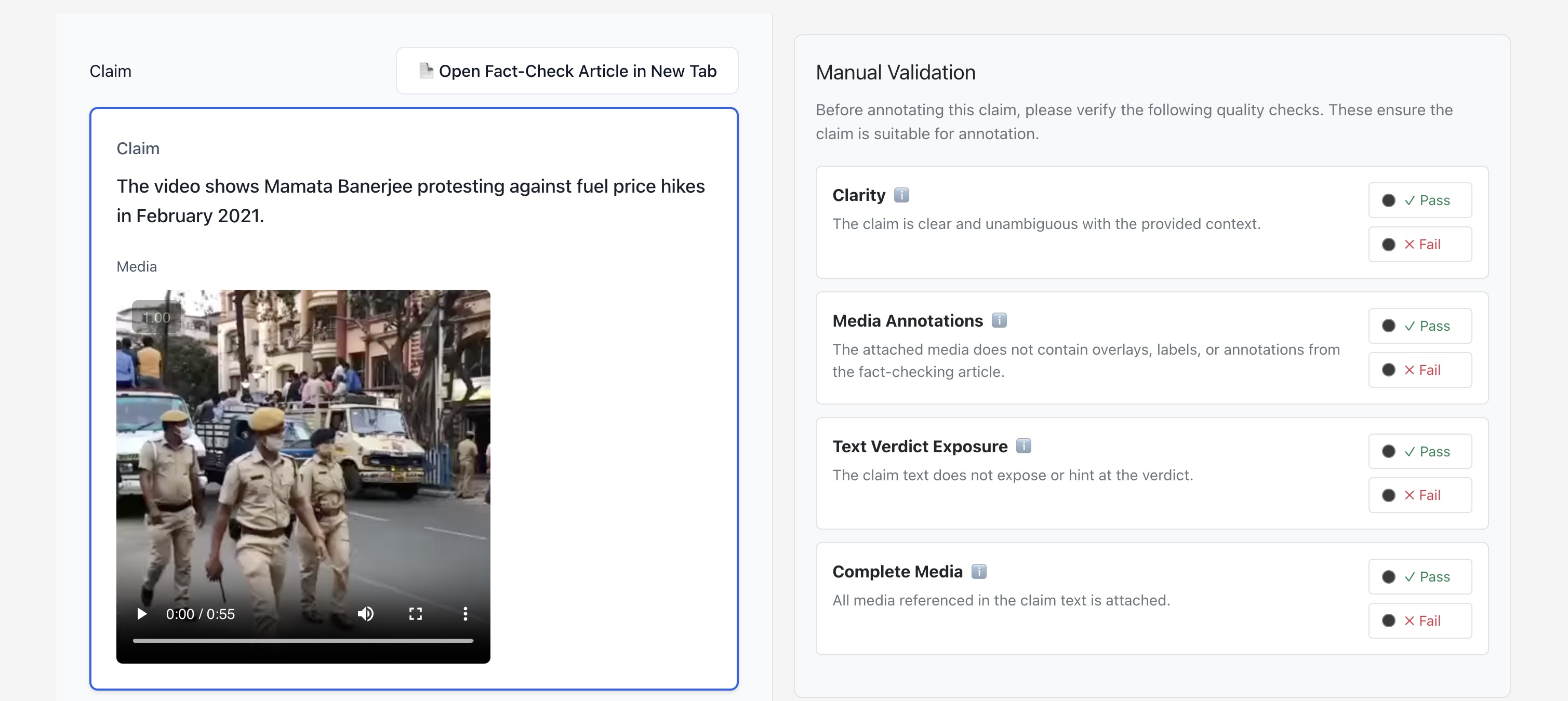}
        \caption{Step 1: Validate that claim fulfills all quality requirements.}
    \end{subfigure}
    
    \begin{subfigure}[t]{0.46\textwidth}
        \centering
        \includegraphics[width=\textwidth]{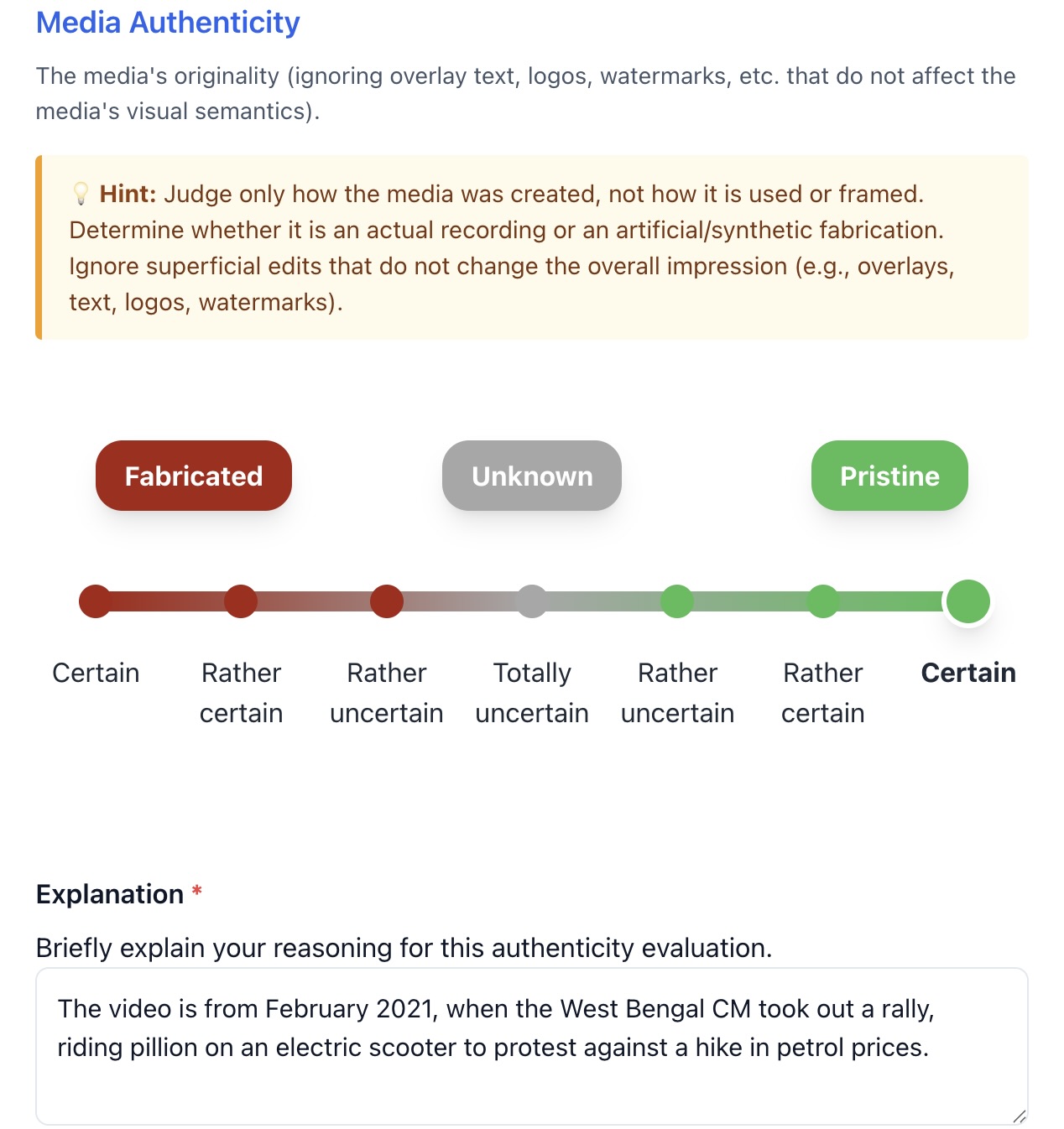}
        \caption{Step 2: Assess media authenticity.}
    \end{subfigure}
    \hfill
    \begin{subfigure}[t]{0.46\textwidth}
        \centering
        \includegraphics[width=\textwidth]{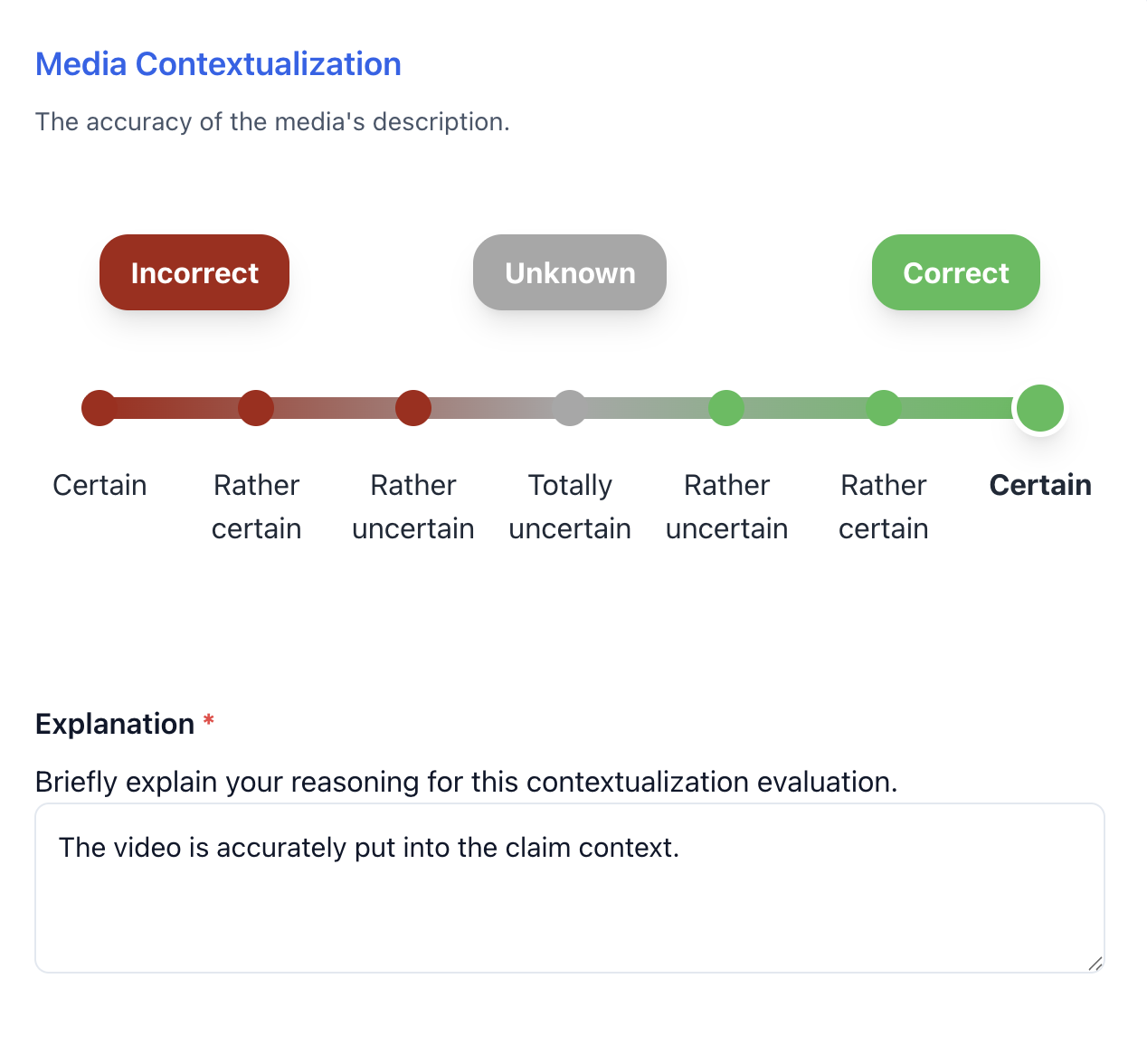}
        \caption{Step 3: Assess media contextualization. Terminate annotation if the contextualization of at least one media item is wrong.}
    \end{subfigure}

    \begin{subfigure}[t]{0.46\textwidth}
        \centering
        \includegraphics[width=\textwidth]{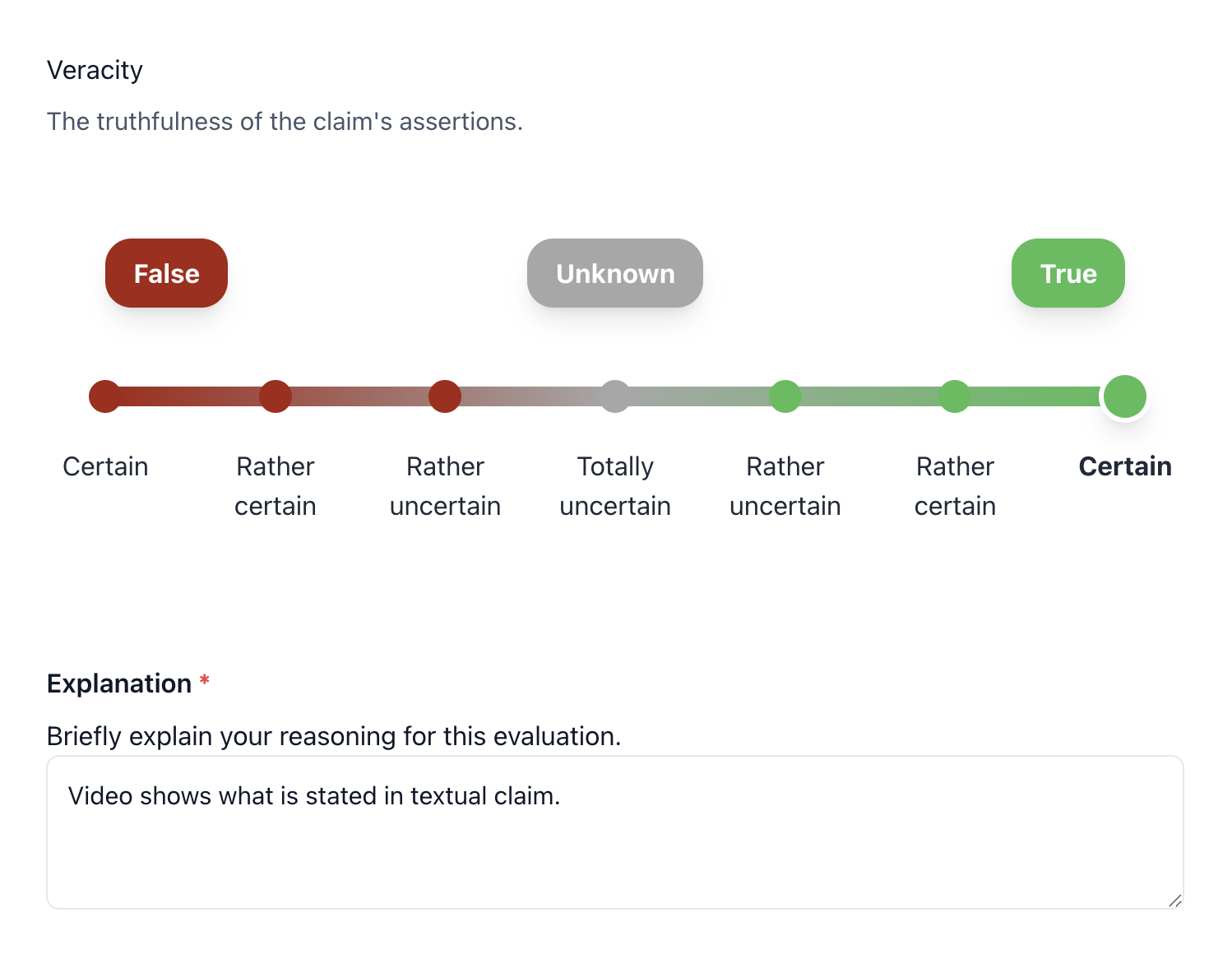}
        \caption{Step 4: Assess claim veracity. Terminate annotation if the veracity is false.}
    \end{subfigure}
    \hfill
    \begin{subfigure}[t]{0.46\textwidth}
        \centering
        \includegraphics[width=\textwidth]{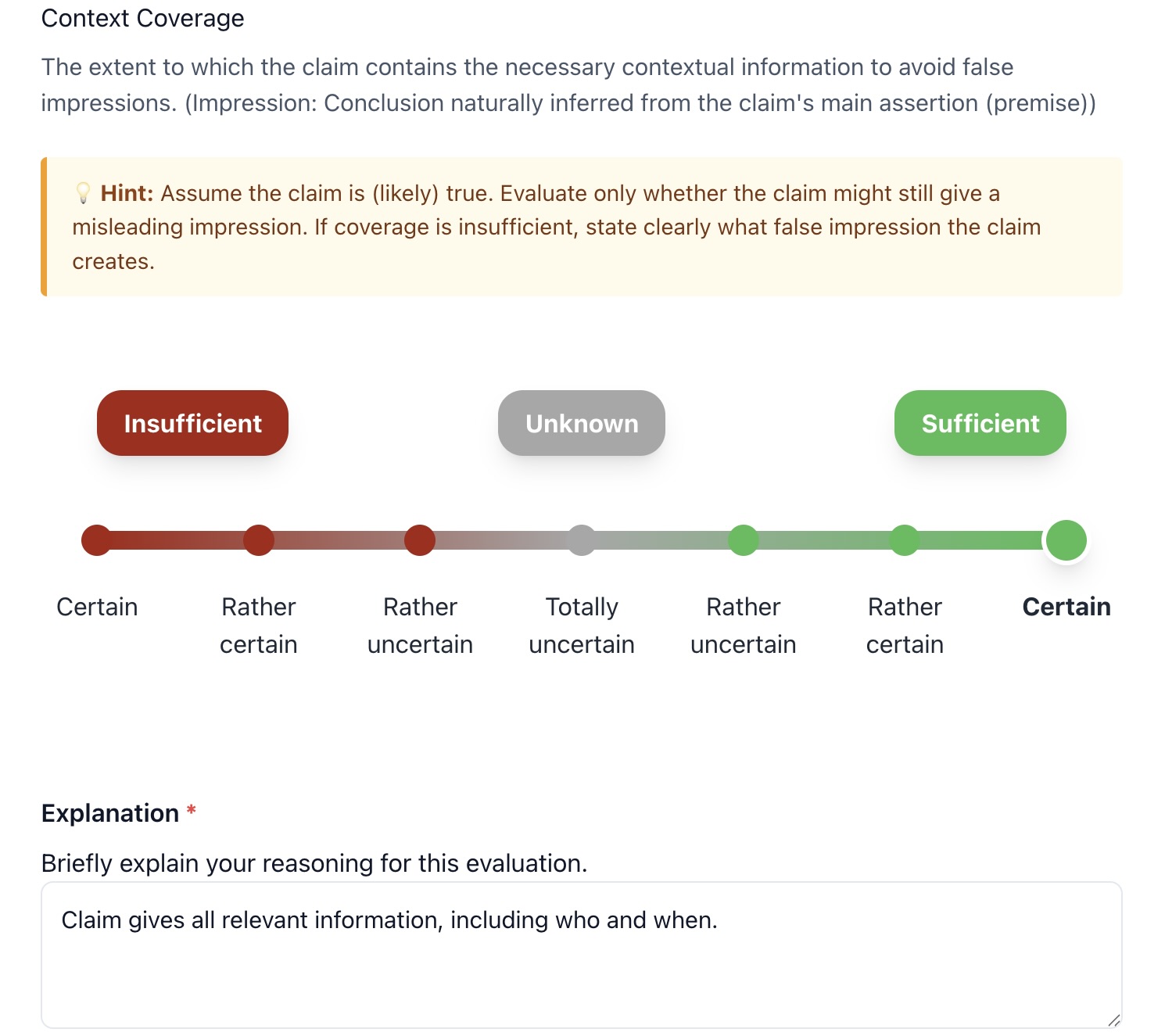}
        \caption{Step 5: Assess context coverage.}
    \end{subfigure}
    \caption{Overview of the human annotation process: After validating quality requirements, annotators assess all properties sequentially with early termination. (b)-(e) only show the right column of the annotation view, the left column always shows the claim and metadata like in (a).}
    \label{fig:human-annotation-process}
\end{figure*}

\subsection{Additional Results}
\label{app:human_eval_additional_results}

\begin{table}[]
    \centering
    \resizebox{\linewidth}{!}{
    \begin{tabular}{l|c|cc|cc}
        \toprule
        \textbf{Property} & $N$ & \multicolumn{2}{c}{\textbf{Error Rates (↓)}} & \multicolumn{2}{c}{\textbf{Accuracy (↑)}} \\
        & & MSE & MAE & $7$-bin & $3$-bin \\
        \midrule
        Authenticity      & 173 & 0.285 & 0.371 & 37.0 & 72.8 \\
        Contextualization & 173 & 0.121 & 0.152 & 74.0 & 91.3 \\
        Veracity          & 138 & 0.021 & 0.072 & 74.6 & 97.8 \\
        Context Coverage  & 116 & 0.030 & 0.085 & 77.6 & 99.1 \\
        Integrity         & 204 & 0.034 & 0.102 & 69.1 & 97.5 \\
        \midrule
        All Properties    & 600 & 0.128 & 0.184 & 64.2 & 89.0 \\
        \bottomrule
    \end{tabular}
    }
    \caption{Human evaluation results over $N$ claims. The scores reflect agreement between human annotation and automated \method annotations. \textbf{MSE} = Mean Squared Error, \textbf{MAE} = Mean Absolute Error, \textbf{Accuracy} = share of matches of $n$-bin discretized scores (in \%).}
    \label{tab:human_eval_extra}
\end{table}

Tab.~\ref{tab:human_eval_extra} and Fig.~\ref{fig:human_eval_confusion} compare human annotation with automated annotation for all $5$ properties, including an aggregated statistic averaging all properties.

\subsection{Discussion of Human Evaluation Disagreements}

While error rates for \textit{Media Authenticity} and \textit{Media Contextualization} are relatively high, these do not directly translate to errors in the overall \textit{Integrity} assessment. Since \textit{Integrity} aggregates multiple properties, disagreements at the property level do not necessarily imply incorrect end-to-end judgments. To better understand these discrepancies, we manually analyzed cases where human annotations and \method predictions diverge.

\paragraph{Ambiguities in \textit{Authenticity} Judgments.}
\textit{Media Authenticity} exhibits numerous edge cases that make consistent annotation challenging for both humans and models. The boundary between \pristine and \fabricated media is often not clear-cut. For instance, benign modifications such as logos or watermarks are not considered manipulations, as they do not introduce misleading impressions. More challenging are staged scenes that are recorded without alteration: While the media itself is \pristine, correct contextualization requires revealing the artificial setup. Similarly, screenshots of user interfaces containing \fabricated images raise subtle questions. Under our definition, the digital environment is part of the real world, and thus such screenshots can be considered \pristine; however, this may be misleading when the claim's text refers to the fabricated content within the image rather than the screenshot itself. Related ambiguities arise for digital artwork, photographs of physical artwork, movies, or images containing textual overlays that convey false claims. These examples highlight that authenticity is inherently context-dependent and subject to interpretation. We leave further disambiguation to future work.

\paragraph{Cross-Property Interactions in \textit{Contextualization}.}
We further observe that errors in \textit{Media Contextualization} are often linked to interactions with textual claims. In cases where the media is only loosely related to the textual assertion (e.g., symbolic or illustrative images), the correct \textit{Contextualization} label is \nei. However, \method occasionally evaluates the textual claim during the \textit{Contextualization} step, effectively refuting it instead of focusing on the media. While this behavior increases the measured error rate for \textit{Contextualization}, it often still leads to a correct overall \textit{Integrity} assessment. This suggests that some apparent errors arise from deviations in task decomposition rather than failures in factual reasoning. We will address this issue in future iterations of \method.

\begin{figure*}
    \centering
    
    \begin{subfigure}[t]{0.48\textwidth}
        \centering
        \includegraphics[width=\textwidth]{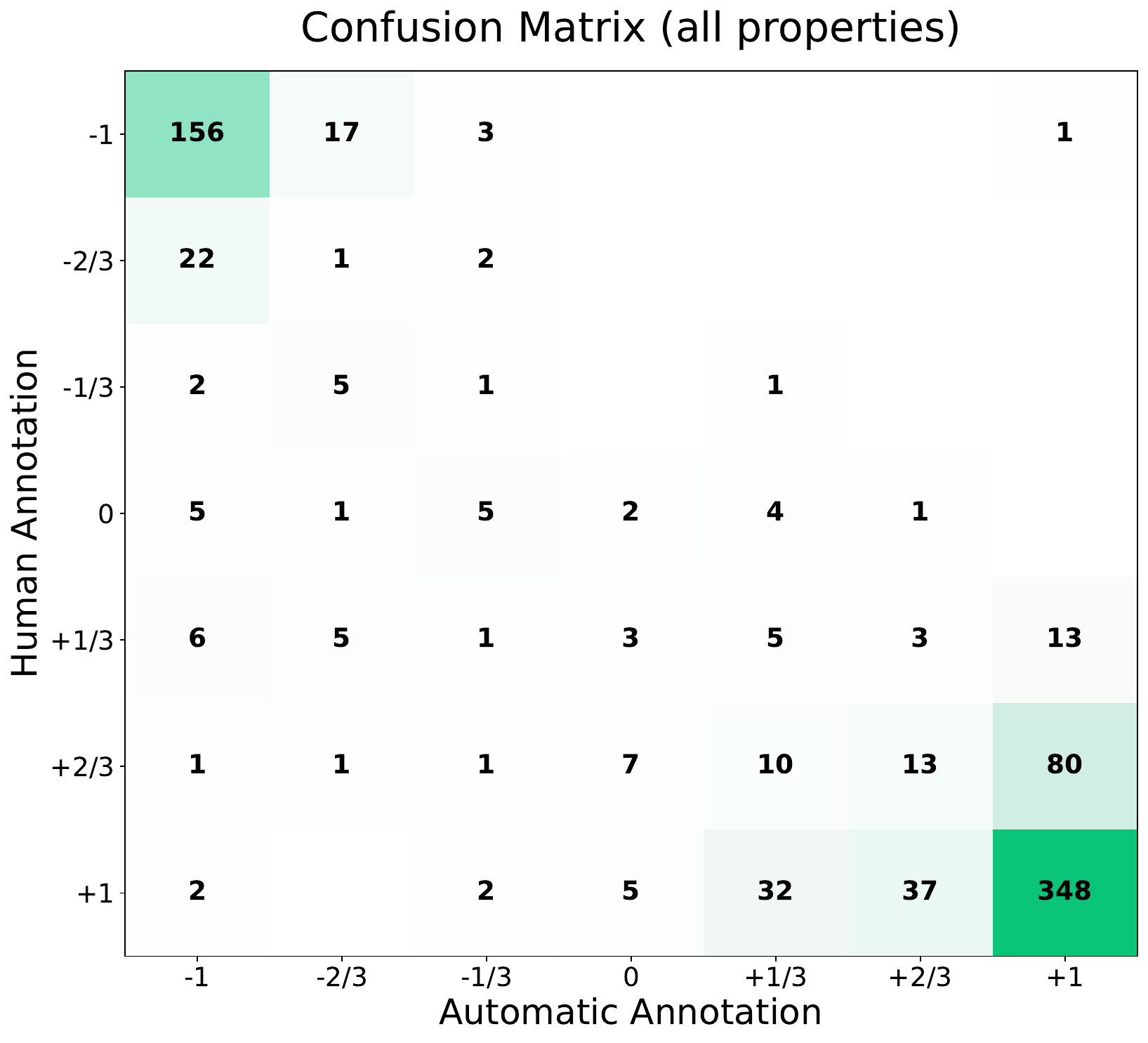}
    \end{subfigure}
    \hfill
    \begin{subfigure}[t]{0.48\textwidth}
        \centering
        \includegraphics[width=\textwidth]{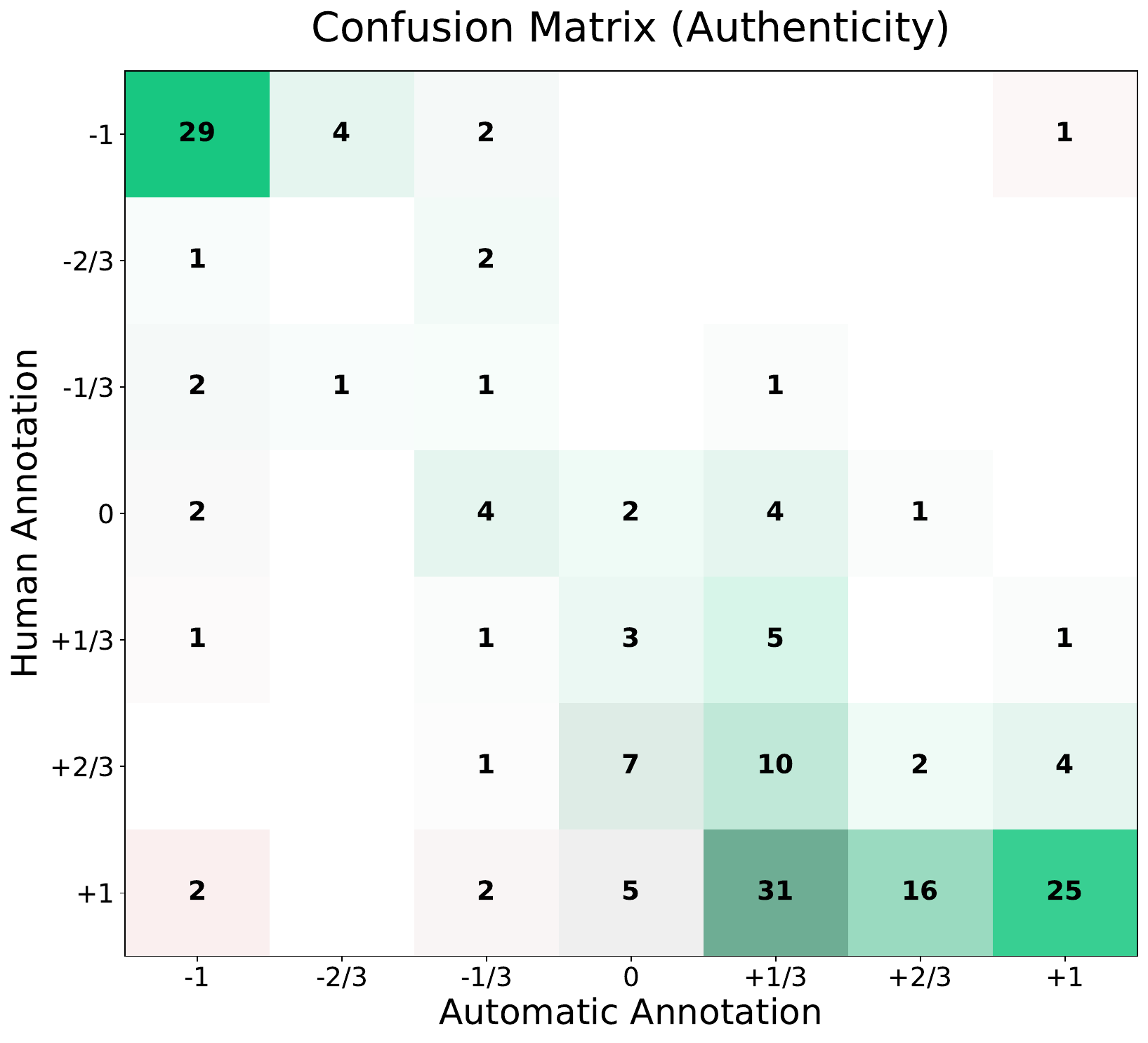}
    \end{subfigure}\vspace{0.5em}
    
    \begin{subfigure}[t]{0.48\textwidth}
        \centering
        \includegraphics[width=\textwidth]{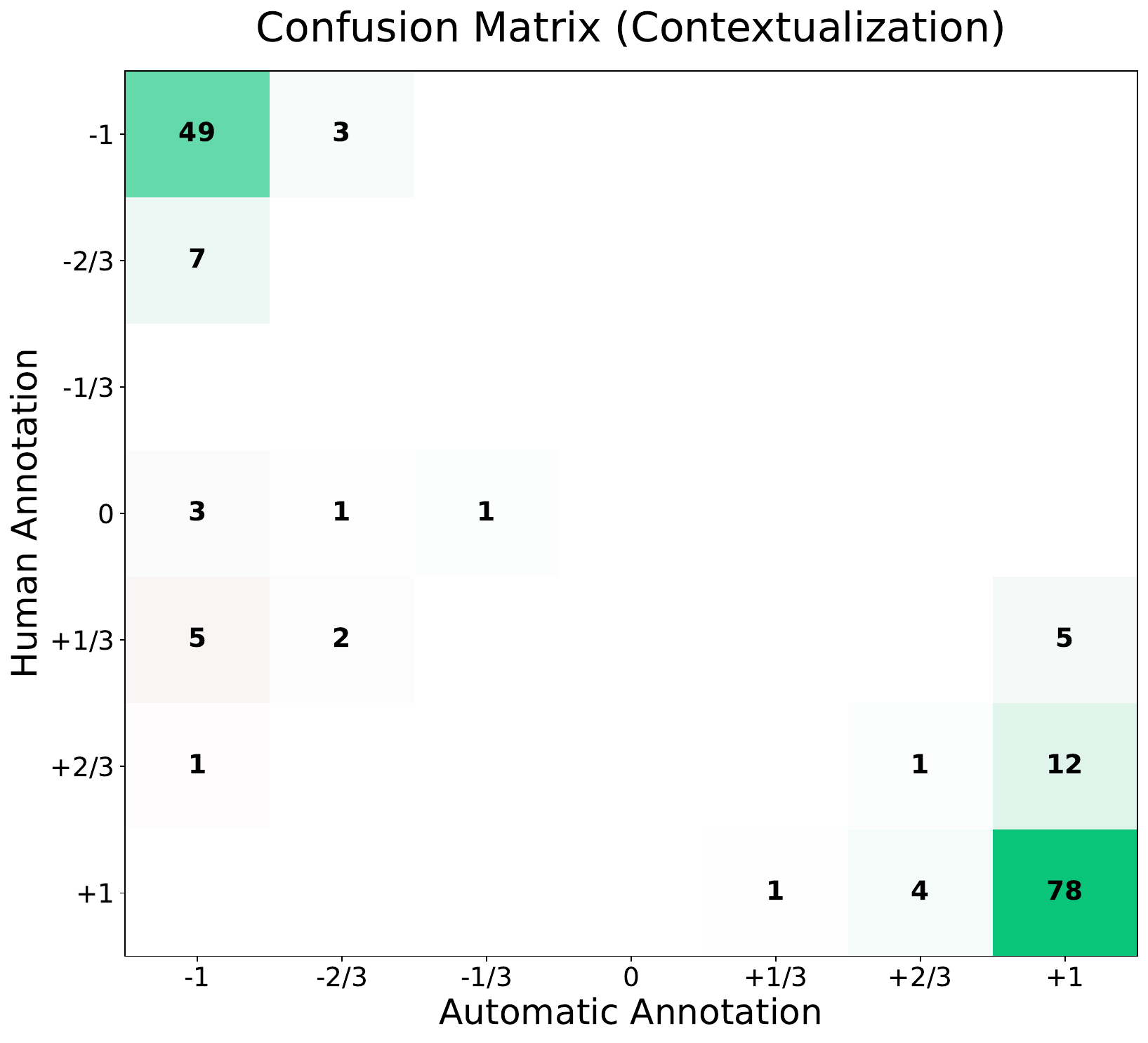}
    \end{subfigure}
    \hfill
    \begin{subfigure}[t]{0.48\textwidth}
        \centering
        \includegraphics[width=\textwidth]{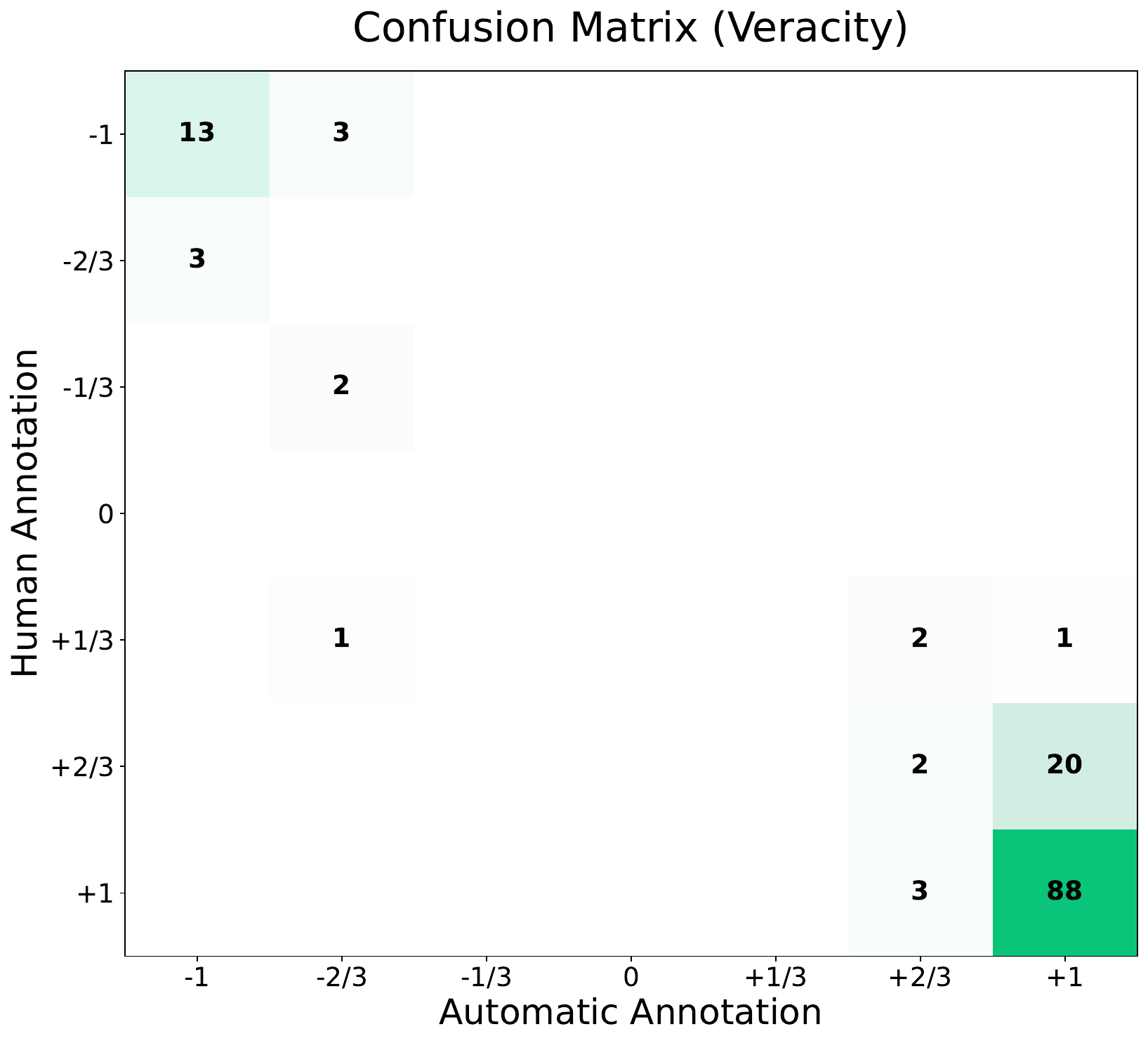}
    \end{subfigure}\vspace{0.5em}
    
    \begin{subfigure}[t]{0.48\textwidth}
        \centering
        \includegraphics[width=\textwidth]{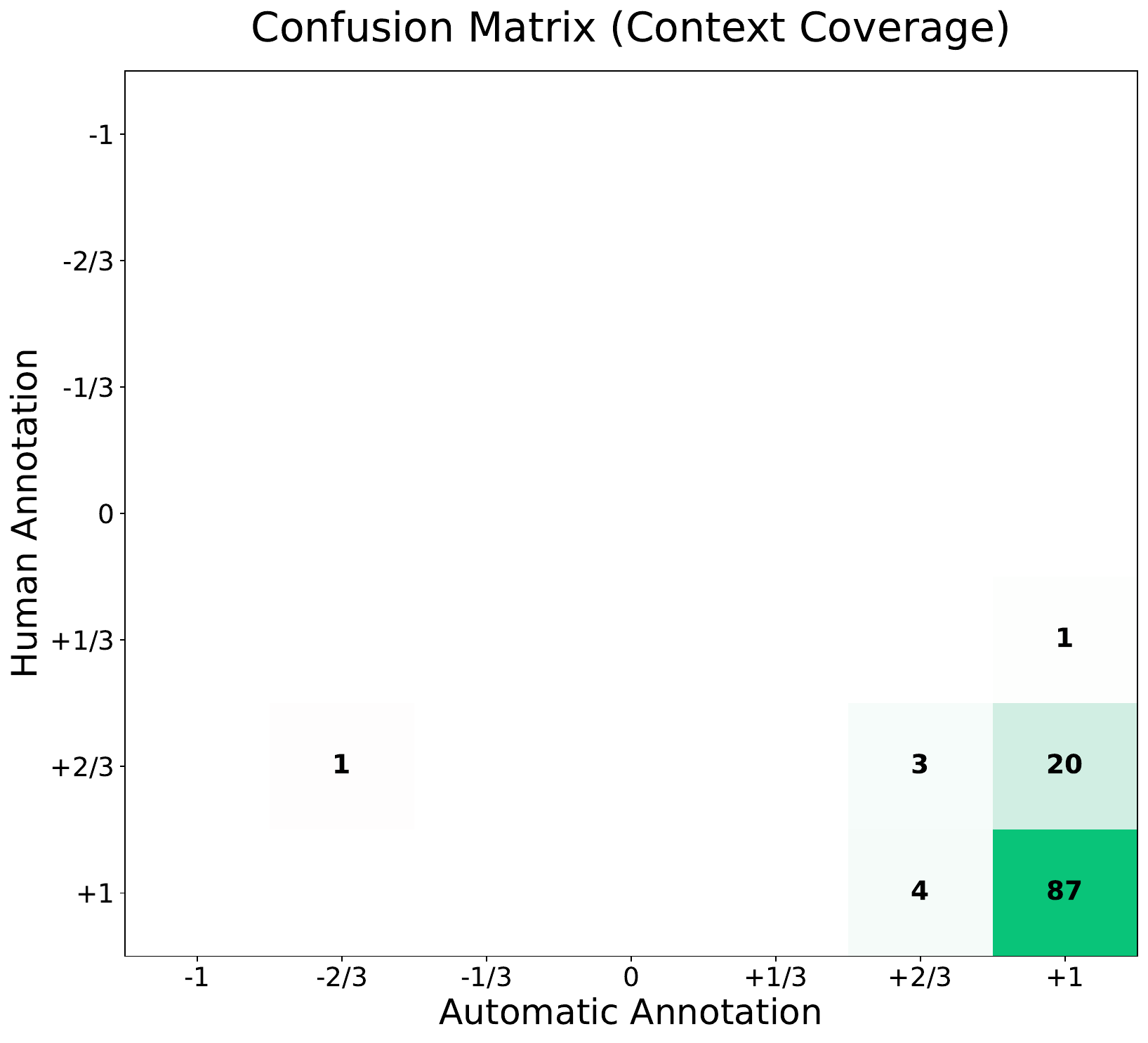}
    \end{subfigure}
    \hfill
    \begin{subfigure}[t]{0.48\textwidth}
        \centering
        \includegraphics[width=\textwidth]{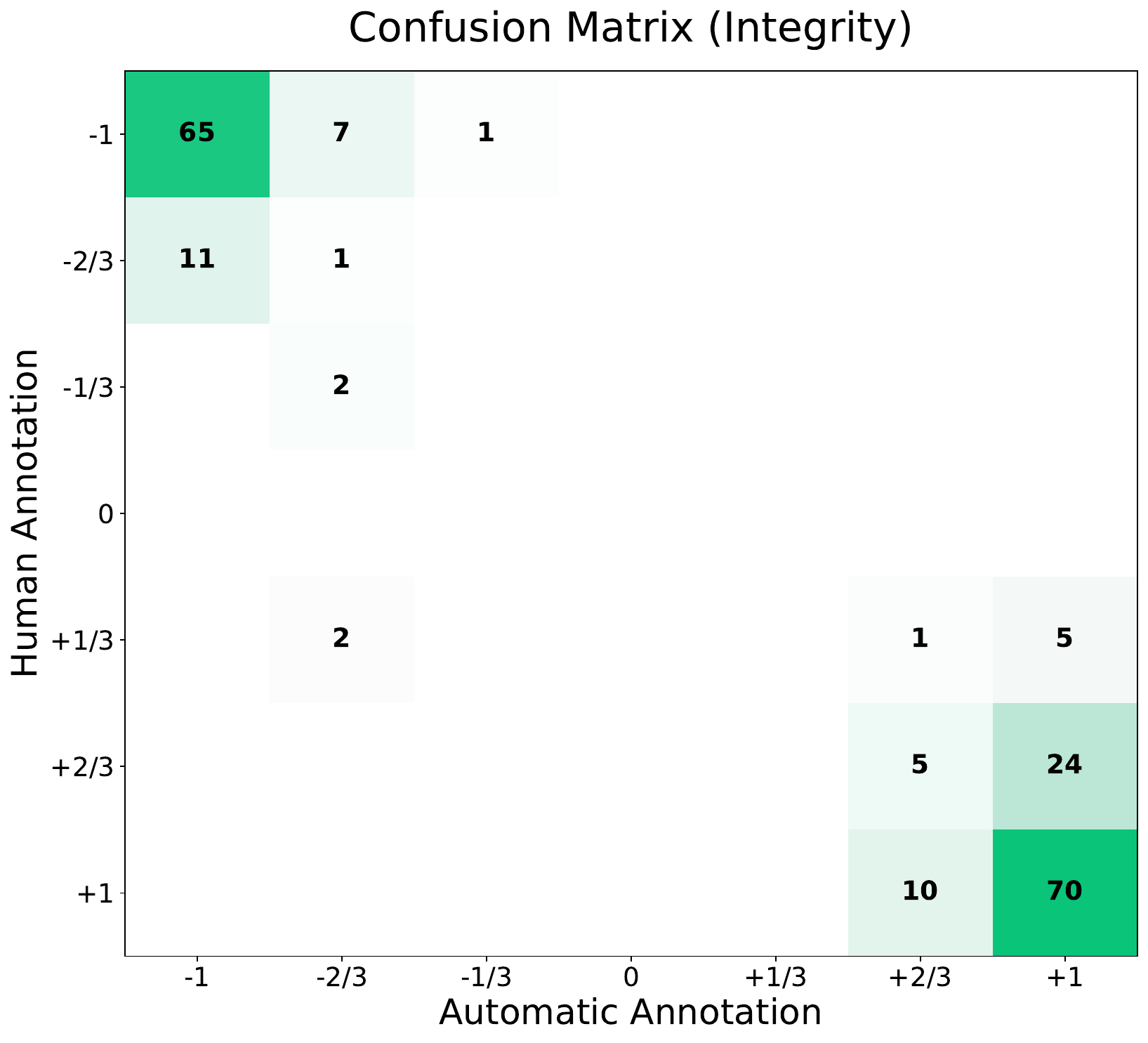}
    \end{subfigure}\vspace{0.5em}
    
    \caption{Confusion matrices comparing human annotations with \method' automated annotations.}
    \label{fig:human_eval_confusion}
\end{figure*}

\section{Claim Browser UI}
\label{app:claims_ui}
To facilitate exploration of the VeriTaS dataset, we provide a Claim Browser that allows admitted researchers to browse, filter and inspect claims through a graphical user interface, see Fig.~\ref{fig:claim-browser-ui}. It offers access to all gathered $104$\,K, not only the released, including also dismissed claims. Claims can be filtered along more than twenty dimensions, including free-text search over claim text, language, publication date range, verdict properties and validation status.

Selecting a claim opens a detailed view information on (1) the claim text, date, and associated media, (2) all model-individual label assignments with justifications and per-medium authenticity and contextualization scores, (3) appearances where the claim was observed, (4) link to the original fact-checker article, and (5) outcomes of the automated validation checks.

\begin{figure*}
    \centering
    \begin{subfigure}[t]{0.94\textwidth}
        \centering
        \includegraphics[width=\textwidth]{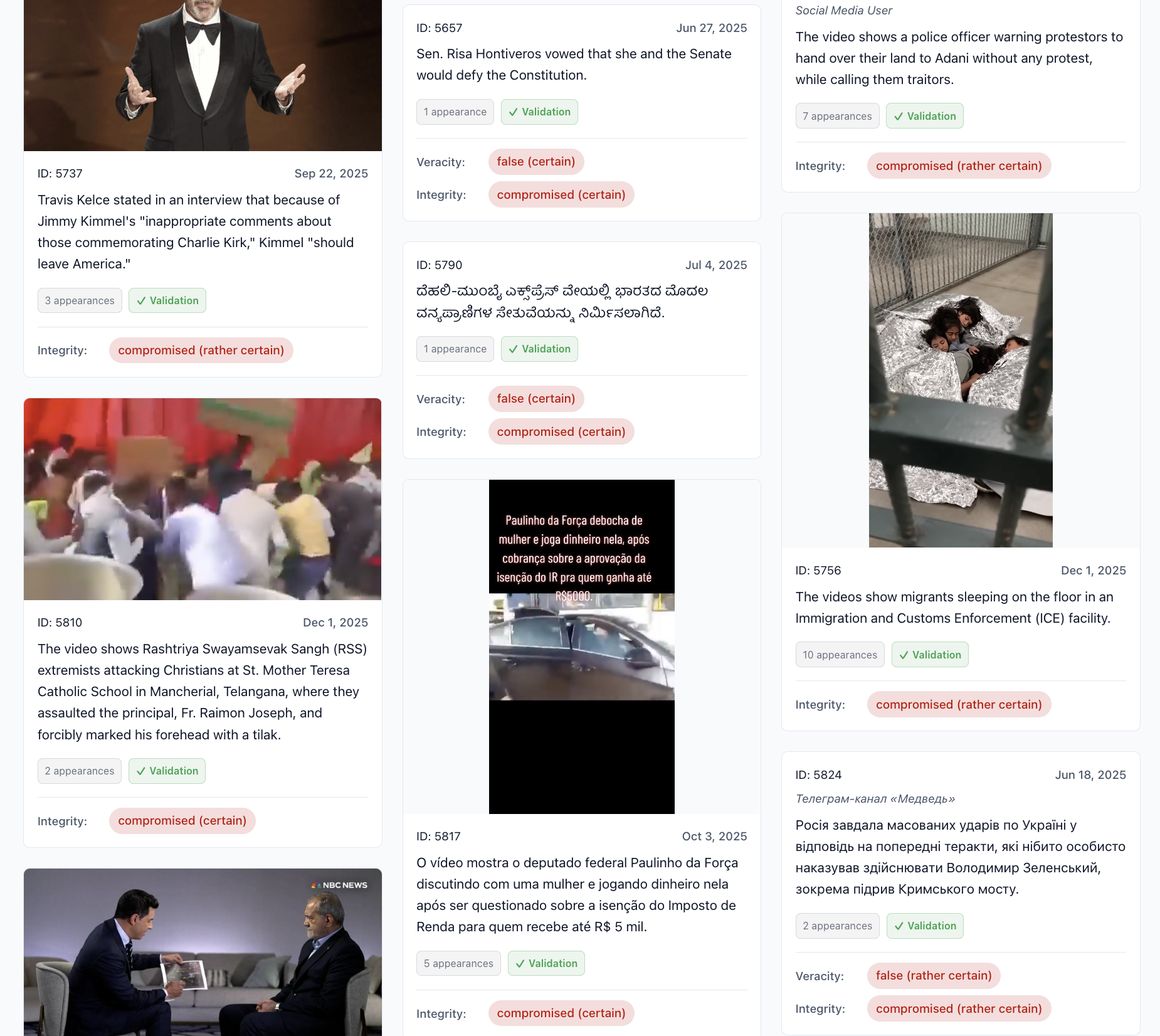}
        \caption{Screenshot of the Claim Browser UI}
    \end{subfigure}
    \begin{subfigure}[t]{0.94\textwidth}
        \centering
        \includegraphics[width=\textwidth]{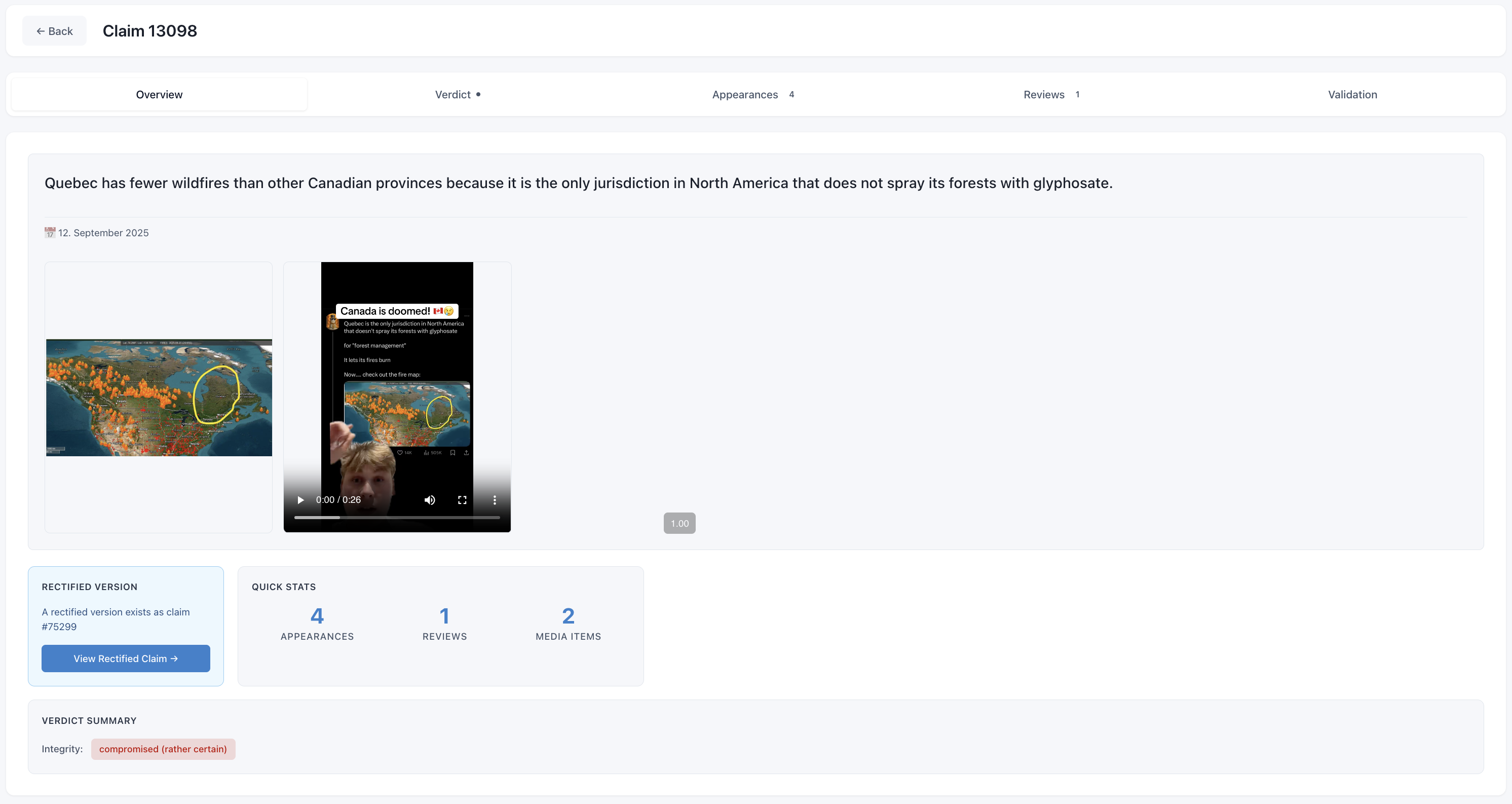}
        \caption{Screenshot of the detail-page of a claim from the Claim Browser UI}
    \end{subfigure}
    \caption{Screenshots from the Claim Browser UI showing the claim overview and details for a specific claim.}
    \label{fig:claim-browser-ui}
\end{figure*}

\end{document}